%% file: main.tex
\documentclass[a4paper,UKenglish,cleveref, autoref, thm-restate]{lipics-v2021}
\nolinenumbers
\usepackage{fancyvrb}
\usepackage{graphicx}
\usepackage{xspace}
\usepackage{enumerate}
\usepackage[inline]{enumitem}
\usepackage{xcolor}
\usepackage{pbox}
\usepackage{hyperref}
\usepackage{listings}
\usepackage{float}
\usepackage{subcaption}
\usepackage{amsmath}
\usepackage{cleveref}
\usepackage{tikz}
\usepackage{inconsolata}
\usetikzlibrary{trees}
\usepackage{algpseudocode}
\usepackage[linesnumbered,ruled,vlined]{algorithm2e}
\usepackage{textcomp}
\usepackage{makecell}
\lstdefinelanguage{PTX}{
    morekeywords={.version, .target, .address_size, .visible, .global, .align, .u32, .entry, .param, .u64, .reg, .pred, .f32, .b32, .b64, .s64, .f64, ld, cvta, to, global, bra, ret, not, max, add, mul, lo, s64, u64, selp, b64, mov, and, setp, lt, gt, eq, .pragma, .loc, .file},
    sensitive=false, 
    morestring=[b]" 
}
\lstdefinestyle{ptx}{
    language=PTX,
    basicstyle=\footnotesize\ttfamily,
    breaklines=true,
    frame=single,
    numbers=left,
    numberstyle=\tiny,
    tabsize=4,
    escapeinside={(*@}{@*)},
    literate={\_}{\textunderscore}1
            {\$}{\$}1
             {\#}{\#}1
             {\%}{\%}1
             {\^}{\^{}}1
             {\&}{\&}1
               {\_\_}{\_\_}1
             {\~}{\textasciitilde}1
             {\{}{\{}1
             {\}}{\}}1
   {0}{{0\allowbreak}}1 {1}{{1\allowbreak}}1 {2}{{2\allowbreak}}1
    {3}{{3\allowbreak}}1 {4}{{4\allowbreak}}1 {5}{{5\allowbreak}}1
    {6}{{6\allowbreak}}1 {7}{{7\allowbreak}}1 {8}{{8\allowbreak}}1
    {9}{{9\allowbreak}}1
    {a}{{a\allowbreak}}1 {b}{{b\allowbreak}}1 {c}{{c\allowbreak}}1
    {d}{{d\allowbreak}}1 {e}{{e\allowbreak}}1 {f}{{f\allowbreak}}1
    {g}{{g\allowbreak}}1 {h}{{h\allowbreak}}1 {i}{{i\allowbreak}}1
    {j}{{j\allowbreak}}1 {k}{{k\allowbreak}}1 {l}{{l\allowbreak}}1
    {m}{{m\allowbreak}}1 {n}{{n\allowbreak}}1 {o}{{o\allowbreak}}1
    {p}{{p\allowbreak}}1 {q}{{q\allowbreak}}1 {r}{{r\allowbreak}}1
    {s}{{s\allowbreak}}1 {t}{{t\allowbreak}}1 {u}{{u\allowbreak}}1
    {v}{{v\allowbreak}}1 {w}{{w\allowbreak}}1 {x}{{x\allowbreak}}1
    {y}{{y\allowbreak}}1 {z}{{z\allowbreak}}1
    {A}{{A\allowbreak}}1 {B}{{B\allowbreak}}1 {C}{{C\allowbreak}}1
    {D}{{D\allowbreak}}1 {E}{{E\allowbreak}}1 {F}{{F\allowbreak}}1
    {G}{{G\allowbreak}}1 {H}{{H\allowbreak}}1 {I}{{I\allowbreak}}1
    {J}{{J\allowbreak}}1 {K}{{K\allowbreak}}1 {L}{{L\allowbreak}}1
    {M}{{M\allowbreak}}1 {N}{{N\allowbreak}}1 {O}{{O\allowbreak}}1
    {P}{{P\allowbreak}}1 {Q}{{Q\allowbreak}}1 {R}{{R\allowbreak}}1
    {S}{{S\allowbreak}}1 {T}{{T\allowbreak}}1 {U}{{U\allowbreak}}1
    {V}{{V\allowbreak}}1 {W}{{W\allowbreak}}1 {X}{{X\allowbreak}}1
    {Y}{{Y\allowbreak}}1 {Z}{{Z\allowbreak}}1,
    postbreak=\mbox{\textcolor{red}{$\hookrightarrow$}\space},
}

\author{Jie Qiu}{Duolingo, USA}{jieq@berkeley.edu}{}{}
\author{Colin Cai}{University of California, Berkeley, USA}{cai9@berkeley.edu}{}{}
\author{Sahil Bhatia}{University of California, Berkeley, USA}{sahilbhatia@berkeley.edu}{}{}
\author{Niranjan Hasabnis}{Intel Labs, USA}{niranjan.hasabnis@intel.com}{}{}
\author{Sanjit A. Seshia}{University of California, Berkeley, USA}{sseshia@berkeley.edu}{}{}
\author{Alvin Cheung}{University of California, Berkeley, USA}{akcheung@berkeley.edu}{}{}

\Copyright{Jie Qiu, Colin Cai , Sahil Bhatia, Niranjan Hasabnis, Sanjit A. Seshia and Alvin Cheung}
\authorrunning{J.~Qiu, C.~Cai, S.~Bhatia, N.~Hasabnis, S.~A.~Seshia and A.~Cheung}
\relatedversion{}\relatedversiondetails{Full Version}{https://arxiv.org/abs/2404.18249}
\category{}
\hideLIPIcs

\acknowledgements{We would like to thank Jayaram Bobba and Zhongkai Zhang from Intel's Habana team for inputs on Gaudi architecture, \tpcc programming model, and obtaining high-performance from \tpckernel{}s. We would like to thank Hasan Genc and Sophia Shao for helpful insights into Gemmini code generation.
This work was supported in part by DARPA Contract FA8750-23-C-0080, a Google BAIR Commons project, 
NSF grants IIS-1955488, IIS-2027575, ARO W911NF2110339, ONR N00014-21-1-2724, and DOE award DE-SC0016260, DE-SC0021982, and the Sloan Foundation.
}

\EventEditors{John Q. Open and Joan R. Access}
\EventNoEds{2}
\EventLongTitle{42nd Conference on Very Important Topics (CVIT 2016)}
\EventShortTitle{CVIT 2016}
\EventAcronym{CVIT}
\EventYear{2016}
\EventDate{December 24--27, 2016}
\EventLocation{Little Whinging, United Kingdom}
\EventLogo{}
\SeriesVolume{42}
\ArticleNo{23}

\begin{document}
\title{Tenspiler: A Verified-Lifting-Based Compiler for Tensor Operations}

\newcommand{\compiler}{\textsc{Tenspiler}\xspace}
\newcommand{\ir}{\textsc{Tensir}\xspace}
\newcommand{\src}{{\tt S}}
\newcommand{\metalift}{\textsc{MetaLift}\xspace}
\newcommand{\ctotaco}{C2TACO\xspace}
\newcommand{\myregistered}{\textsuperscript{\textregistered}}
\newcommand{\inv}{$inv$}
\newcommand{\ps}{$PS$\xspace}
\newcommand{\tpckernel}{TPC kernel\xspace}
\newcommand{\tpcc}{TPC-C\xspace}
\newcommand{\gaudi}{Gaudi\xspace}
\newcommand{\gemmini}{Gemmini\xspace}
\newcommand{\mlx}{MLX\xspace}
\newcommand{\pytorch}{PyTorch\xspace}
\newcommand{\numpy}{NumPy\xspace}
\newcommand{\tensorflow}{TensorFlow\xspace}

\newif\ifcomments
\commentstrue
\ifcomments
    \providecommand{\sahil}[1]{{\protect\color{red}{\bf [sahil: #1]}}}
    \providecommand{\alvin}[1]{{\protect\color{purple}{\bf [alvin: #1]}}}
    \providecommand{\jie}[1]{{\protect\color{teal}{\bf [jie: #1]}}}
    \providecommand{\colin}[1]{{\protect\color{orange}{\bf [colin: #1]}}}
    \providecommand{\yuto}[1]{{\protect\color{blue}{\bf [yuto: #1]}}}
    \providecommand{\niranjan}[1]{{\protect\color{brown}{\bf [niranjan: #1]}}}
    \providecommand{\sanjit}[1]
    {{\protect\color{orange}{\bf [sanjit: #1]}}}
\else
    \providecommand{\sahil}[1]{}
    \providecommand{\alvin}[1]{}
    \providecommand{\jie}[1]{}
    \providecommand{\colin}[1]{}
    \providecommand{\yuto[1]}{}
    \providecommand{\niranjan[1]}{}
    \providecommand{\sanjit[1]}{}
\fi

\crefname{section}{Sec.}{Sec.}
\crefname{figure}{Fig.}{Fig.}
\Crefname{figure}{Fig.}{Fig.}
\crefname{table}{Tab.}{Tab.}
\Crefname{table}{Tab.}{Tab.}
\Crefname{equation}{Eq.}{Eq.}
\crefname{equation}{Eq.}{Eq.}
\crefname{appendix}{Appendix}{Appendix}

\maketitle

\begin{abstract}
Tensor processing infrastructures such as deep learning frameworks and specialized hardware accelerators have revolutionized how computationally intensive code from domains such as deep learning and image processing is executed and optimized. These infrastructures provide powerful and expressive abstractions while ensuring high performance. However, to utilize them, code must be written specifically using the APIs / ISAs of such software frameworks or hardware accelerators. Importantly, given the fast pace of innovation in these domains, code written today quickly becomes legacy as new frameworks and accelerators are developed, and migrating such legacy code manually is a considerable effort. 

To enable developers in leveraging such DSLs while preserving their current programming paradigm, we present \compiler, a verified-lifting-based compiler that uses program synthesis to translate sequential programs written in general-purpose programming languages (e.g., C++ or Python code that does not leverage any specialized framework or accelerator) into tensor operations. Central to \compiler is our carefully crafted yet simple intermediate language, named \ir, that expresses tensor operations. \ir enables efficient lifting, verification, and code generation. Unlike classical pattern-matching-based compilers, \compiler uses program synthesis to translate input code into \ir, which is then compiled to the target API / ISA. Currently, \compiler already supports \textbf{six} DSLs, spanning a broad spectrum of software and hardware environments. Furthermore, we show that new backends can be easily supported by \compiler by adding simple pattern-matching rules for \ir. Using 10 real-world code benchmark suites, our experimental evaluation shows that by translating code to be executed on 6 different software frameworks and hardware devices, \compiler offers on average \textbf{105$\times$} kernel and \textbf{9.65$\times$} end-to-end execution time improvement over the fully-optimized sequential implementation of the same benchmarks.  

\end{abstract}

\input{sections/introduction}
\input{sections/overview}
\input{sections/ir}

\input{sections/approach}
\input{sections/synthopt}
\input{sections/experiments}

\input{sections/related}
\input{sections/conclusions}

\bibliographystyle{plainurl}
\bibliography{paper}
\newpage
\input{sections/appendix}
\end{document}

%% file: sections/introduction.tex
\section{Introduction}
\label{sec:intro}
We have witnessed an explosion of new computational infrastructures for tensor computation in recent years: from software frameworks such as \tensorflow to specialized hardware accelerators like tensor processing units. Such infrastructures arise due to new application domains such as image processing and training deep learning (DL) models, and they often expose their functionality via various domain-specific languages (DSLs) that range from specialized instruction sets such as vectorized instructions to high-level programming interfaces such as Apple's \mlx~\cite{mlx} or \tensorflow's XLA~\cite{xla}.

To leverage the optimization offered by such infrastructures, applications must be written against the provided programming interfaces: developers must first master each DSL's programming model to write new code, and existing applications must be rewritten. This problem is recurring as new DSLs keep appearing targeting different application domains. Manually rewriting existing applications is tedious and increases the likelihood of introducing bugs. The classical way of addressing such issues is to build transpilers~\cite{qbs,casper,dexter,stng,c2taco,ngst} that translate code from paradigms developers are familiar with (e.g., C++ code using the STL library) to the one provided by the target DSL (e.g., \numpy API). Nonetheless, building such a transpiler is resource-intensive, error-prone, and each one is specialized to a specific target DSL. For instance, existing compilers such as Dexter~\cite{dexter}, STNG~\cite{stng}, and C2TACO~\cite{c2taco} target specific DSLs like Halide and TACO, and are not easily extensible to support new operators or backends.  While recently developed DL models such as GPT have shown promise in code translation, they do not provide any guarantees on the correctness of output. Moreover, GPT fails to generate even syntactically correct code for DSLs it has not seen in training data, limiting its applicability to new or less popular DSLs.

In this paper, we describe a tensor compiler that addresses these challenges. We introduce \compiler---a compiler designed to automate the transpilation of code to {\em multiple} tensor processing frameworks and hardware accelerators. \compiler uses verified lifting~\cite{metalift} (VL), a technique using inductive program synthesis to infer provably equivalent program summaries expressed using a user-defined intermediate representation (IR), and generate executable code from the synthesized summary to the target DSL. In contrast to conventional compilers that rely on pattern-matching to compile code, VL uses a search-based technique for the translation process. The two key steps of VL are:
\begin{itemize} 
    \item \textbf{Search Phase:} This stage lifts the input code to an equivalent program written using a user-provided IR, where the IR is used to model the functionality of each operator in the target DSL. Lifting is formulated as a syntax-guided synthesis~\cite{sygus} problem.
    \item \textbf{Verification:} Once lifted, a theorem prover is used to validate if the synthesized summary is functionally equivalent to the input. If so, executable code is produced by calling the user-provided code generator from the summary; otherwise, another summary is generated by the search phase.
\end{itemize}

The key to making lifting efficient lies in the design of the IR (i.e., how the target DSL is modeled). In prior work~\cite{casper,dexter,rake,domino,katara}, each function or instruction exposed by the target DSL is modeled explicitly. While doing so makes the search efficient, such explicit modeling makes the compiler hard to extend to other DSLs. With \compiler, we introduce, for the first time, a {\em single unified} IR, \ir, that is designed for tensor operations and can easily generate code to {\em multiple} tensor processing software frameworks and hardware accelerators. Surprisingly, \ir is a small language based on tensor algebra that includes commonly used vector and matrix operations. While other unifying IR exists (e.g., MLIR~\cite{mlir}), they are targeted for classical pattern-matching compilers. As we will discuss in \cref{sec:synthopt} and \cref{sec:verif}, \ir instead is designed for synthesis-based compilers and thus aims to enable both efficient search and verification.

In summary, this paper makes the following contributions: 
\begin{enumerate}
    \item We describe the design of \ir for transpiling code to tensor processing DSLs. \ir is simple yet expressive enough to model the functionalities provided by different software frameworks and hardware accelerators, and enables efficient code transpilation using verified lifting, as detailed in~\cref{sec:ir}.
    \item Based on \ir, we devise various optimization techniques to make synthesis and verification tractable, and scale to real-world programs in~\cref{sec:synthopt}.
\item We implement \compiler, a verified lifting-driven transpiler built using \ir as the modeling language. We demonstrate the effectiveness of \compiler by using it to lift real-world code from 10 different suites to \textbf{6} different open-source and commercially-available tensor processing software frameworks and hardware accelerators. We illustrate the ease of constructing such transpilers by building one for \mlx, a new tensor processing framework that was released only four months ago, using less than \textbf{200} lines of code in \cref{sec:exps}.
\end{enumerate}

We have released Tenspiler's code on \url{https://github.com/tenspiler/tenspiler}.

%% file: sections/overview.tex
\section{Overview}
\label{sec:overview}

\begin{figure}[!ht]
\begin{subfigure}{\textwidth}
    \begin{lstlisting}[language=C++,basicstyle=\ttfamily\scriptsize, breaklines=true]
inline uint8_t screen_8x8 (uint8_t a, uint8_t b) { return a + b - (a * b) / 255; }
vector<vector<int>> screen_blend(vector<vector<int>> b, vector<vector<int>> a) {
  vector<vector<int>> out; int m = b.size(); int n = b[0].size();
  for (int row = 0; row < m; row++) {
    vector<int> r_v;
    for (int col = 0; col < n; col++) 
      r_v.push_back(screen_8x8(b(col, row), a(col, row)));
    out.push_back(r_v);}
  return out;}
\end{lstlisting}
\caption{Original Blend function in C++.}
\label{fig:unopt_code}
\end{subfigure}

\begin{subfigure}{\textwidth}
    \begin{lstlisting}[language=python,basicstyle=\ttfamily\scriptsize, breaklines=true]
def t_t(x, y, operation):
  if len(x) < 1 or len(x) != len(y): return []
  else: return [operation(x[0], y[0])] + t_t(x[1:], y[1:], operation)

def t_s(x, a, operation):
  if len(x) < 1: return []
  else: return [operation(x[0]), a] + t_s(x[1:], a, operation)
\end{lstlisting}
\caption{Operators in \ir. We represent {\tt tensor\_scalar} as {\tt t\_s} and {\tt tensor\_tensor} as {\tt t\_t}.}
\label{fig:semantic_func}
\end{subfigure}

\begin{subfigure}{\textwidth}
    \begin{lstlisting}[language=Python,basicstyle=\ttfamily\scriptsize, breaklines=true]
def inner_loop(row, col, b, a, r_v, out):
  return col >= 0 and col <= len(b[0]) and row >= 0 and row < len(b) and 
    r_v == t_t(t_t(b[row][:col], a[row][:col], +), 
               t_s(t_t(b[row][:col], a[row][:col], *), 255, /), -) and
    out == t_t(t_t(b[:row], a[:row], +), t_s(t_t(b[:row], a[:row], *), 255, /), -)

def outer_loop(row, col, b, a, row_vec, out):
  return row >= 0 and row < len(b) and 
    out == t_t(t_t(b[:row], a[:row], +), t_s(t_t(b[:row], a[:row], *), 255, /), -)
    \end{lstlisting}
\caption{Synthesized loop invariants.}
\label{fig:invariants}
\end{subfigure}

\begin{subfigure}{\textwidth}
    \begin{lstlisting}[language=Python,basicstyle=\ttfamily\scriptsize, breaklines=true]
def screen_blend(b, a): return b + a - b * a // 255 # NumPy/TensorFlow/PyTorch/MLX
uchar256 Screen8x8(uchar256 a, uchar256 b) { # TPC-C implementation for Gaudi
  uchar256 c = v_u8_mul_b(a, b) * v_reciprocal_fast_f32(255);
  uchar256 d = v_u8_add_b(a, v_u8_sub_b(b, c));
  return d; }
\end{lstlisting}
\caption{Generated executable code for different tensor processing DSL.}
\label{fig:codegen}
\end{subfigure}
\caption{End-to-End example of using \compiler to transpile code.}
\end{figure}

\compiler takes in C/C++ or Python code as input\footnote{\compiler currently supports a subset of the C/C++ and Python language (in particular it does not support code that uses pointers or objects, which we have not encountered such use in our benchmarks). It also expects any external libraries used in the input to be functionally modeled, which is how \compiler currently supports code that uses the {\tt STL::vector} library.} 
and transpiles it to a functionally equivalent program that leverages different software frameworks and hardware accelerators (details described in ~\cref{sec:backend}) for tensor computation. As mentioned, \compiler uses verified lifting to first translate the source program into \ir. Unlike traditional pattern-matching compilers, \compiler formulates code translation as a search for a program expressed in \ir that is provably semantic-equivalent to the input. Doing so avoids the need to devise pattern-matching rules and prove their correctness. To make the search scalable, instead of directly searching within the DSL exposed by each target, we designed a high-level IR called \ir that abstracts away the low-level implementation details of each DSL operator and captures only their semantics, unifying various DSLs into a common set of tensor operators. \compiler uses a program synthesizer (currently Rosette~\cite{rosette}, a synthesizer for finite domain theories) to lift the input code to \ir during the search phase. The synthesized output is then verified using a theorem prover (currently, an SMT solver, CVC5~\cite{cvc5}) for the unbounded domain. ``Unbounded domain'' means the verification is performed for all possible program states, not just a bounded set of states (e.g., all states where integers are represented using 8 bits) that Rosette considers during the synthesis phase. Once verified, \compiler then translates the \ir program to the concrete syntax of the target DSL using simple pattern-matching rules.

We illustrate \compiler using the example in~\cref{fig:unopt_code} as our \src{} (source), where \src{} implements blending, a common image processing operation. It lightens the base color by iterating over each pixel, implemented as a nested loop over all the {\tt row}s and {\tt col}s in the image. Our goal is to transpile this code to the target DSLs supported by \compiler as shown in \cref{fig:codegen}.

\compiler first translates the input code to \ir. To be discussed in~\cref{sec:approach}, \ir consists of several operators that model common tensor algebra operations, two of which are shown in \cref{fig:semantic_func}. The {\tt t\_t} function performs element-wise operations (one of $+, -, *, /, \%$) on tensors {\tt x} and {\tt y} and is defined recursively on each element. Meanwhile, {\tt t\_s} performs element-wise scalar operations on tensor {\tt x} using the scalar value {\tt a} and is similarly defined. Importantly, both operators are purely functional models of the tensor operations that lack implementation details that a specific target might leverage (e.g., tiling, vectorization, etc). The idea is that if \src{} can be expressed using only these operators via lifting, then the lifted program can be easily translated to the targeted backends. 

In \compiler, lifting is formulated as a Syntax-Guided Synthesis (SyGuS)~\cite{sygus} problem, where the goal is to synthesize a semantically equivalent program summary (\ps), represented as a sequence of operators from our \ir, with the input code as the specification. A search space (specified using grammar) describes the set of potential candidate programs for the given specification. An input program \src{} is semantically equivalent to the synthesized expression {\tt S'} if for all possible program inputs {\tt i}, {\tt \src(i) = S'(i)}. 

\compiler uses symbolic search to solve the synthesis problem. Symbolic search is typically implemented through enumerative or deductive search, and using constraint-solving approaches which often rely on domain-specific heuristics to scale. As a SyGuS problem, symbolic search is implemented as enumerating different expressions over a user-provided grammar, where the grammar encodes all possible combinations of operators in the target DSL up to a specified depth. However, as the depth increases, the number of choices grows exponentially, making the search intractable.
As we will discuss in~\cref{sec:synthopt}, \ir is designed to make synthesis scalable. For \src{}, the synthesis phase returns the following solution:
\begin{lstlisting}[language=python,basicstyle=\ttfamily\scriptsize, numbers=none, breaklines=true]
def lifted_program(b, a): return t_t(t_t(b, a, +), t_s(t_t(b, a, *), 255, /), -)
\end{lstlisting}

As \compiler's synthesizer currently can only reason about finite domains, all synthesized solutions are checked for full functional equivalence using an automated theorem prover. Since \src{} has loops, checking equivalence with the generated program on all inputs requires loop invariants. Such invariants are synthesized during the synthesis phase by constructing a grammar similar to the \ps grammar. For instance, for \src{}, the synthesis phase yields two loop invariants (one for each loop) alongside a \ps. As shown in~\cref{fig:invariants}, these loop invariants are not arbitrary; within the loop invariants, the output variable, \texttt{out}, is expressed as a combination of operators from the \ir that help prove the synthesized \ps. We will leverage this to improve synthesis efficiency, to be explained in \cref{sec:synthopt}

With the synthesized solution expressed in \ir, the final step is to translate it into the concrete syntax of the target DSL(s). In \compiler, this is done via simple pattern-matching rules. In \cref{fig:codegen}, we present the translated code for different supported DSLs. As we will discuss in \cref{sec:backend}, \ir is designed such that generating executable code is straightforward. In fact, the code generators for the different tensor processing infrastructures supported by \compiler are highly similar to each other, as we will discuss in~\cref{sec:exps}.

%% file: sections/ir.tex
\section{The {\mdseries\ir} Intermediate Representation}
\label{sec:ir}
We now discuss our intermediate representation, \ir, which plays a pivotal role in \compiler. While prior lifting-based compilers all use search to compile programs, their IRs are specialized for their use cases. Those IRs consist of operators from the target languages, describing their high-level semantics while avoiding low-level implementation details. For example, Casper~\cite{casper} was built to translate sequential Java to MapReduce programs. The MapReduce framework consists of several versions of {\tt map} that differ by their input types. However, Casper only defines one operator in its IR that models {\tt map}'s functionality, and decides on the implementation to use during code generation. While doing so makes synthesis tractable, it also makes the compiler inflexible as adding another target (e.g., a hardware accelerator that supports map over tensors) will require modeling its functionality, which may be incompatible with the existing {\tt map} from Casper's IR. 

To address this challenge, we designed a novel IR, \ir, by studying the DSLs provided by various software frameworks and hardware accelerators for tensor computation. \ir is rooted in tensor algebra and is designed for flexibility, allowing translation to both software (deep learning frameworks, vector processing libraries) and hardware environments (machine learning accelerators) as to be discussed in~\cref{sec:backend}. This flexibility enables developers to select which target to execute the translated code based on availability and specific performance requirements. Given the dynamic nature of tensor processing infrastructures, \ir can be modified easily in terms of both adding support for new tensor operators and new target backends. This is illustrated in~\cref{sec:exps}, where it only took \textbf{200} lines of code for \compiler to support Apple's recently introduced \mlx framework~\cite{mlx}.

\noindent\textit{Comparison with MLIR.} MLIR~\cite{mlir} is a compiler infrastructure that enables the representation and transformation of code at various levels of abstraction. The core idea behind MLIR is to provide a unified IR that can capture the semantics of the program at different levels of detail (dialects), from high-level abstractions down to low-level, target-specific instructions. Developers can use MLIR by progressively lowering the code through different dialects until it reaches a level suitable for the target hardware. While MLIR and its dialects offer a powerful infrastructure for progressively targeting multiple hardware backends, we found that the existing dialects do not fully support all the operators required for our use case. Independently, both the {\tt linalg} and {\tt tensor} dialects do not support all the operators \ir supports. For example, the $select$ operator, which is crucial for image processing kernels that apply operations conditioned on pixel values, is not supported by any MLIR dialects. Additionally, unifying these dialects can be challenging for developers. Instead of unifying, recent work such as mlirSynth~\cite{mlirsynth} has explored using program synthesis to translate between different MLIR dialects. In contrast, \ir is designed to be flexible and easily extensible. Developers can add new operators to \ir by simply describing their high-level semantics, and new backend support can be incorporated by defining simple pattern-matching rules. This approach allows developers to extend \ir without going into the intricacies of MLIR. Moreover, \ir can practically be compiled into different MLIR dialects, providing developers the flexibility to leverage the MLIR infrastructure if desired.

\subsection{Language Definition}
\label{sec:irDef}

The operators and grammar of \ir are shown in~\cref{fig:hardgram}. \ir operates on tensors\footnote{In the grammar, the $Tensor \; Literal$ refers to 1D or 2D tensors as we did not encounter higher dimensional tensors in our benchmarks.} and includes various operations. The core strength of \ir lies in {\tt tensorOp}, which forms the backbone of tensor operations, including a diverse range of manipulations on tensors, such as element-wise operations, tensor vector multiplication, and reductions. These are grouped into different categories: 
\begin{itemize}
\item {\tt tensor\_scalar} operations describe element-wise operations involving tensors and scalars, such as scalar multiplication of each element in a matrix.
\item {\tt tensor\_tensor} operations perform element-wise operations between two tensors, such as element-wise multiplication of two tensors.
\item Tensor reshaping such as {\tt transpose}.
\item {\tt tensor\_vec\_prod} operation denotes tensor-vector products, enabling operations like matrix-vector multiplication.\footnote{While we can also define a tensor-tensor product operator, we did not encounter such benchmarks in our evaluation and hence omitted it from \ir's grammar.} 
\item Tensor reductions such as {\tt reduce\_max} and {\tt reduce\_sum}, which focus on aggregating tensor values, with the former determining the maximum element and the latter computing the sum across specified dimensions. 
\end{itemize}

\input{figures/hardir}

The recursive nature of \ir's grammar allows tensor operations to be composed, facilitating the expression of complex algorithms encountered in source code. \ir extends its expressiveness beyond tensor operations by also including a control flow operator ({\tt controlflowOp}). It integrates control flow through the {\tt ite} operator, enabling conditional logic into tensor computation. This operator is crucial for translating real-world loopy programs that contain branches. 

\ir is notable not only for its diversity of operations but also for the granularity of each operator, which significantly enhances its utility in translating code. 
The fine-grained nature of operations, from basic element-wise computations to advanced tensor reductions and control flow constructs, allows the grammar to capture the diverse tensor computation present in the input. Moreover, the selected set of operations aligns with the core functionalities supported by most tensor processing infrastructures. This ensures that \ir can seamlessly integrate with various frameworks and accelerators, offering flexibility in supporting multiple target DSLs. This comprehensive yet concise grammar serves as a bridge between traditional loop-based programming paradigms and the highly parallelizable world of tensor computation, providing a clear and expressive language for describing mathematical operations on tensors.

Besides tensor operations, \ir also supports tensor accessing and manipulation: 
\begin{itemize}
\item {\tt take(t, n)} extracts {\tt n} elements from the beginning. 
\item {\tt tail(t, n)} returns all the elements in tensor {\tt t} after the first {\tt n} elements. 
\item {\tt slice(t, l, s, e)} extracts a contiguous sub-tensor from {\tt t} from indices {\tt s} (inclusive) to {\tt e} (exclusive) along dimension {\tt l}. 
\item {\tt size(t, l)} returns the size of tensor {\tt t} in dimension {\tt l}.
\end{itemize}
Such functions are used to express the loop invariants and program summaries in the synthesis phase, as we describe next.

%% file: figures/hardir.tex
\begin{figure}
\begin{eqnarray*}
p \in Op &:=& s_{comp} \; \mid t_{comp} \; \mid \; c_{control}  \\ 
t_{comp} \in tensorOp &:=& tensor\_scalar(t, l, o) \; \mid \; tensor\_tensor(t, t, o) \; \mid \; \\ && transpose(t) \; \mid \; tensor\_vec\_prod(t, t) \; \mid \; a  \\
s_{comp} \in scalarOp &:=& reduce\_max(t) \; \mid \; reduce\_sum(t) \\
o \in op &:=& + \; \mid \; - \; \mid \; / \; \mid \; * \; \mid \; \% \\
c_{control} \in controlflowOp &:=& ite(cond, i, i) \\
cond \in boolExpr &:=& i \; rop \; i \\
i \in inp &:=& l \; \mid \;  t \\
rop \in relOp  &:=&  > \; \mid \; < \; \mid \; == \; \mid \; \neg \\ 
a \in accessOp &:=& take(t, l) \; \mid \; tail(t, l) \; \mid \;  slice(t, l, l_1, l_2) \\
l &:=& Integer \; Literal \; \mid \; size(t, l) \; \mid \;  s_{comp} \\
t &:=& Tensor \; Literal \; \mid \;  t_{comp} \\
\end{eqnarray*}
\caption{\ir grammar.}
\label{fig:hardgram}
\end{figure} 

%% file: sections/approach.tex
\section{Transpiling Code Using \compiler}
\label{sec:approach}
\input{figures/framework}

As shown in~\cref{fig:framework}, \compiler is designed to translate a program in high-level languages, source (\src{}), into another program that leverages different tensor processing infrastructures. \compiler currently support a vector processing library (\numpy) for CPU execution, DL frameworks (\pytorch, \tensorflow, \mlx) for GPU execution, and ISAs for specialized hardware accelerators (\gaudi, \gemmini). \compiler is a verified-lifting-based compiler, leveraging search to find a program within the target domain. Instead of relying on traditional pattern-matching rules, \compiler translates source programs with a 3-phase workflow:
\begin{enumerate*}
    \item synthesis,
    \item verification, and
    \item backend code generation.
\end{enumerate*}
\compiler uses a single IR, \ir, to facilitate all 3 phases of its workflow. \ir is designed to include tensor processing operators common to all target backends. As shown in the figure, the synthesis phase takes in \src{} and generates a program summary expressed using \ir. Then, in the verification phase, \compiler verifies the generated summaries to ensure their semantic equivalence with \src. Finally, in the code generation phase, the \ir program is translated to the concrete syntax of the target DSL(s).

\subsection{Synthesis}
The objective of this phase is to search for a program expressed using the operators in \ir, and to ensure that the generated program is semantically equivalent to \src{}. We formulate the search as a SyGuS problem~\cite{sygus} characterized by three parameters:
\begin{enumerate*}
    \item the \textbf{specification} describing the property the synthesized \ir expression should satisfy,
    \item the \textbf{search space} that describes the space of possible solutions, 
    \item the \textbf{search algorithm} which searches for the candidate programs.
\end{enumerate*}

For \compiler, the specification is to find a functionally equivalent program to \src{}. Various methods exist to express this specification, such as using input-output examples, bounded-model checking, and verification conditions (VC)~\cite{hoare-logic}. 

In \compiler, we use VCs as the specification for the synthesis phase as it provides full guarantees (i.e., for all program states up to a bound, e.g., states where all integers are encoded using 8 bits) on the equivalence of \src{} and the translated program. VCs are logical expressions encoding the correctness properties of \src{}. 

Specifically, given a program $P$ with $vars$ representing all the variables appearing in $P$, and $pre$, $post$, $inv$ representing the pre-conditions, the post-condition, and the invariant(s), respectively, the VCs for a program with loops consist of the following clauses:
\begin{enumerate}
    \item \textbf{Initial Condition}: $\forall \; vars. \; pre(vars) \rightarrow inv(vars)$: loop invariants must hold before the loop begins its execution.
    \item \textbf{Loop Preservation}: $\forall \; vars, vars'. \;  inv(vars) \;\land\; P(vars, vars') \rightarrow inv(vars')$: if the invariant holds before a loop iteration, they should continue to hold after that iteration.
    \item \textbf{Post-Condition}: $\forall \; vars. \; inv(vars) \rightarrow post(vars)$: invariants should hold once the loop has completed its execution.
\end{enumerate}
There exist standard techniques for generating VCs from a given source program~\cite{wp}. In \compiler, the \ps and invariants in the VCs are generated as placeholders as \src{} is analyzed, with their bodies to be synthesized during the synthesis phase. 

Next, we define the search space for synthesis. This space outlines the potential solutions for both the \ps and invariant(s), describing the solutions that could potentially satisfy the VC. Expressed as a context-free grammar (CFG), the search space imposes syntactic constraints on the structure of the outputs. In \compiler, the goal is not to find any \ps or invariants but ones that represent the output variables in \src{} as some sequence of operators from \ir, expressed mathematically as:

\begin{equation}
  \forall \; o \in outputVars. \; o = p, \textrm{where} \; p \in Op \; \textrm{as defined in \cref{fig:hardgram}}.   
\end{equation}

This states that all return variables in \src{} should be expressed as a program from \ir{}. With the specification in the form of VCs, $p$ expressed using \ir, and the search space for the \ps and invariants, the synthesis problem can be formally defined as:

\begin{equation} 
    \exists \; inv_0, \; inv_1, \; \ldots, \; PS \in G. \; \forall \sigma. \; VC(S, \; inv_0, \; inv_1, \; ..., \; PS, \sigma)
    \label{eqn:sygus-full}
\end{equation}

The goal of synthesis is then to find expressions from the search space $G$ for \ps and \inv{}s such that, for all program states $\sigma$, they satisfy the VC. 

For \compiler, we use an off-the-shelf symbolic search engine, Rosette~\cite{rosette}, which uses constraint solving to address the synthesis problem. In a constraint-solving-based approach, the specification $\phi$ (i.e., VC) and the search space $G$ are encoded as a single formula, and an SMT solver is then utilized to find a model that satisfies the formula. As a constraint-based solver, increasing the number of constraints makes the problem more challenging. Given that $\phi$ is fixed for a particular benchmark, the design of the $G$ becomes crucial. In \cref{sec:synthopt}, we discuss how the design of \ir helps keep the grammar size reasonable and scales the synthesis process.

\subsection{Verification}
\label{sec:verif}
During the synthesis phase, as the \ps and loop invariant(s) are validated against the VC only for a bounded set of program states,\footnote{We are unaware of any SyGuS solvers that can validate against an unbounded set of program states efficiently, including state of the art solvers such as Z3 and CVC5.} it is essential to check their validity for all program states. \compiler uses an SMT solver to do so by negating the VC in program verification i.e., checking if $\;\neg VC(S, \; inv_1, \; inv_2, \; ..., \; PS, \; \sigma)$ 
is satisfiable for some $\sigma$. The placeholders in the VC are substituted with the synthesized bodies of \ps and \inv{}s. 
If the solver cannot find any such $\sigma$, then the generated \ps and \inv{}s are correct for all possible program states, thus proving \ps and \inv{}s hold for all program states. If a $\sigma$ is found, then \compiler will iterate back to the synthesis phase in search of another candidate expression.

Besides using SMT solvers for \cref{eqn:sygus-full}, \compiler also leverages SMT solvers' support of algebraic data types (ADT) to allow users to define common data structures such as lists and tuples. Internally, \compiler models tensors using the list data structure defined using ADTs. We use ADT's accessor and constructor functions to retrieve and create new tensors. All the tensor accessing functions like {\tt slice}, {\tt take} are modeled as recursive functions over the list data structure. Currently, while image processing kernels use integers and deep learning kernels operate over floats, we verify all the benchmarks using the theory of integers and reals, due to poor solver support for reasoning about floats.

Since the verification of loop invariants is undecidable in general, we define additional {\em axioms} for the operators in \ir to aid verification. These axioms describe the behavior of functions that cannot be automatically deduced by the solver. 
Identifying the axioms requires an understanding of the program's semantics and the properties that need verification. Such axioms describe simple attributes such as distributivity, associativity, and commutativity of the tensor operators. In \cref{fig:axioms}, we show an inductive axiom for the {\tt tensor\_scalar} operator which states that the result for a given index is determined by the product of the first element of the tensor and an integer, plus the result for the remaining sub-tensor up to that index. As shown, having tensors as first-class objects in \ir greatly simplifies the task of defining these properties. Instead of defining these properties using low-level SMT-LIB list data structures, \ir enables users to define them at the tensor level, abstracting away the low-level solver-related details. This high-level representation greatly simplifies the task of defining these properties and makes the axioms more readable and maintainable.

\begin{figure}
\centering
\begin{lstlisting}[language=python,basicstyle=\ttfamily\scriptsize, breaklines=true]
(assert (forall ((data (Tensor Int)) (a Int) (idx Int))
(=> (>= index 0) $\land$ (<= index len(data)) 
(= t_s(data[:idx],a,*) (+ [data[0]*a] t_s(data[1:idx], a, *))))))
\end{lstlisting}
\caption{Example of an inductive axiom for the {\tt tensor\_scalar} operator in \ir described using SMT-LIB. ``+'' corresponds to the concat operator.}
\label{fig:axioms}
\end{figure}

\subsection{Code Generation}
\label{sec:codegen}

\label{sec:backend}
After successfully verifying the synthesized \ir program, the final stage in \compiler's workflow is to translate the \ir program into the concrete syntax of a target DSL. \ir makes this easy as it inherently represents tensor operations supported by all the target DSLs, and code can be generated using simple syntax-driven rules that map \ir operators to their DSL-specific counterparts. 

To translate the \ir expression into an executable DSL program, our code generation step recursively processes each part of the \ir expression.  \cref{fig:codegen-example} illustrates a portion of the code generation function for \pytorch. The function maps \ir variables to their names (line 3), literals to their values (line 5), and function calls to their \pytorch equivalents based on function signatures (lines 6-15).

Consider the running example in~\cref{fig:unopt_code}, where the synthesized \ir solution is {\tt t\_t(t\_t(b, a, +), t\_s(t\_t(b, a, *), 255, /), -)}. This expression represents a {\tt t\_t} function call with the {\tt -} operator, which maps to {\tt torch.subtract} as shown in line 13. Next, the {\tt codegen} function is called recursively on the two arguments, {\tt t\_t(b, a, +)} and {\tt t\_s(t\_t(b, a, *), 255, /)}. This results in the final translated \pytorch expression as {\tt torch.subtract(torch.add(b, a), torch.divide(torch.add(b, a), 255))}.

To extend support for a new backend, one simply needs to replace the DSL operator names in lines 11, 15, 17, and 19. For example, in \mlx's codegen, {\tt torch.add} on line 11 would be replaced by {\tt mlx.core.add}.

\begin{figure}[!ht]
\begin{lstlisting}[language=python,basicstyle=\ttfamily\scriptsize, breaklines=true]
def codegen(expr: Expr):
  if isinstance(expr, Var):
      return expr.name()
  elif isinstance(expr, Lit):
      return expr.val()
  elif isinstance(expr, Call):
    f_name, args = expr.name(), expr.arguments()
    if f_name in {"t_t", "t_s"}:
      op = args[-1]
      if op == "+":
          return f"torch.add({codegen(args[0])},{codegen(args[1])})"
          #corresponding MLX return statement
          #return f"mlx.core.add({codegen(args[0])},{codegen(args[1])})"
      elif op == "-":
          return f"torch.subtract({codegen(args[0])},{codegen(args[1])})"
      elif op == "*":
          return f"torch.multiply({codegen(args[0])},{codegen(args[1])})"
      elif op == "/":
          return f"torch.divide({codegen(args[0])},{codegen(args[1])})"
      ...
\end{lstlisting}
\caption{Code generation for the element-wise add operator to different targets.}
\label{fig:codegen-example}
\vspace{-0.1in}
\end{figure}

This direct and syntactic translation simplifies the integration of new tensor-based target DSL into \compiler, as one would only need to add simple translation rules in the code generation process. For instance, we add support for \mlx by changing only 65 lines of code to an existing 200-line template, as its API closely follows that of \numpy. 

As a part of this work, we have implemented support for \textbf{six} different target DSLs in our code generator: \textbf{\numpy}, \textbf{\tensorflow}, \textbf{\pytorch}, \textbf{\mlx} (an ML framework for Apple silicon), \textbf{\tpcc} (C-based programming language for Intel's \gaudi processor), and \textbf{\gemmini} (an open-source neural network accelerator generator).\footnote{We provide further details of these DSLs in \cref{sec:app_back} in the Appendix.}

%% file: figures/framework.tex
\begin{figure}[!t]
\centering
\includegraphics[width=\linewidth]{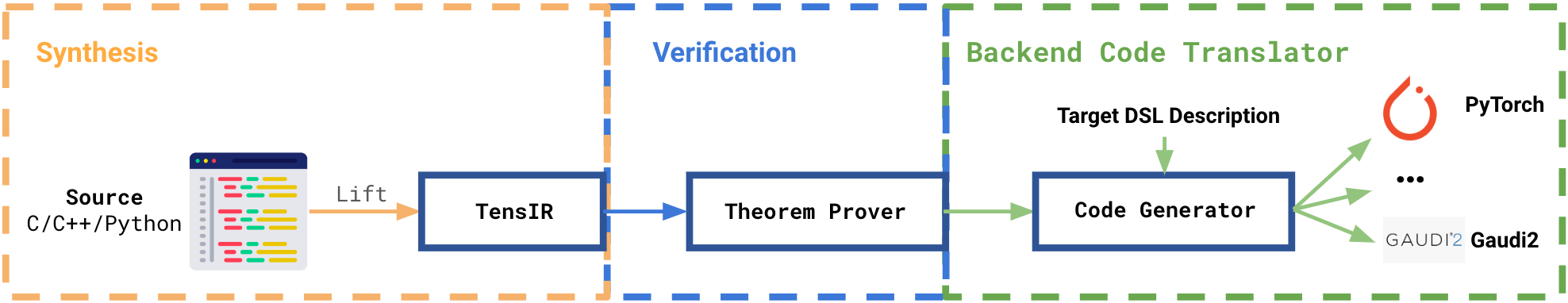}
\caption{An overview of the \compiler Framework.}
\label{fig:framework}
\end{figure}

%% file: sections/synthopt.tex
\section{Synthesis Optimizations}
\label{sec:synthopt}

\input{figures/depth5_general_grammar}

A naive approach to constructing the grammar for search space is to enumerate all possible combinations of \ir expressions up to a fixed depth. For \ir as defined in \cref{fig:hardgram}, if we focus solely on the compute operators, a depth-4 grammar (i.e., sequence of 4 operators) results in a search space of $\sim$200k expressions, since it grows exponentially with the depth and the number of operations. In \cref{fig:depth_5_general_grammar}, we show a small part of the depth-4 grammar. We have devised several optimizations to reduce the search space and make the search tractable.

\subsection{Restricting Operators} 
\label{sec:restrict_op}
First, we generate the grammar based on types, i.e., we only include the operators whose output types match with the expected return type. In the case of \src{} in \cref{fig:unopt_code}, since the return type is {\tt vector$\langle$vector$\langle$int$\rangle$$\rangle$}, all reduction operations are excluded. In \cref{fig:depth_5_general_grammar}, the operators in $v_4$ and $l_4$ will be removed (shown in \textcolor{red}{red}). These correspond to operators returning 1-D vectors and integers respectively.

\subsection{Restricting Program States} 
\label{sec:restrict_prog_state}
We further optimize the search space by restricting the set of program states in \cref{eqn:sygus-full}. Instead of satisfying the VC for all $\sigma$, we find \ps and \inv{}s that satisfy a bounded set. Bounded synthesis is crucial because most SyGuS solvers have limited support for recursive function definitions and require SMT solvers for validation. However, SMT solvers lack inherent support for reasoning about \ir operators that are not covered by the standard theories defined in SMT-LIB~\cite{smt-lib2} and require additional axioms to be defined. We instead integrate bounded synthesis by restricting the maximum unrollings of recursive operators, thereby eliminating the need for additional axioms. Specifically, we restrict the program states by limiting the lengths of the data structures and the sizes of the data types. For instance, in \compiler, we constrain all 1D tensors to length 2 and the integers to 6 bits or less for the first rounds of synthesis. In \cref{fig:depth_5_general_grammar}, all the tensor literals in $t_1$ are changed from an unbounded length ``n'' (shown in \textcolor{orange}{orange}) to length 2 (shown in \textcolor{blue}{blue}). If the synthesized choices fail to verify, we then increase the bounds in subsequent rounds. Note that since the synthesized solutions only work for a restricted set of program states, we invoke the theorem prover for subsequent verification to check if \ps and \inv{}s are valid for all states.

\subsection{Leveraging Expression Trees} 
\label{sec:leverage_expr_tree}
Despite the above two optimizations, the synthesis search space remains large. For example, in the context of \src{} in \cref{fig:unopt_code} for which we need to synthesize two \inv{}s and one \ps, a depth-4 grammar, after removing the reduction operations, still presents around 100k potential solutions just for \ps. \ir plays a significant role in the further pruning of this search space. The design of \ir operators effectively bridges the gap between high-level tensor operations and the loop-based paradigm commonly used for computing on tensors.  This property of \ir allows us to leverage the expression-tree-based filtering approach, which we describe next, to efficiently prune the synthesis search space. 

Our approach starts with the static analysis of \src{} to identify the computations performed; the static analysis pass emits an expression tree that represents the computation. For example, the pre-order traversal of the expression tree for \src{} from \cref{fig:unopt_code} is: \texttt{(\textcolor{blue}{-} (\textcolor{blue}{+} b a) (\textcolor{blue}{/} (\textcolor{blue}{*} b a) 255))}. In \cref{fig:depth_5_general_grammar}, this results in the pruning of {\tt tensor\_scalar} and {\tt ite(cond5, m4, m4)} at the top-level (shown in \textcolor{teal}{teal}) and similarly operators at other depths ($m_4, m_3$) are filtered. The generated expression tree is then transformed into an abstract expression tree, where variables and constants are replaced with placeholders, resulting in a synthesis template. 

The abstract expression tree for \src{} is then: \texttt{(\textcolor{blue}{-} (\textcolor{blue}{+} var var) (\textcolor{blue}{/} (\textcolor{blue}{*} var var) lit))}. This abstract expression tree guides \compiler in identifying the sequence of \ir operators. In this example, \compiler deduces the sequence of operators from the tree as: \texttt{t\_t(t\_t(var, var, +), t\_s(t\_t(var, var, *), lit, /), -)}, where {\tt var} and {\tt lit} are variables and literals to be synthesized, respectively.

Our expression trees are amenable to vectorized operations, which simultaneously perform the same computation on multiple data elements. Specifically, each level of the tree corresponds to an operation with the branches indicating data flow. In the example expression shown above, we see element-wise subtraction, addition, scalar division, and element-wise multiplication orchestrated such that it aligns with the vectorized execution of the original computation. 

This approach is not confined to specific operators but is adaptable to a range of operations in \ir. It can identify constructs like if-else blocks, where $ite$ arguments are determined using the same expression tree strategy, allowing the synthesis process to determine the optimal sequence of operators within the constructs. This flexibility extends to reduction operators and other complex operations, aiding in the synthesis of efficient operational sequences.

\subsection{Constraining Variables}
The final optimization is to 
pinpoint specific variables (\texttt{var}s such as {\tt a,b}  in~\cref{fig:unopt_code}) and literals (\texttt{lit}s such as {\tt 255} in~\cref{fig:unopt_code}) to be used in the grammar. Specifically, we constrain the variables to the set of live variables and also constrain constants to the set of constants that have appeared in the program. This strategy simplifies the computational task, avoiding the complexity of synthesizing a complete depth-4 operator sequence. By leveraging our expression tree-based approach, the search space reduces to 64 expressions, and the synthesizer promptly yields the correct solution within 76 secs: \texttt{t\_t(t\_t(b, a, +), t\_s(t\_t(b, a, *), 255, /), -)}.

\subsection{Overall Synthesis Algorithm}
The algorithm described in \cref{fig:synth} summarizes the synthesis phase in \compiler. This phase is used to search for the bodies of \ps and invariants which satisfy the VC. The synthesis is an iterative process conducted over multiple rounds, assuming a filtered search space leveraging type-based and expression tree optimizations described earlier. We start with the tensor bound size set to 2 which corresponds to restricting the program states optimization. In each round, we invoke Rosette's search algorithm (line 5) to generate candidates for \ps and \inv{}s. Upon obtaining a solution, the candidate undergoes validation against the VC for all program states, as the synthesis phase only checks within a constrained set of program states. We invoke a verifier (CVC5) (line 8) to perform this check. If the verifier yields ``UNSAT,'' the generated candidates are correct. Conversely, if ``SAT'' is returned, indicating incorrect candidates (line 11), the VC is augmented with blocking constraints. These constraints state that the generated \ps or \inv{}s in next round should differ from those in the previous rounds. This iterative process continues for a specified number of rounds (\texttt{max\_rounds}) before incrementing the tensor bound sizes. In cases where Rosette's search algorithm does not produce a solution initially (line 13), indicating an overly restrictive grammar, the initial grammar is expanded to include additional options for both \ps and invariants, such as choices for loop bounds, indexing, and operator sequences. We keep a separate timer (not shown in~\cref{fig:synth}) that maintains a maximum time bound for the entire synthesis process.

 \begin{figure}
    \begin{lstlisting}[language=python,basicstyle=\ttfamily\scriptsize, breaklines=true]
def synthesis_algorithm(spec, tensor_size_bound, holing_grammar, search_algorithm, verifier, max_rounds, timeout):
  r = 0 # rounds within one list bound
  #bounded synthesis optimization
  while r < max_rounds:
    ps_inv = search_algorithm(spec, holing_grammar) #rosette 
    if ps_inv is not None:  
      ps_r, inv_r = ps_inv
      if verifier(specification, ps_r, inv_r) == "UNSAT": return ps_r, inv_r
      else:
        spec = spec and (ps != ps_r) and (inv != inv_r) #add blocking constraint
        r += 1
    else:
      expand_holing_grammar(holing_grammar)
  # Increment tensor size bound
  return synthesis_algorithm(spec, tensor_size_bound + 1, holing_grammar, search_algorithm, verifier, max_rounds, timeout)

\end{lstlisting}
\caption{\compiler synthesis algorithm.}
\label{fig:synth}
\end{figure}

%% file: figures/depth5_general_grammar.tex
\begin{figure}
\begin{eqnarray*}
out &:=& m_4 \; \mid \; \textcolor{red}{v_4} \; \mid \; \textcolor{red}{l_4} \\
m_4 &:=& \textcolor{teal}
{tensor\_scalar(m_3,\; l_4,\; o)} \; \mid \; tensor\_tensor(m_3,\; m_3,\; o) \; \mid \; \\ && transpose(m_3) \; \mid \; \textcolor{teal}{ite(cond_4,\; m_3,\; m_3)}\; \mid \; m_3 \\
v_4 &:=& tensor\_scalar(v_3,\; l_4,\; o) \; \mid \; tensor\_tensor(v_3,\; v_3,\; o) \; \mid \; \\ && tensor\_vec\_prod(m_3, v_3) \; \mid \; ite(cond_4,\; v_3,\; v_3)\; \mid \; v_3 \\
l_4 &:=& reduce\_max(l_3) \; \mid \; reduce\_sum(l_3) \; \mid \; l_3 \\
cond_4 &:=& l_3\; rop\; l_3 \\
... \\
l_1 &:=& 255\; \mid \; size(t_1) \\
t_1 &:=& \color{orange}{a\langle a_1,\; a_2\; ...\; a_n \rangle}\; \mid \; \color{orange}{b\langle b_1,\; b_2\; ...\; b_n \rangle} \; \; \color{blue}{a\langle a_1,\; a_2\;\rangle}\; \; \mid \; \color{blue}{b\langle b_1,\; b_2 \rangle}    \\
rop &:=& > \; \mid \; < \; \mid \; == \; \mid \; \neg \\ 
o &:=& + \; \mid \; - \; \mid \; / \; \mid \; * \;
\end{eqnarray*}
\caption{A depth 4 general synthesis grammar for the source in \cref{fig:unopt_code}.}
\label{fig:depth_5_general_grammar}
\end{figure} 

%% file: sections/experiments.tex
\section{Experiments}
\label{sec:exps}
We evaluate \compiler's effectiveness in converting code into various tensor processing infrastructures using 10 loop-based real-world benchmark sets: \textbf{blend}, \textbf{Llama}~\cite{llamacpp},  \textbf{blas}, \textbf{darknet}, \textbf{dsp}, \textbf{dspstone}, \textbf{makespeare}, \textbf{mathfu}, \textbf{simpl\_array}, and \textbf{utdsp}. The \textbf{blend} benchmarks focus on image processing kernels, the \textbf{Llama} benchmarks contain traditional deep learning applications, and the rest 8 are all sourced from various existing software libraries, such as the BLAS linear algebra library and the TI signal processing library, and are used recently to evaluate C to TACO translations ~\cite{c2taco}. This combination of benchmarks allows for a comprehensive assessment of \compiler's effectiveness and adaptability across diverse domains and programming paradigms.

\begin{enumerate}[nosep,leftmargin=1.5em,labelwidth=*,align=left]
    
    \item Used in prior work~\cite{dexter}, the \textbf{blend} benchmarks consist of 12 functions dedicated to point-wise image blending operations --- a fundamental aspect of image processing known for diverse visual effects. These functions span 180 lines of C++ code, and 10 are characterized by doubly-nested loops, which are common in image processing algorithms. 
    \item \textbf{Llama} benchmarks are derived from the C++ based inference code of Llama2~\cite{llamacpp}, an open-source LLM from Meta. We labeled the portion of code to be lifted from the source code without doing any extensive syntax or code logic edits. These benchmarks include 11 functions capturing essential operations like computing activations, attention mechanisms, and layer norms. They total around 106 lines of code, with 2 functions incorporating doubly-nested loops.
    \item \textbf{blas}: 2 benchmarks from the BLAS~\cite{blas} linear algebra library.
    \item \textbf{darknet}: 10 neural network functions sourced from the Darknet~\cite{darknet} deep learning framework.
    \item \textbf{dsp}: 12 signal processing functions from the TI library~\cite{DSP}.
    \item \textbf{dspstone}: Kernels from the DSPStone suites~\cite{dspstone} that target digital signal architectures.
    \item \textbf{makespeare}: Programs originally from Rosin~\cite{makespeare} that manipulate integer arrays.
    \item \textbf{mathfu}: 11 mathematical functions extracted from the Mathfu library~\cite{mathfu}.
    \item \textbf{simpl\_array}: 5 functions for computations on integer arrays originally from prior work~\cite{simpl_array}.
    \item \textbf{utdsp}: Kernels from the UTDSP suite~\cite{utdsp} that targets digital signal architectures.

\end{enumerate}

\subsection{Evaluation Setup}
The synthesis and verification phases for all benchmarks are executed on a MacBook Pro 2 GHz Quad-Core Intel Core i5 Processor with a timeout of 1 hour. After lifting the code to \ir, the code generator, as explained in \cref{sec:backend}, generates executable code for each target backend. In the next section, we first describe the datasets used for evaluating the performance of lifted benchmarks. Then, we describe each target backend used for executing these benchmarks.

\subsubsection{Datasets for Evaluation}
\label{sec:execute}
To mimic real-world settings, we evaluate the translated code on actual datasets instead of generating random inputs.

For the \textbf{blend} image processing benchmarks, \textbf{blas}, \textbf{darknet}, \textbf{dsp}, \textbf{dspstone}, \textbf{makespeare}, \textbf{mathfu}, \textbf{simpl\_array}, and \textbf{utdsp}, we source images from ImageNet~\cite{imagenet}, a large-scale image dataset. We process these images as grayscale to ensure pixel values fall within the appropriate range. For benchmarks with 1-D tensor inputs, we flatten the images before feeding them as inputs and pass them in as they are for 2-D tensor inputs. For the blending layers in the blend benchmarks, we generate random pixel values from 0 to 255. We randomly select a set of 10,000 images from the dataset for evaluation.

For the \textbf{Llama} benchmarks, we evaluate the synthesized code using weights from Vicuna~\cite{vicuna}, an open-source LLM with similar model size as Llama2.\footnote{We did not use Llama2 model weights as they are not publicly available.} Some kernels operate over model weights, for which we directly use the weight matrices from Vicuna. For kernels operating over inputs, we simulate embeddings by creating random 32-bit float vectors within the range [0, 1). The evaluation primarily uses the 33B-parameter Vicuna model; however, for evaluating the \mlx framework, the 7B-parameter version is used due to memory limitations.

\subsubsection{Target Software Frameworks and Hardware Accelerators}
\label{sec:targets}
The core objective of \compiler is to translate sequential programs to a spectrum of diverse target DSLs, which can then be executed on conventional CPUs, GPUs, or specialized hardware accelerators. Although finding the optimal target DSL for the given input program would be an interesting feature in \compiler, currently \compiler is designed to provide users with the flexibility of choosing their preferred environment.

For our experimental evaluation, we choose 6 different target DSLs as we mentioned in \cref{sec:codegen}: \textbf{\numpy}, \textbf{\tensorflow}, \textbf{\pytorch}, \textbf{\mlx}, \textbf{TPC-C}, and \textbf{\gemmini}. We believe that our comprehensive selection of DSLs is necessary to test the robustness of \compiler. 

The \ir design greatly simplified this process as \numpy, \tensorflow, \pytorch, and \mlx have similar APIs, and each of these DSLs uses only 200 lines of code for generating executable code. 

To establish a baseline for execution performance, we compile C++ code for all the benchmarks using \texttt{gcc-8.3.0} with \texttt{-O3} flag and then run them on an Intel Xeon 8380 CPU. Given that each DSL is tailored to enhance performance on specific hardware, we conduct evaluation across five distinct platforms: the Intel Xeon 8380 CPU, Nvidia V100 GPU, Apple M1 Pro, Intel \gaudi2 processor, and the \gemmini accelerator.\footnote{Due to lack of physical device, \gemmini evaluations are done on a simulator with limited computing power and no file system support. Thus, it is compared with smaller random inputs.} In all, we utilized 7 different DSL-hardware device combinations for our experiments: (1) \numpy-CPU, (2) \tensorflow-V100, (3) \pytorch-V100, (4) \mlx-Apple M1 (5) \tpcc-\gaudi, (6) \pytorch-\gaudi, and (7) \gemmini. This comprehensive mapping of each backend to its corresponding device enables us to accurately assess the benefits \compiler provided through lifting.  \footnote{The exact configurations of all the hardware devices and their software environments are described in \cref{sec:hardware_config} in the Appendix.}.

\subsection{Synthesis Timings}
In this section, we present the time \compiler takes to synthesize and verify solutions for each of our benchmarks. During the synthesis phase, we apply all the optimizations mentioned in \cref{sec:synthopt} with a 1 hour timeout. \compiler synthesizes the correct translations for all the benchmarks under \textbf{15 mins}.

\input{figures/synthesis_timings}
\cref{fig:overall_synth_time} illustrates the synthesis performance for our 10 benchmark suites.
\cref{fig:blend_synth_time} illustrates the synthesis performance for the \textbf{blend} benchmarks. All but three benchmarks are synthesized in one synthesis (and verification) iteration, with an average synthesis time of 40.7 seconds and a median synthesis time of 2.357 seconds. Single-loop benchmarks synthesize with an average of 2.17 seconds and a median of 1.92 seconds. Double-loop benchmarks, on the other hand, have an average of 128.91 seconds and a median of 22.74 seconds for synthesis.

The three benchmarks that take more than one round of synthesis are {\tt softmax1}, {\tt transformer1}, and {\tt transformer2} from the \textbf{LLama} suite. {\tt softmax1} fails to synthesize within one round because the initial grammar is overly restrictive for its loop invariant. {\tt transformer1} and {\tt transformer2} involve complex indexing constraints for their invariants, initially leading to spurious solutions with tensor size limit of 2. {\tt transformer2} finds the correct solution after 6 tries, while {\tt transformer1} exhausts the maximum number of tries (10) with tensor size 2. We then increase the tensor size limit to 3 for {\tt transformer1} and a correct solution is generated within 3 rounds.

\subsubsection{Analysis}
We observe that synthesis difficulty is correlated to both the number of loops and the complexity of the \ir solution. For example, the {\tt dot} benchmark in the blas benchmark set has a single loop. Its \ir solution is {\tt reduce\_sum(t\_t(a, b, *))}, which has two operators and only two arguments, {\tt a} and {\tt b}, to be synthesized. This benchmark synthesizes in around two seconds. On the other hand, the {\tt transformer\_part1} benchmark from the \textbf{LLama} benchmark set has a doubly-nested loop and a complex solution with six operators. All arguments to these operators must be synthesized, with some requiring 3 expressions such as {\tt head * head\_size + head\_size} and {\tt sqrt(head\_size * 1)}. This benchmark takes around 1300 seconds to synthesize.

In addition to synthesis challenges, we recognize that the tree approach may restrict the ability of \compiler to synthesize different solutions. However, through manual verification, we confirm that \compiler consistently generates optimal solutions across our benchmarks. We evaluate the solutions using expression length as the cost function. Generally, shorter expressions mean fewer function calls and thus lower execution cost. To illustrate, we use the {\tt linear\_dodge} example from the \textbf{blend} benchmarks for which the synthesized solution is as follows:
\begin{lstlisting}[language=python,basicstyle=\ttfamily\scriptsize, numbers=none, breaklines=true]
def linear_dodge(a, b): t_t(a, b, +)
\end{lstlisting}
After relaxing the constraint on the tree structure, we were able to synthesize a different solution as shown below:
\begin{lstlisting}[language=python,basicstyle=\ttfamily\scriptsize, numbers=none, breaklines=true]
def linear_dodge(a, b): t_t(a, t_s(b, -1, *), -)
\end{lstlisting}
The latter solution is longer in length, and involves \textbf{2} tensor operations -- a {\tt tensor\_tensor} operation and a {\tt tensor\_scalar} operation -- as opposed to 1 {\tt tensor\_tensor} operation synthesized using the tree approach. Therefore, it is less optimized. In addition, the tree approach also speeds up the synthesis process as shorter expressions are easier to synthesize. 

\input{figures/timings}

\subsection{Performance Timings}
In this section, we evaluate the performance of the original input code—sequential C++ programs compiled with {\tt gcc -O3} —by comparing them with their translated versions executed across different target backends as detailed in Sections \ref{sec:targets} and \ref{sec:execute}.

\textbf{Kernel Performance.} Kernel performance focuses on computation time excluding data transfer overhead. We see significant improvements as illustrated in Figures \ref{fig:kernel_time}, with an average speedup of \textbf{105.1$\times$} across all benchmarks. Notably, the \gaudi2 processor demonstrates an exceptional speedup of \textbf{241$\times$}, highlighting the advantages of migrating legacy code to newer hardware platforms. For other backends such \tensorflow, \pytorch, \mlx and \numpy we see speedups of \textbf{46$\times$}, \textbf{244$\times$}, \textbf{10.5$\times$} and \textbf{26.12$\times$} respectively. 

However, compatibility issues can emerge with certain backends. For example, the \gemmini accelerator does not support certain operations in our \ir like {\tt tensor\_tensor} element-wise multiplication, {\tt slice}, and {\tt tail}. To address this, we only translate supported \ir operations from the synthesized \ps, and default to running the unsupported operations using sequential C on CPUs. Out of 69 benchmarks, 41 are translatable to \gemmini’s instruction set architecture (ISA), yet only 10 can be fully expressed using \gemmini instructions alone. Challenges are notable in benchmarks like {\tt screen\_blend} (\cref{fig:blend_kernel_time}), where element-wise vector multiplication must fallback to execution on a less powerful CPU. Furthermore, most \gemmini's instructions require square matrices inputs. This means that we need to pad vector inputs to square matrices before being able to utilize \gemmini's instructions, effectively squaring the data volume to be processed. This results in varied performance as shown in \cref{fig:llama_kernel_time} and \cref{fig:darknet_kernel_time}.

\input{figures/timing_e2e}
\textbf{End-to-end Performance.} While frameworks and accelerators deliver substantial kernel performance enhancements, a comprehensive assessment must account for end-to-end benchmark times, encompassing initial setup and data movement between the host (CPU) and the accelerator device. Our focus here is on data transfer (\tensorflow, \pytorch, and \gaudi processor) and memory management (C++). As illustrated in ~\cref{fig:e2e_performance}, we again observe an overall speedup, averaging \textbf{9.7$\times$}. In particular, CPU libraries like \numpy and \mlx show more improvements with the notable advantage of avoiding transferring data to specialized hardware. These benchmarks, involving the processing of 1D or 2D character vectors, benefit largely from C++'s efficient handling of contiguous data structures. Meanwhile, \gaudi2 drivers encounter performance bottlenecks due to the overheads associated with hardware initialization and frequent small data transfers. This significant upfront cost, especially pronounced in small-scale data operations, leads to a much less announced speedup. We believe such a phenomenon is uncommon in real-world use cases such as training deep learning models, due to techniques like batch processing or pipelining to minimize data transfers or to overlap computations with communications, thereby reducing or hiding transfer overhead and enhancing overall efficiency.

\begin{figure}[!ht]
\begin{subfigure}{\textwidth}
    \begin{lstlisting}[language=C++,basicstyle=\ttfamily\scriptsize, breaklines=true]
vector<float> matmul(vector<vector<float>> weight, vector<float> input) {
    vector<float> output;
    int m = weight.size();
    int n = input.size();
    for (int row = 0; row < m; row++) {
        float curr = 0;
        for (int col = 0; col < n; col++) {
            curr += weight[row][col] * input[col];}
        output.push_back(curr);}
    return output;}
\end{lstlisting}
\caption{Original {\tt matmul} function in C++.}
\label{fig:orig_code}
\end{subfigure}

\begin{subfigure}{\textwidth}
    \begin{lstlisting}[language=Python,basicstyle=\ttfamily\scriptsize, breaklines=true]
@cuda.jit()
def matmul (weight, input, res):
    m = len(weight)
    n = len(input)
    for i in range(m):
        curr = 0
        for j in range(n):
            curr += weight[i][j] * input[j]
        res[i] = curr
\end{lstlisting}
\caption{Numba kernel annotated version of {\tt matmul}.}
\label{fig:numba_code}
\end{subfigure}
\caption{Manually rewritten Numba example.}
\label{fig:orig_numba_rewrite}
\end{figure}
\textbf{Compare Against Pattern Matching-Based Compilers.} As outlined in \cref{sec:intro}, \compiler is designed to address the limitations inherent in traditional transpilers that rely on pattern matching to compile. Such compilers are resource-intensive to develop and prone to errors. To the best of our knowledge, no existing compiler matches the breadth of DSL support offered by \compiler. However, specialized compilers, such as Numba~\cite{Numba}, have been introduced for accelerators like GPUs. Numba leverages LLVM IR to generate GPU-accelerated code from Python code, making it a suitable candidate for comparison. 

For benchmarking purposes, we utilize the same datasets, test cases, and setup described previously in \cref{sec:exps}. Benchmarks are rewritten in Python and adapted to conform to CUDA kernel requirements by removing return statements, as shown in \cref{fig:orig_numba_rewrite}. Additionally, relevant data are cast to \numpy arrays as Numba focuses on optimizing code written against \numpy's API. These syntactic requirements represent a limitation of Numba's approach. In contrast, \compiler operates directly on the original benchmark implementations. 

Experimental results demonstrate that GPU-based \pytorch and \tensorflow code generated by \compiler performs, on average, 1.87$\times$ faster than code annotated with Numba. Remarkably, while Numba benefits from years of development by expert engineers, \compiler achieves superior performance with only \textbf{200} additional lines of code dedicated to code generation. A closer examination of the compiled PTX assembly code for the {\tt matmul} benchmark, which shows a 2.6$\times$ speedup, reveals that the Numba-generated code lacks the use of advanced instructions and techniques such as fused multiply-add (FMA), tiled-based computation models, or shared memory,\footnote{See \cref{sec:numba_ptx} in the Appendix for the PTX code.} which are crucial for peak performance. These techniques are standards in \pytorch and \tensorflow with optimized kernels. In contrast, Numba requires extensive manual tuning to implement, evident in the more complex and faster {\tt matmul} example in its documentation.\footnote{For the detailed example of an optimized {\tt matmul} function with shared memory for Numba, see {\tt https://numba.readthedocs.io/en/stable/cuda/examples.html\#id30}} \compiler, by automatically recognizing and translating matrix multiplication operations to leverage the pre-optimized kernels, avoids the complexities of manual code optimization while achieving high performance.

\subsection{Ablation Study}

\input{figures/ablation}

In our ablation study, we evaluate using our benchmark suites the effectiveness of the optimizations (described in \cref{sec:synthopt}) in making synthesis scale.

\noindent\textbf{Bounded Synthesis}. In this experiment, we keep the type-based filtering and tree approach while removing the incremental bounded synthesis optimizations. We start with a static tensor bound of 4 instead of the incremental approach. With this, \textbf{6} of the 12 \textbf{blend} benchmarks time out. In addition, benchmarks involving 2D tensors that do not time out see an average of \textbf{36.75$\times$} slowdown.

\noindent\textbf{Tree Approach}. For this experiment, we include type-based filtering and remove the expression tree approach for grammar filtering. We assume a fixed depth for the grammar, i.e., including all operators up to the specified depth, and increase it upon synthesis failure (starting at depth 1). Without static analysis, no assumptions are made about the operators, slice indices, variables, or constants, necessitating their synthesis. Unlike the tree approach with a fixed number of placeholders, this approach exhibits scalability issues as the number of grammar choices increases exponentially with depth. Therefore, only benchmarks with depth 1 and 2 expressions could be successfully synthesized. 

As illustrated in \cref{fig:ab_time}, without the tree based optimization, 
\textbf{42} out of the total \textbf{69} benchmarks timed out. In particular, all benchmarks of the \textbf{blend}, \textbf{blas}, \textbf{dspstone}, \textbf{makespeare}, and \textbf{utdsp} suites timed out. For the benchmarks that succeed, the synthesis phase slows down by an average \textbf{101.55$\times$} compared to the tree-based grammar filtering approach due to the need to synthesize additional expressions in both \ps and \inv{}s.

\noindent\subsection{Comparison with LLMs}
LLMs have shown promising results in various programming-related tasks, such as code generation, translation, and testing. However, these models suffer from a lack of formal verification of the translated code and face challenges in adapting to new DSLs or backends that are not well-represented in their training corpus. 

To test the capabilities of LLMs in generating code for new or low-resource DSLs, we prompt a state-of-the-art proprietary LLM Claude Opus~\cite{claude} (our evaluation using other LLMs such as GPT4 shows similar results). We selected three backends for this experiment: \mlx, a completely new DSL, \gaudi, and \gemmini, which are not well represented in the training corpus of these models. We prompted Claude Opus to generate code for these DSLs for the {\tt linear\_dodge} benchmark from the \textbf{blend} suite. The prompt instructions are shown in ~\cref{fig:prompt}. In~\cref{fig:llm_generated} we show the code generated by the LLM for the three prompts. 

Upon analysis, we found that all three generated programs were incorrect. The \gemmini-generated code in~\cref{fig:gemmini_code} partially uses the correct APIs ({\tt mvin}, {\tt mvouts}), but the computation with {\tt config\_ex} is incorrect. For the \gaudi-generated code in~\cref{fig:gaudi_code}, the model hallucinates the \tpcc library, which does not exist in the actual \gaudi programming model. The \mlx-generated code in~\cref{fig:mlx_code} has the correct call to the library function add, but the imports are incorrect, making the code non-functional. In addition to the generated code being incorrect, it is challenging to verify these outputs formally as syntactically translating the generated code to SMT-LIB is not trivial. The experiment highlights two significant challenges in generating verified code using LLMs mentioned earlier. In contrast, \compiler, which uses a verified lifting-based approach, can easily handle these challenges. LLMs cannot be directly prompted to generate code in new DSLs. LLMs could potentially be fine-tuned or prompted with few-shot learning to generate code in an IR, which can then be utilized within the \compiler's framework for verification; however, we leave this as future work.
\input{figures/prompt}
\input{figures/generated_code}

\noindent\subsection{Extension to Higher-Dimensional Tensors}
Our benchmarks only involve 1D and 2D tensors, as most operations are performed on images (the \textbf{blend} benchmarks) and weight matrices (the \textbf{LLama} benchmarks). In this section, we demonstrate that \compiler can be extended to support higher-dimensional tensors with the generalizability of \ir and the synthesis optimizations discussed in \cref{sec:synthopt}. Specifically, we extend \ir to accommodate 3D tensors and all corresponding element-wise operations. Additionally, we adapt the operator restriction optimization (introduced in \cref{sec:restrict_op}) to apply to 3D tensors. When the source program returns a 3D tensor, our grammar is restricted to include only element-wise 3D tensor operations. We also retain the program state restriction optimization technique from \cref{sec:restrict_prog_state}. Furthermore, we extend our support to leverage expression trees performed on individual elements in tensors, as detailed in \cref{sec:leverage_expr_tree}, to guide the search for vectorized operations within 3D tensor spaces.

We evaluate \compiler's synthesis optimizations on artificial benchmarks involving 3D tensors. We create these benchmarks by combining random element-wise operations. The maximum depth of these benchmarks is chosen to be 5 to match that of all our existing real-world benchmarks, as described in \cref{sec:exps}. Results in \cref{fig:3d-synth-timing} show that the synthesis time grows linearly with the depths of the benchmarks. The depth-1 benchmark synthesizes the fastest in 6 seconds, while the depth-5 benchmark takes the longest, in 184 seconds.

The sharp increase in timing for depth 5 expressions in \cref{fig:3d-synth-timing} is due to the number of expressions we are synthesizing and their complexity. A benchmark with 3 loops involves synthesizing 3 invariants and 1 post-condition, each with expression sizes up to depth 5. Despite these challenges, we easily extend \compiler's optimizations and synthesize these benchmarks well within the 1 hour timeout. As future work, to further scale \compiler's synthesis algorithm for handling more complex benchmarks, we could explore strategies such as guiding the search process using machine learning techniques, implementing bottom-up synthesis starting with inner loops first, performing bounded synthesis with unrolled loops, and combining these approaches with \compiler's current synthesis optimizations.

\input{figures/3d_synth_timing}

%% file: figures/synthesis_timings.tex
\begin{figure}[!t]
\begin{subfigure}{0.68\textwidth}
\centering
\includegraphics[scale=0.20,trim=0cm 0cm 0cm 0cm, clip]{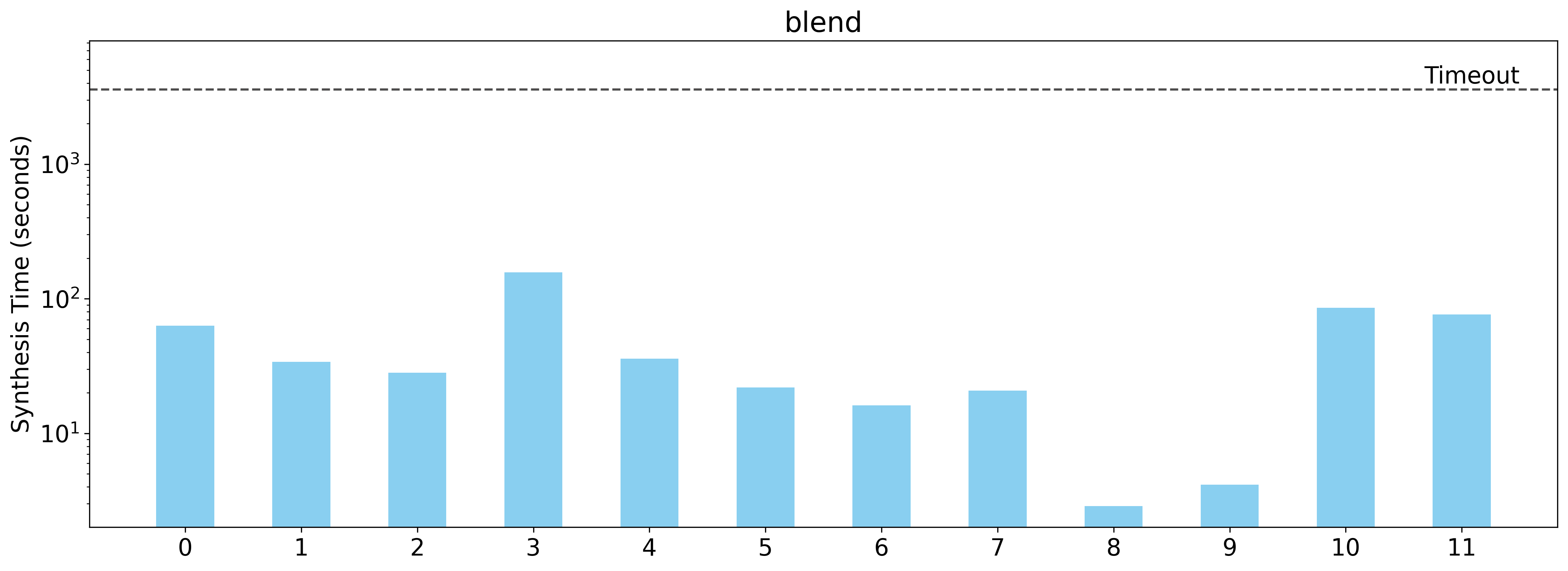}
\caption{\textbf{blend} benchmarks (image processing).}
\label{fig:blend_synth_time} 
\end{subfigure}
\begin{subfigure}{0.30\textwidth}
\centering
\includegraphics[scale=0.20,trim=0cm 0cm 0cm 0cm, clip]{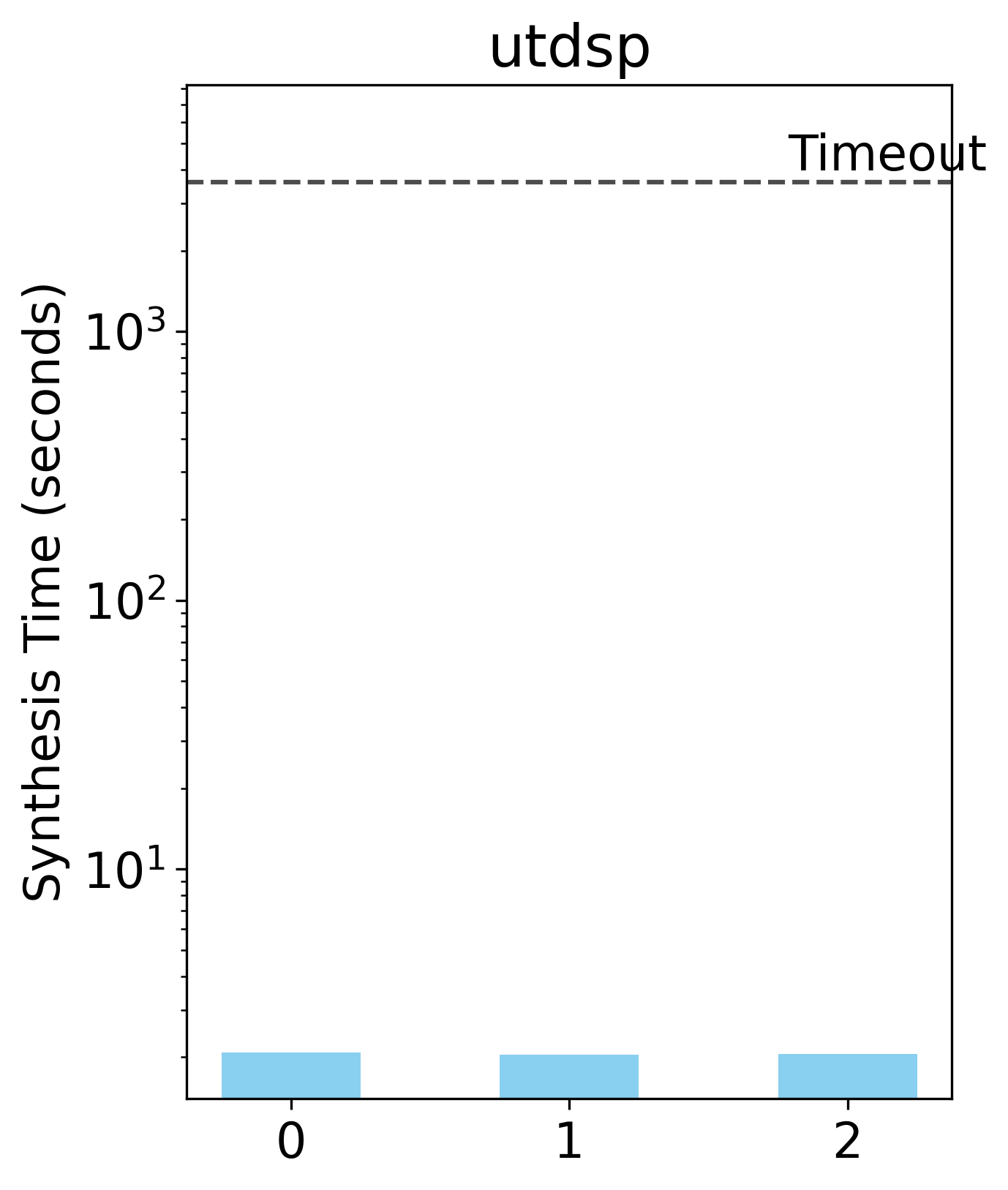}
\caption{\textbf{utdsp} benchmarks.}
\label{fig:utdsp_synth_time} 
\end{subfigure}

\begin{subfigure}{0.68\textwidth}
\centering
\includegraphics[scale=0.2,trim=0cm 0cm 0cm 0cm, clip]{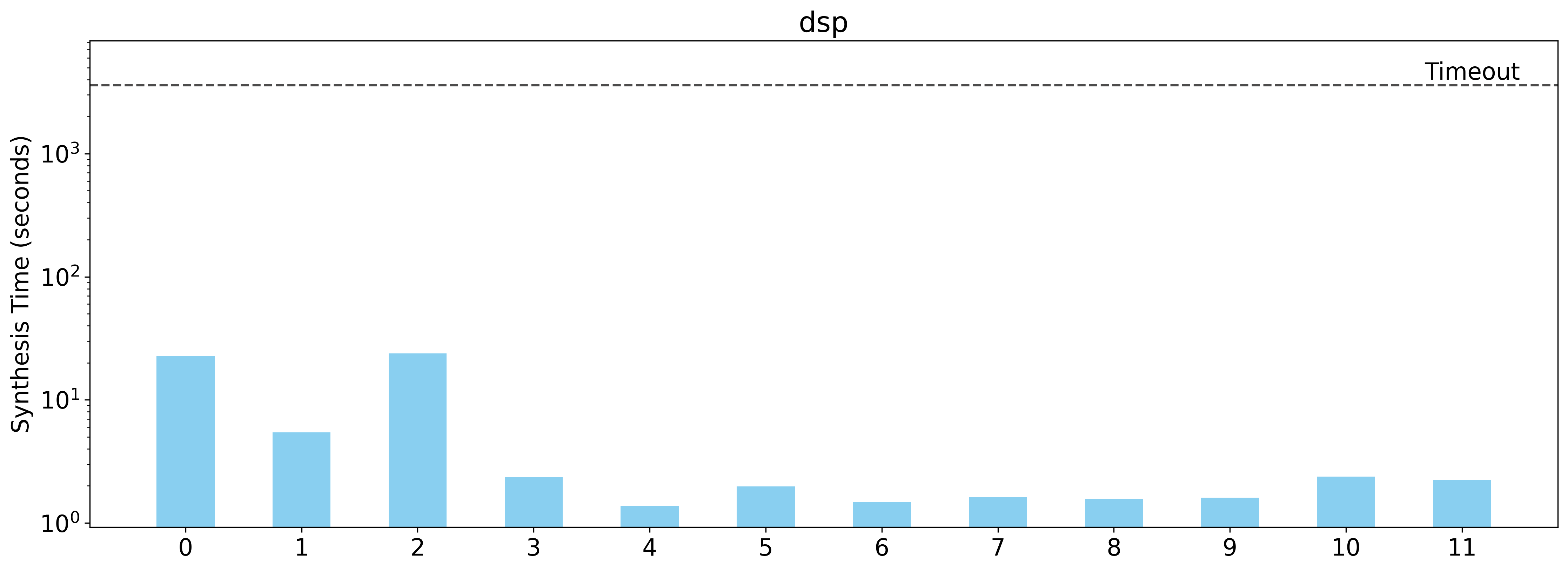}
\caption{\textbf{DSP} benchmarks.}
\label{fig:dsp_time} 
\end{subfigure}
\begin{subfigure}{0.30\textwidth}
\centering
\includegraphics[scale=0.2,trim=0cm 0cm 0cm 0cm, clip]{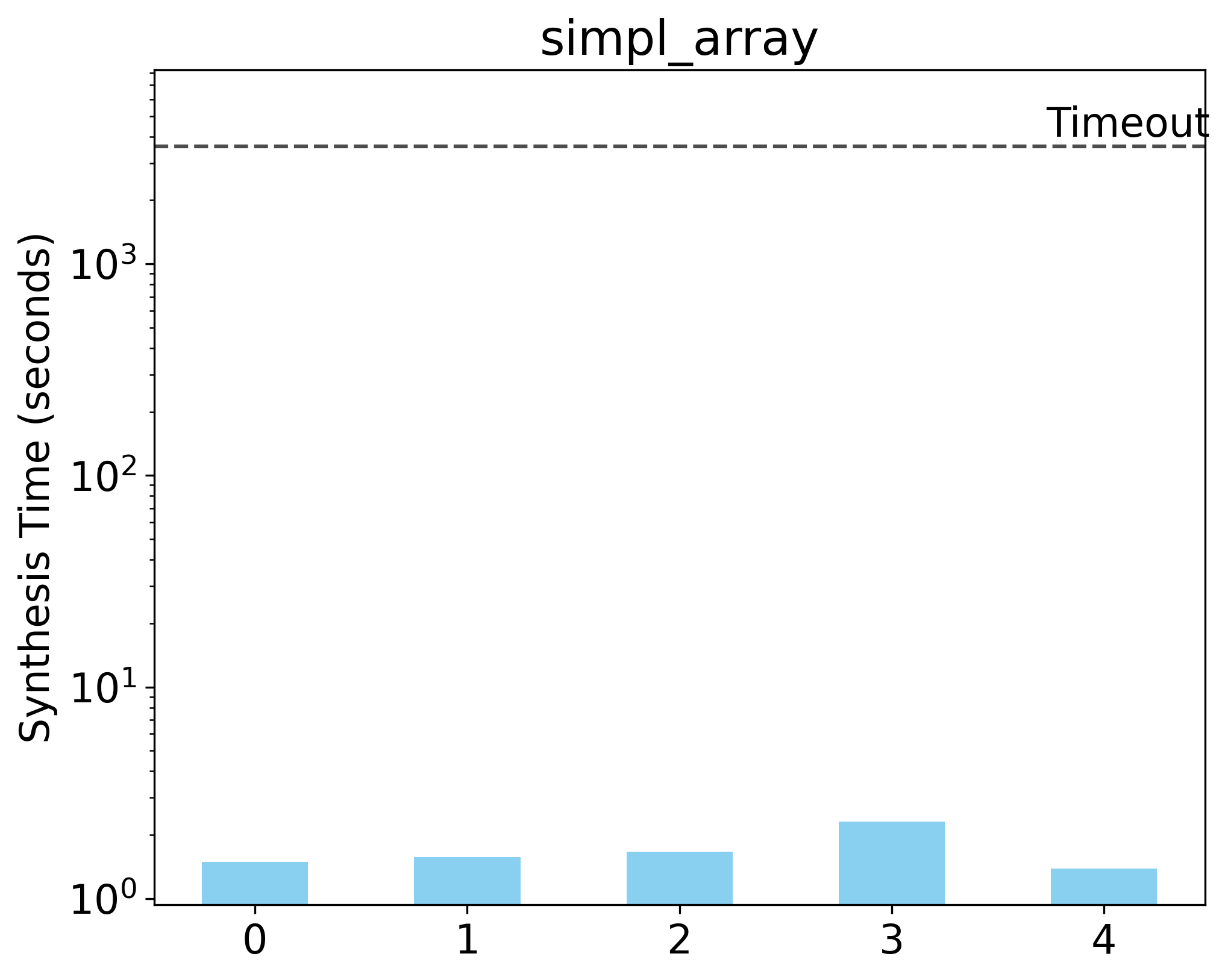}
\caption{\textbf{simpl\_array} benchmarks.}
\label{fig:simpl_array_time} 
\end{subfigure}

\begin{subfigure}{0.68\textwidth}
\centering
\includegraphics[scale=0.2,trim=0cm 0cm 0cm 0cm, clip]{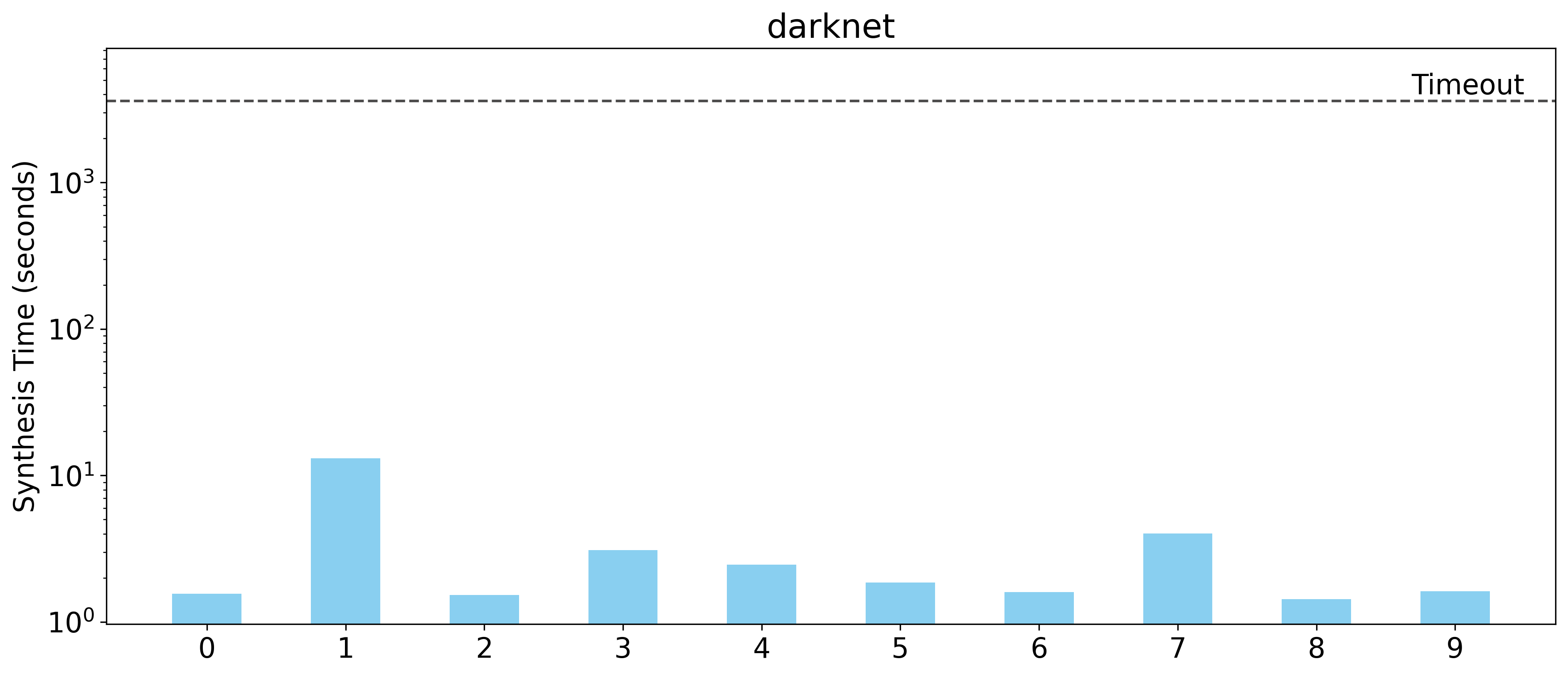}
\caption{\textbf{darknet} benchmarks.}
\label{fig:darknet_time} 
\end{subfigure}
\begin{subfigure}{0.30\textwidth}
\centering
\includegraphics[scale=0.2,trim=0cm 0cm 0cm 0cm, clip]{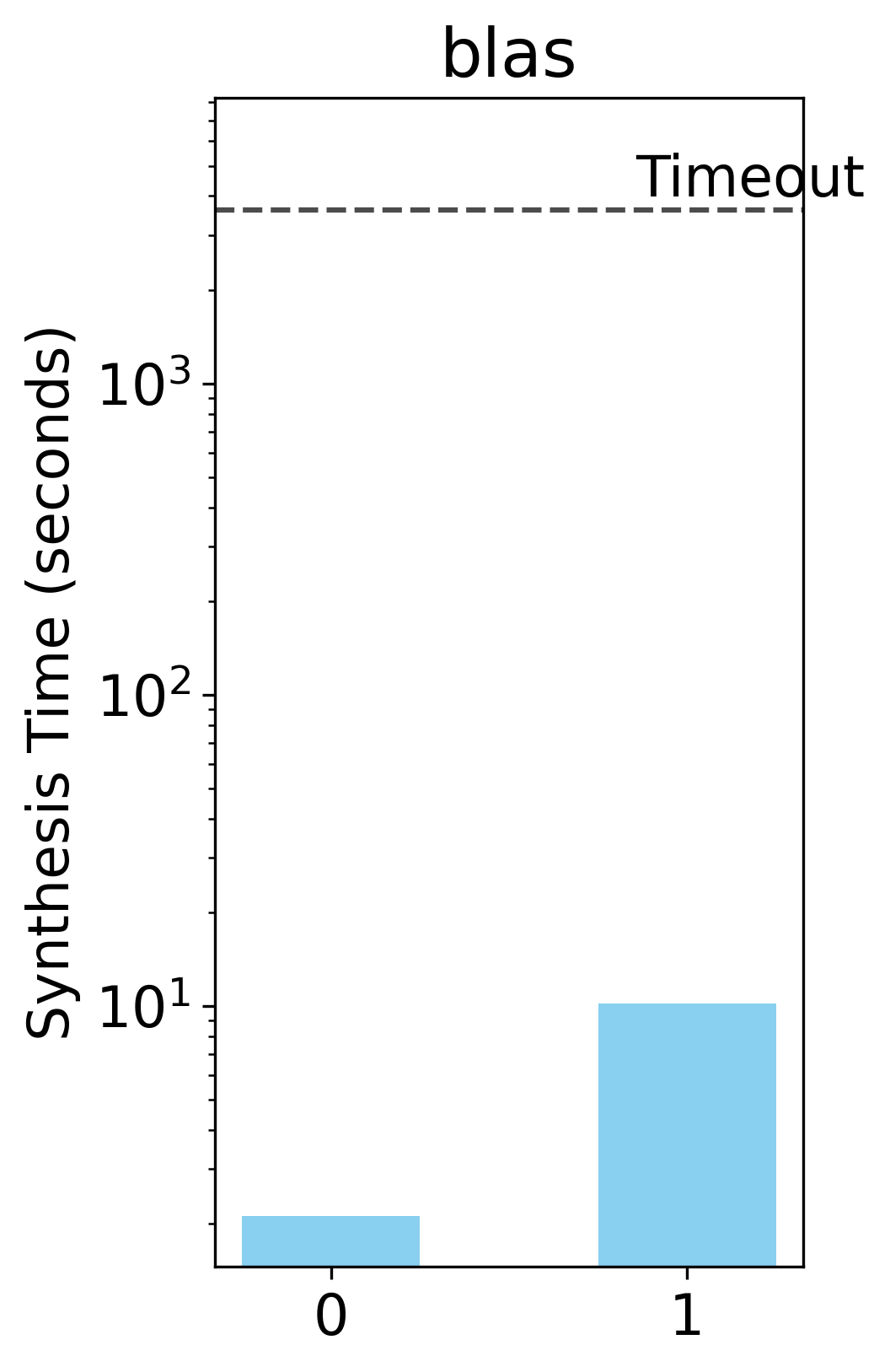}
\caption{\textbf{blas} benchmarks.}
\label{fig:blas_synth_time} 
\end{subfigure}

\begin{subfigure}{0.68\textwidth}
\centering
\includegraphics[scale=0.2,trim=0cm 0cm 0cm 0cm, clip]{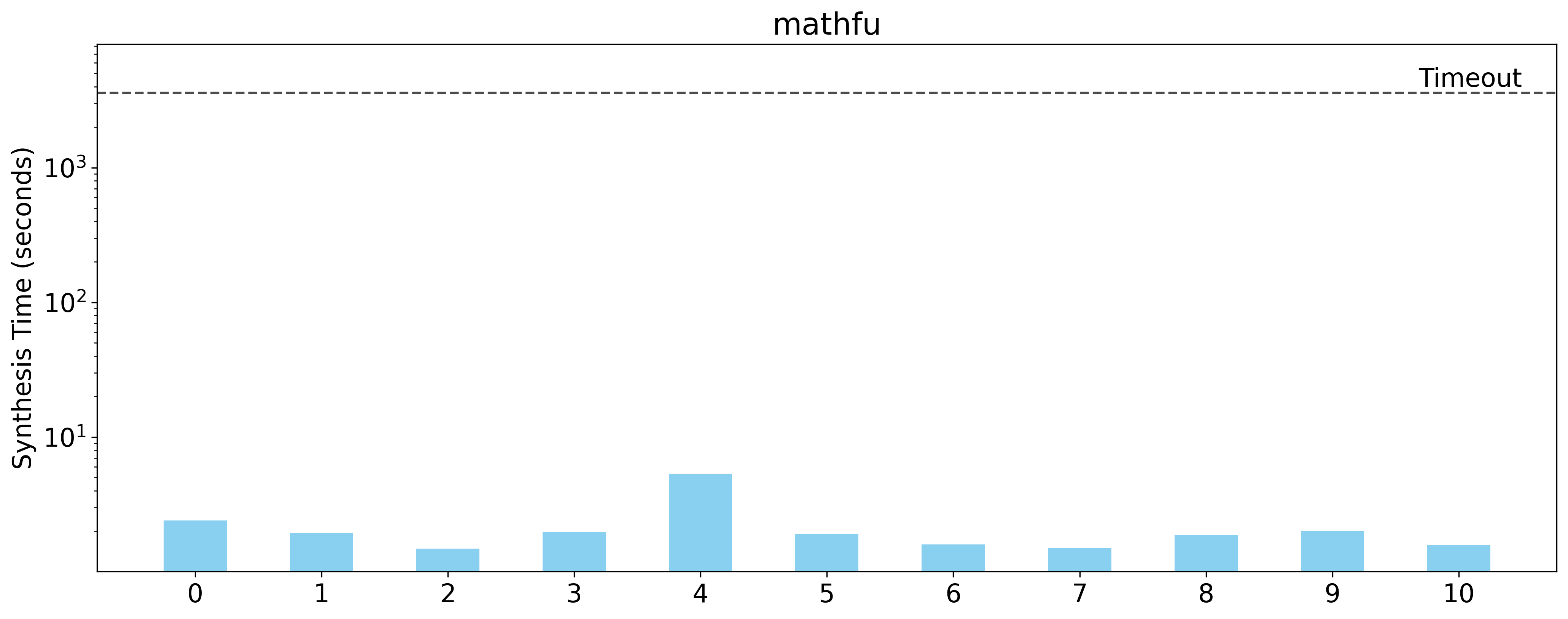}
\caption{\textbf{mathfu} benchmarks.}
\label{fig:mathfu_time} 
\end{subfigure}
\begin{subfigure}{0.30\textwidth}
\centering
\includegraphics[scale=0.2,trim=0cm 0cm 0cm 0cm, clip]{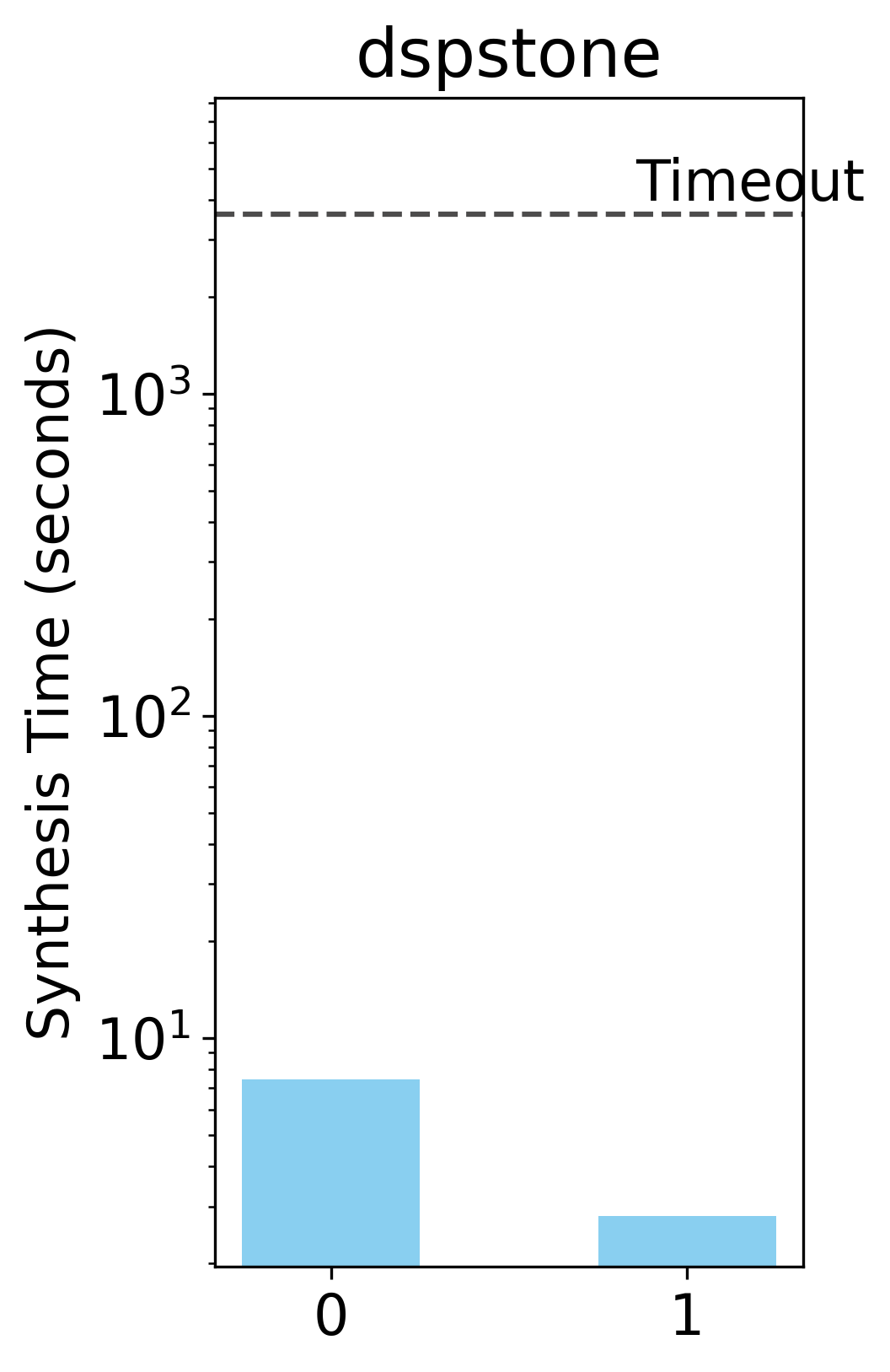}
\caption{\textbf{dspstone} benchmarks.}
\label{fig:dspstone_synth_time}
\end{subfigure}

\begin{subfigure}{0.68\textwidth}
\centering
\includegraphics[scale=0.2,trim=0cm 0cm 0cm 0cm, clip]{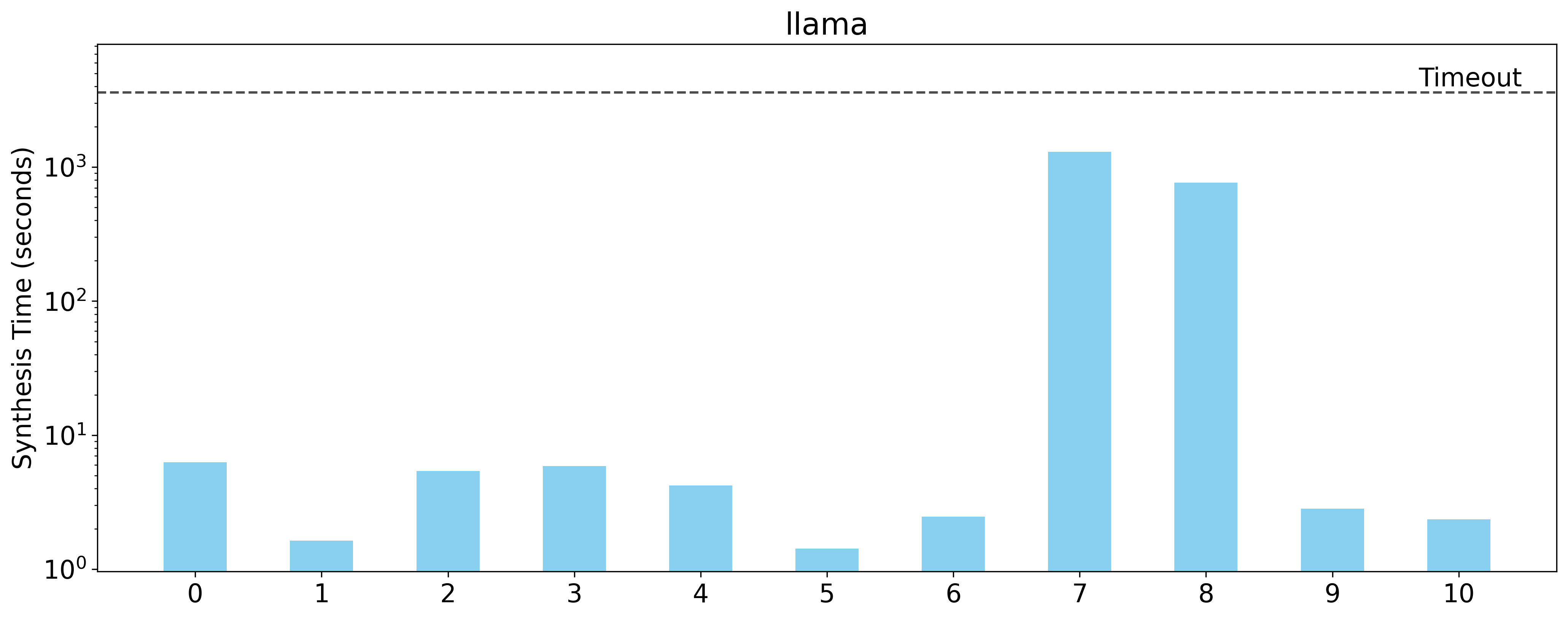}
\caption{\textbf{Llama} benchmarks (ML kernels).}
\label{fig:llama_synth_time} 
\end{subfigure}
\begin{subfigure}{0.3\textwidth}
\centering
\includegraphics[scale=0.2,trim=0cm 0cm 0cm 0cm, clip]{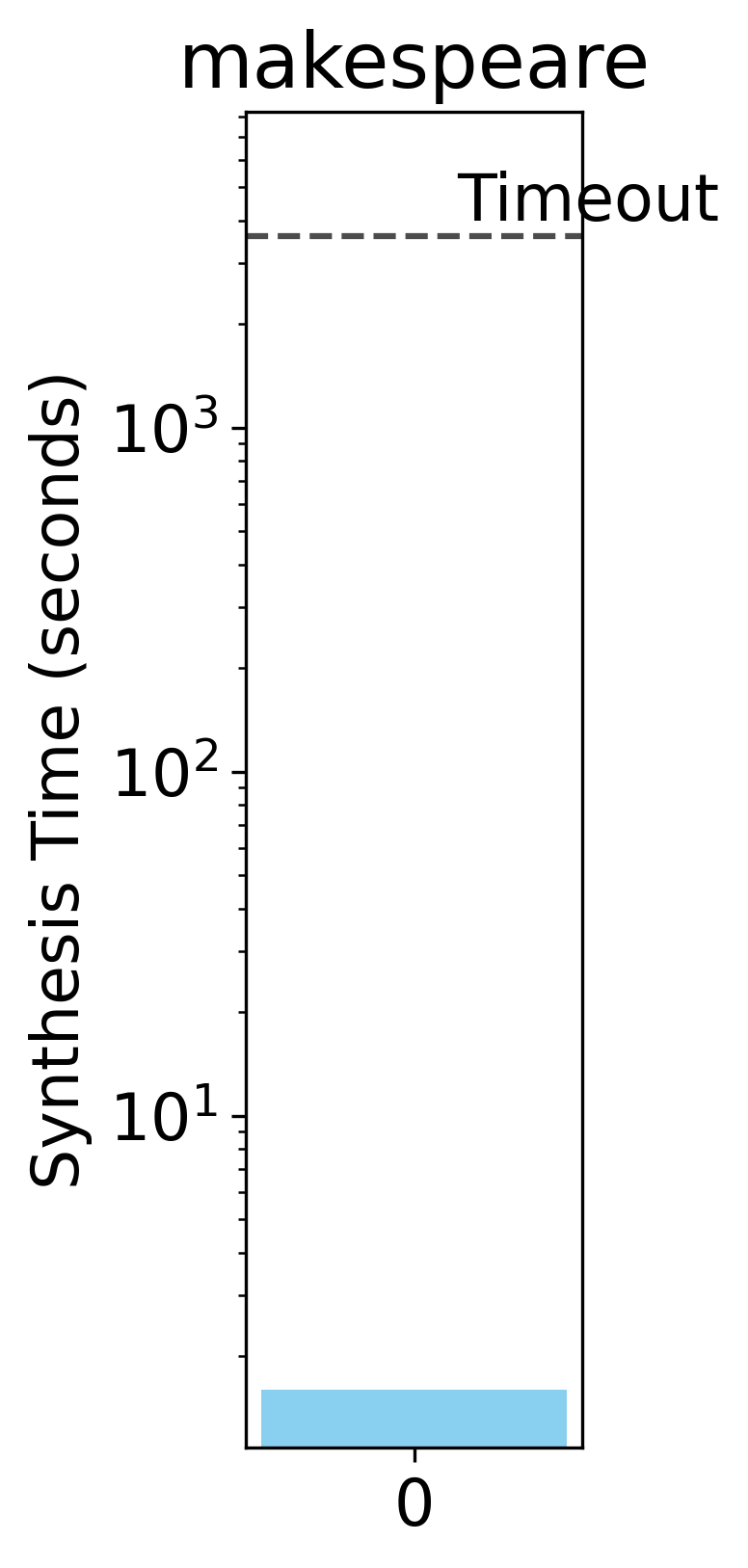}
\caption{\textbf{makespeare} benchmarks.}
\label{fig:makespeare_synth_time} 
\end{subfigure}
\caption{Synthesis timings for all the benchmark suites. Benchmark name legend in \cref{sec:benchmark_name_config} in the Appendix.} 
\label{fig:overall_synth_time}
\end{figure}

%% file: figures/timings.tex
\begin{figure}
\begin{subfigure}{0.7\linewidth}
\centering
\includegraphics[scale=0.37,trim=0cm 0cm 0cm 0cm, clip]{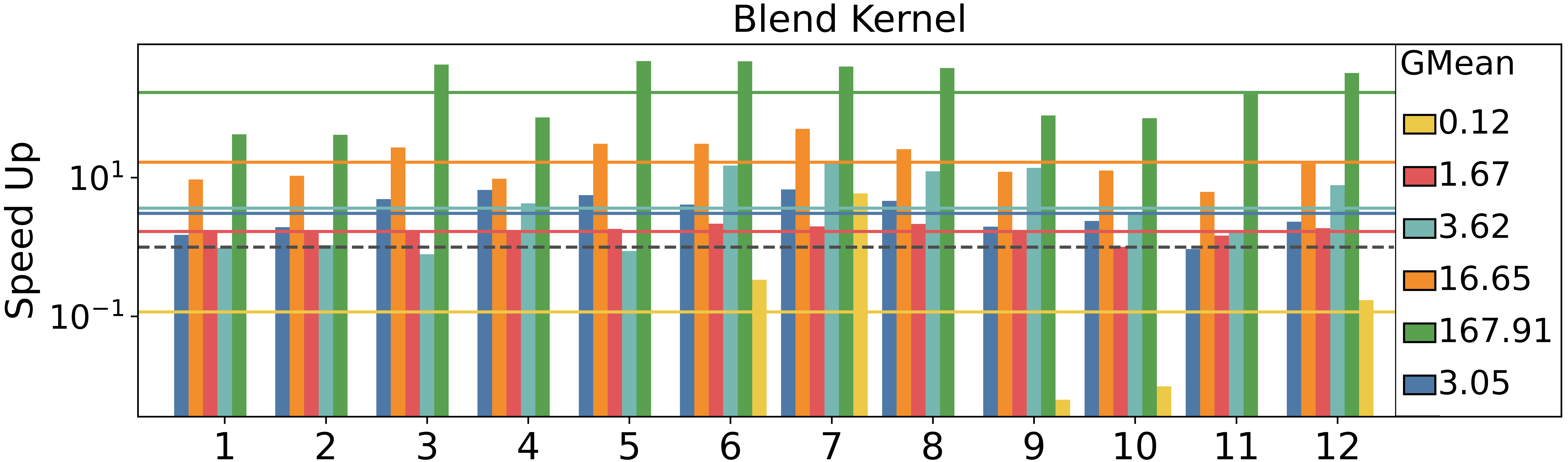}
\caption{\textbf{blend} benchmarks.}
\label{fig:blend_kernel_time} 
\end{subfigure}
\begin{subfigure}{0.29\linewidth}
\centering
\includegraphics[scale=0.37,trim=0cm 0cm 0cm 0cm, clip]{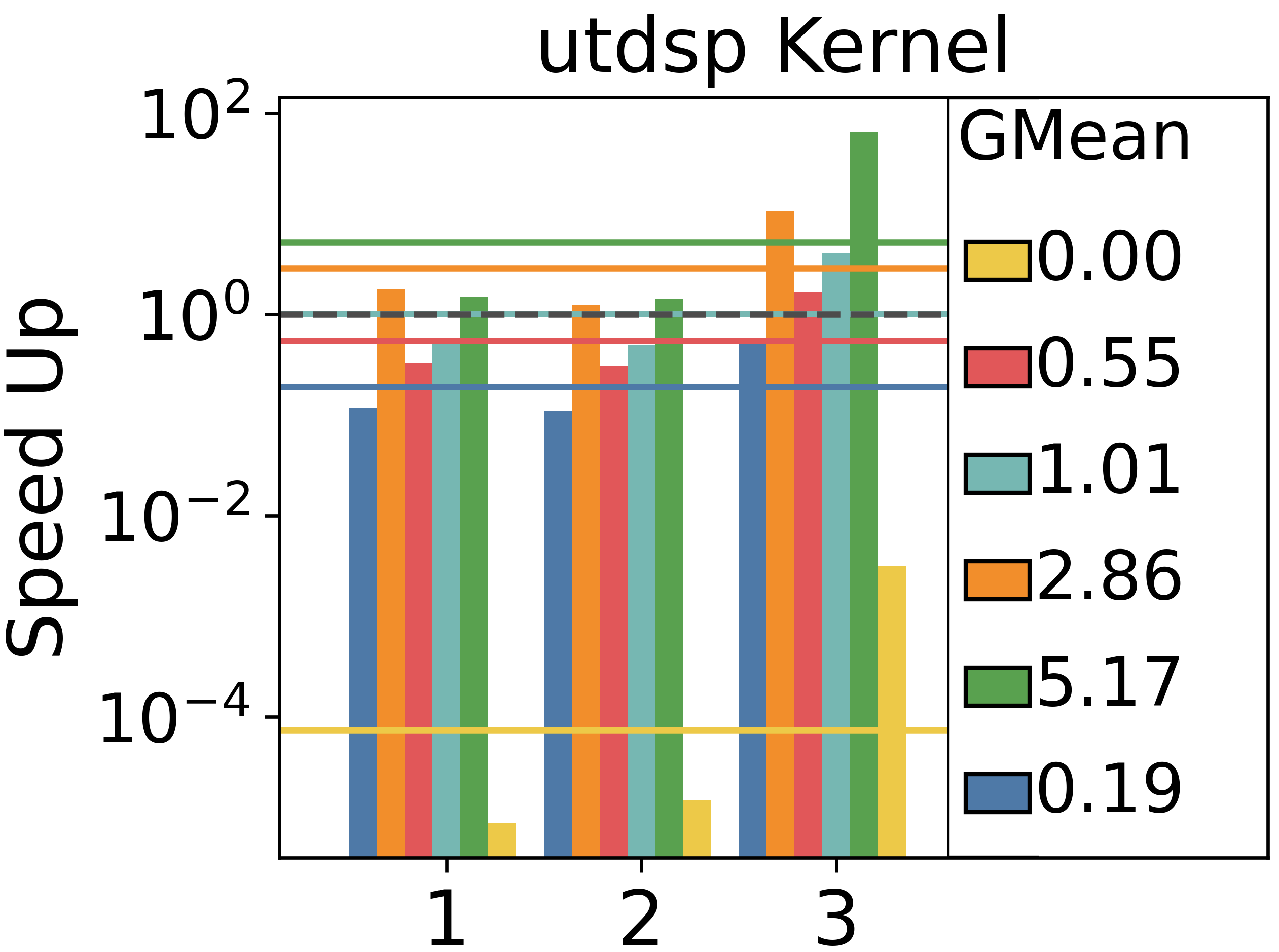}
\caption{\textbf{utdsp} benchmarks.}
\label{fig:utdsp_kernel_time} 
\end{subfigure}

\begin{subfigure}{0.7\linewidth}
\centering
\includegraphics[scale=0.37,trim=0cm 0cm 0cm 0cm, clip]{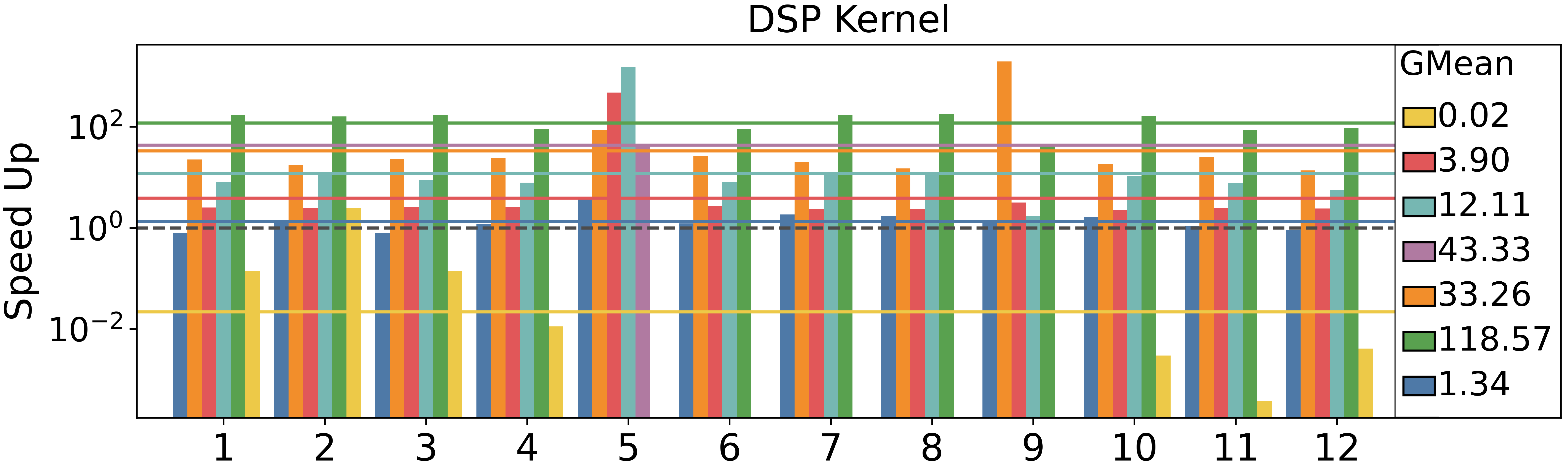}
\caption{\textbf{dsp} benchmarks.}
\label{fig:dsp_kernel_time} 
\end{subfigure}
\begin{subfigure}{0.29\linewidth}
\centering
\includegraphics[scale=0.37,trim=0cm 0cm 0cm 0cm, clip]{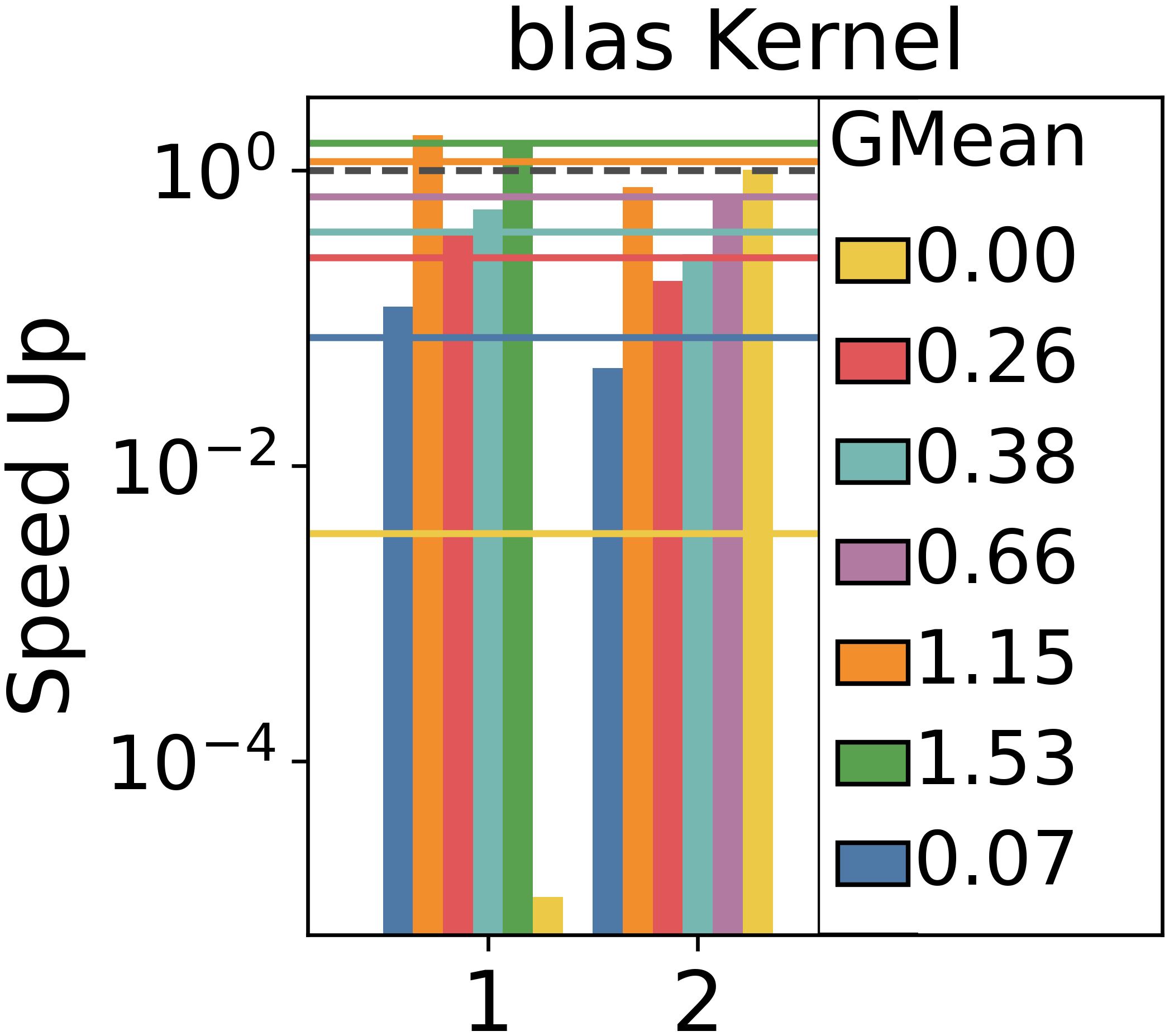}
\caption{\textbf{blas} benchmarks.}
\label{fig:blas_kernel_time} 
\end{subfigure}

\begin{subfigure}{0.62\linewidth}
\centering
\includegraphics[scale=0.37,trim=0cm 0cm 0cm 0cm, clip]{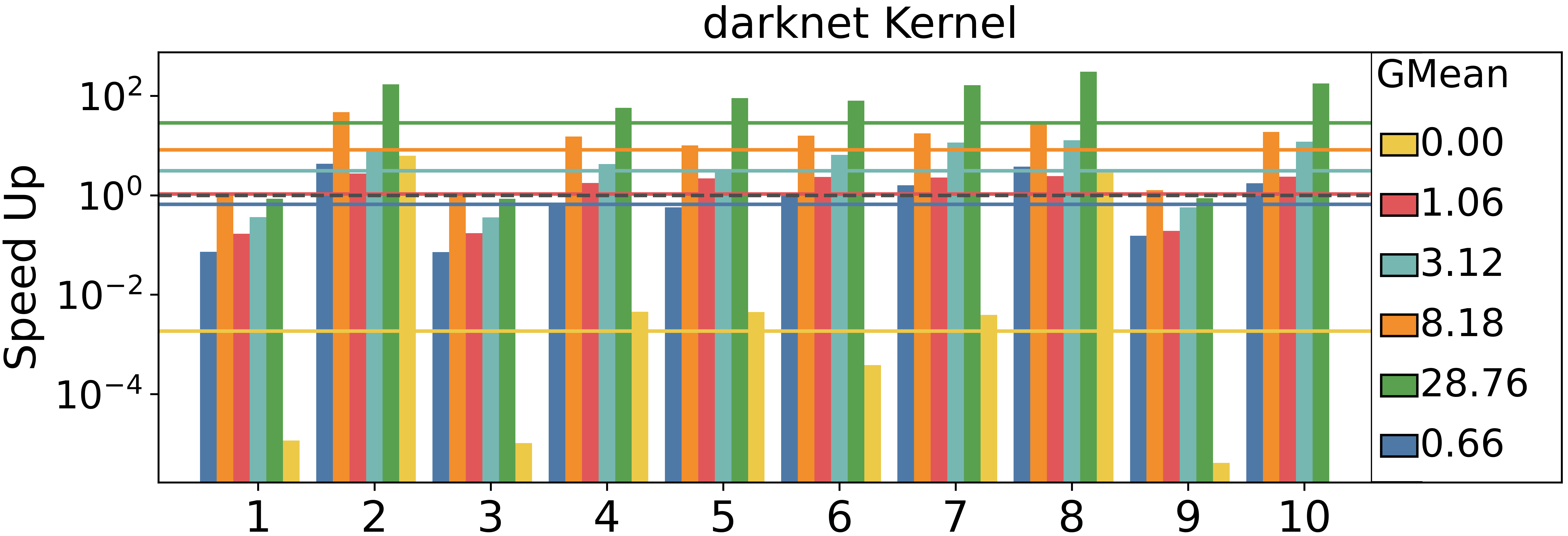}
\caption{\textbf{darknet} benchmarks.}
\label{fig:darknet_kernel_time} 
\end{subfigure}
\begin{subfigure}{0.37\linewidth}
\centering
\includegraphics[scale=0.37,trim=0cm 0cm 0cm 0cm, clip]{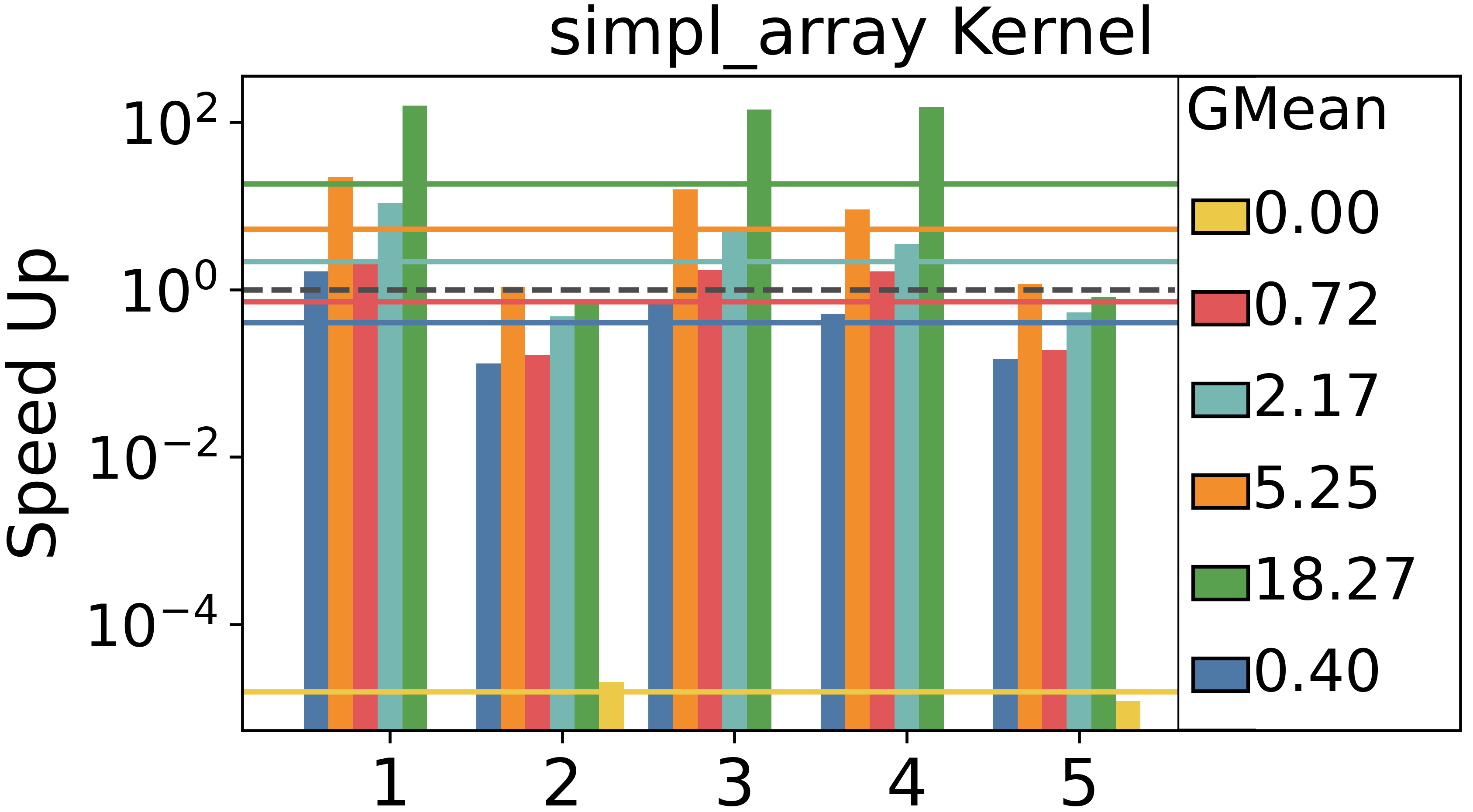}
\caption{\textbf{simpl\_array} benchmarks.}
\label{fig:simpl_array_kernel_time} 
\end{subfigure}

\begin{subfigure}{0.7\linewidth}
\centering
\includegraphics[scale=0.37,trim=0cm 0cm 0cm 0cm, clip]{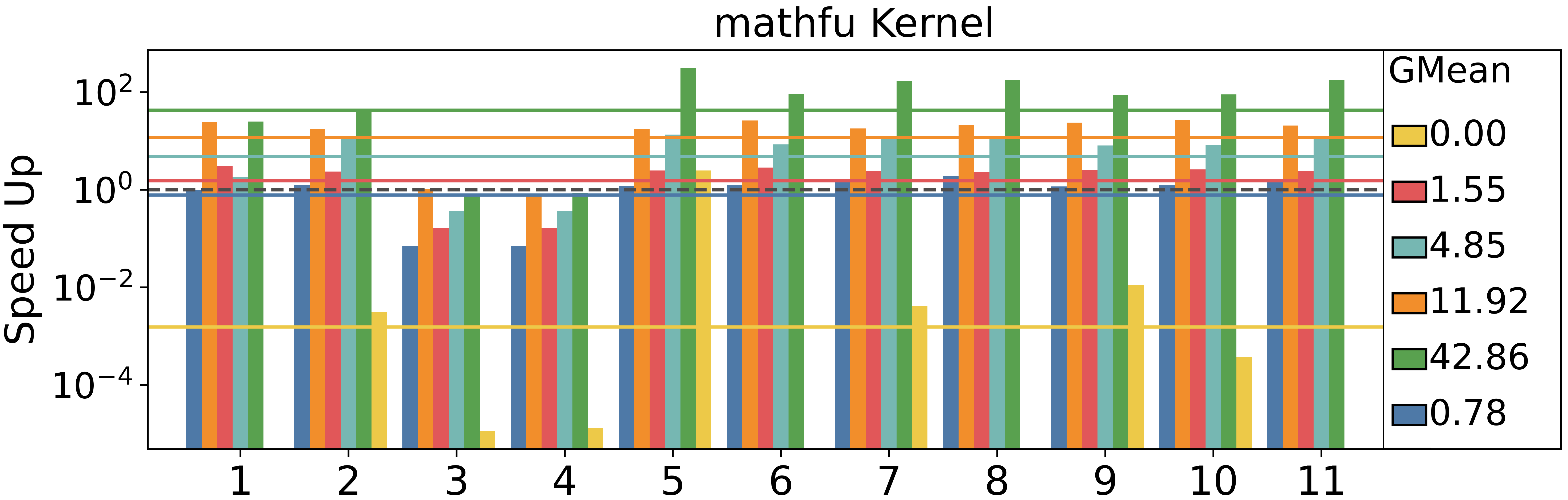}
\caption{\textbf{mathfu} benchmarks.}
\label{fig:mathfu_kernel_time} 
\end{subfigure}
\begin{subfigure}{0.29\linewidth}
\centering
\includegraphics[scale=0.37,trim=0cm 0cm 0cm 0cm, clip]{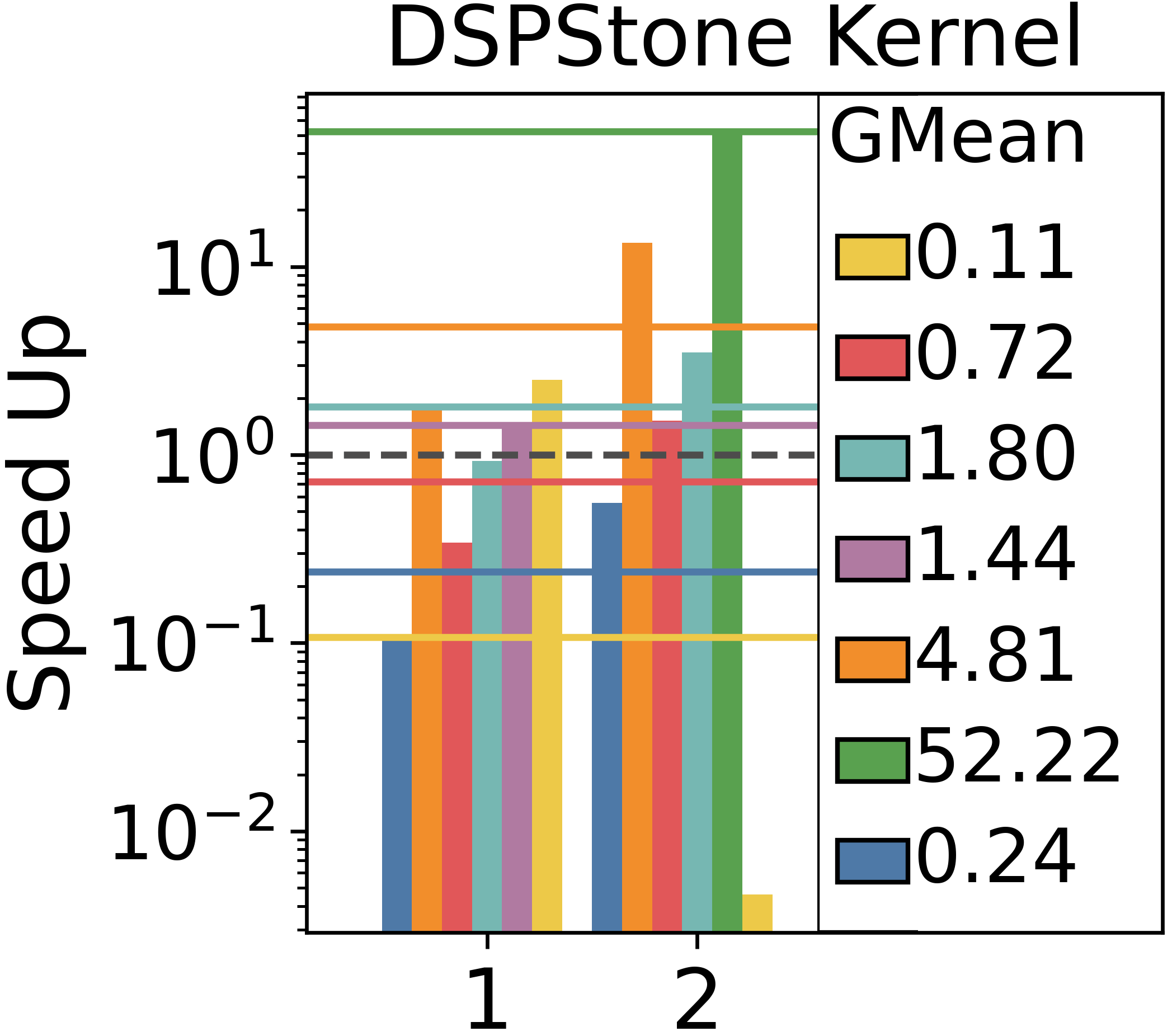}
\caption{\textbf{dspstone} benchmarks.}
\label{fig:dspstone_kernel_time} 
\end{subfigure}

\begin{subfigure}{0.7\linewidth}
\centering
\includegraphics[scale=0.37,trim=0cm 0cm 0cm 0cm, clip]{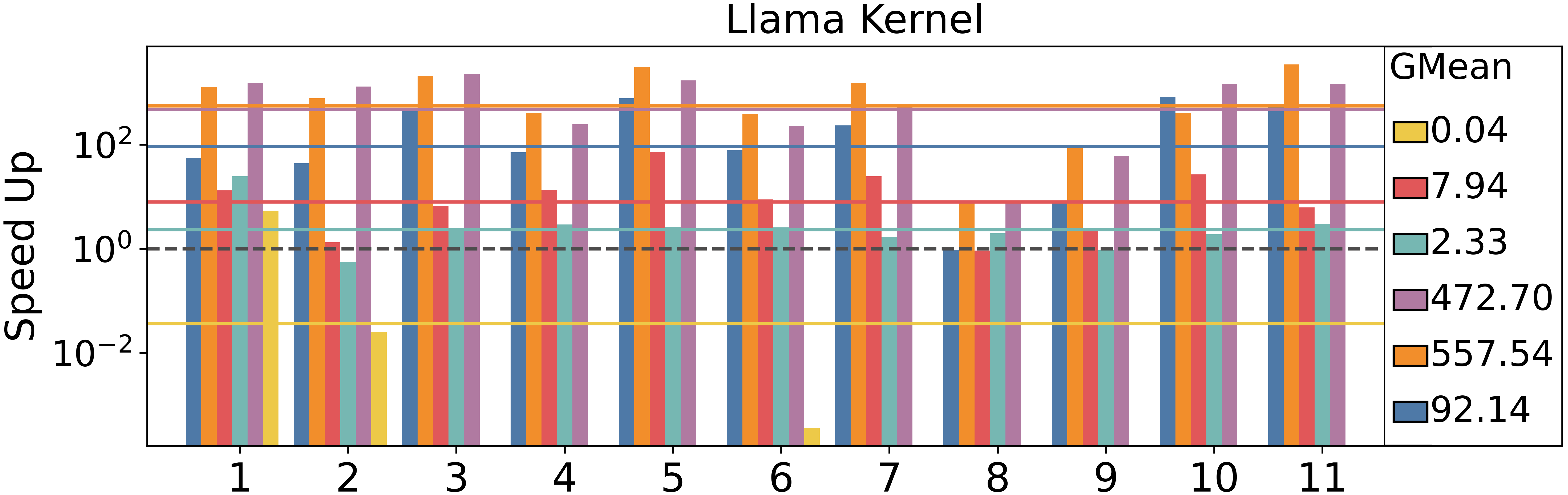}
\caption{\textbf{Llama} benchmarks.}
\label{fig:llama_kernel_time} 
\end{subfigure}
\begin{subfigure}{0.29\linewidth}
\centering
\includegraphics[scale=0.37,trim=0cm 0cm 0cm 0cm, clip]{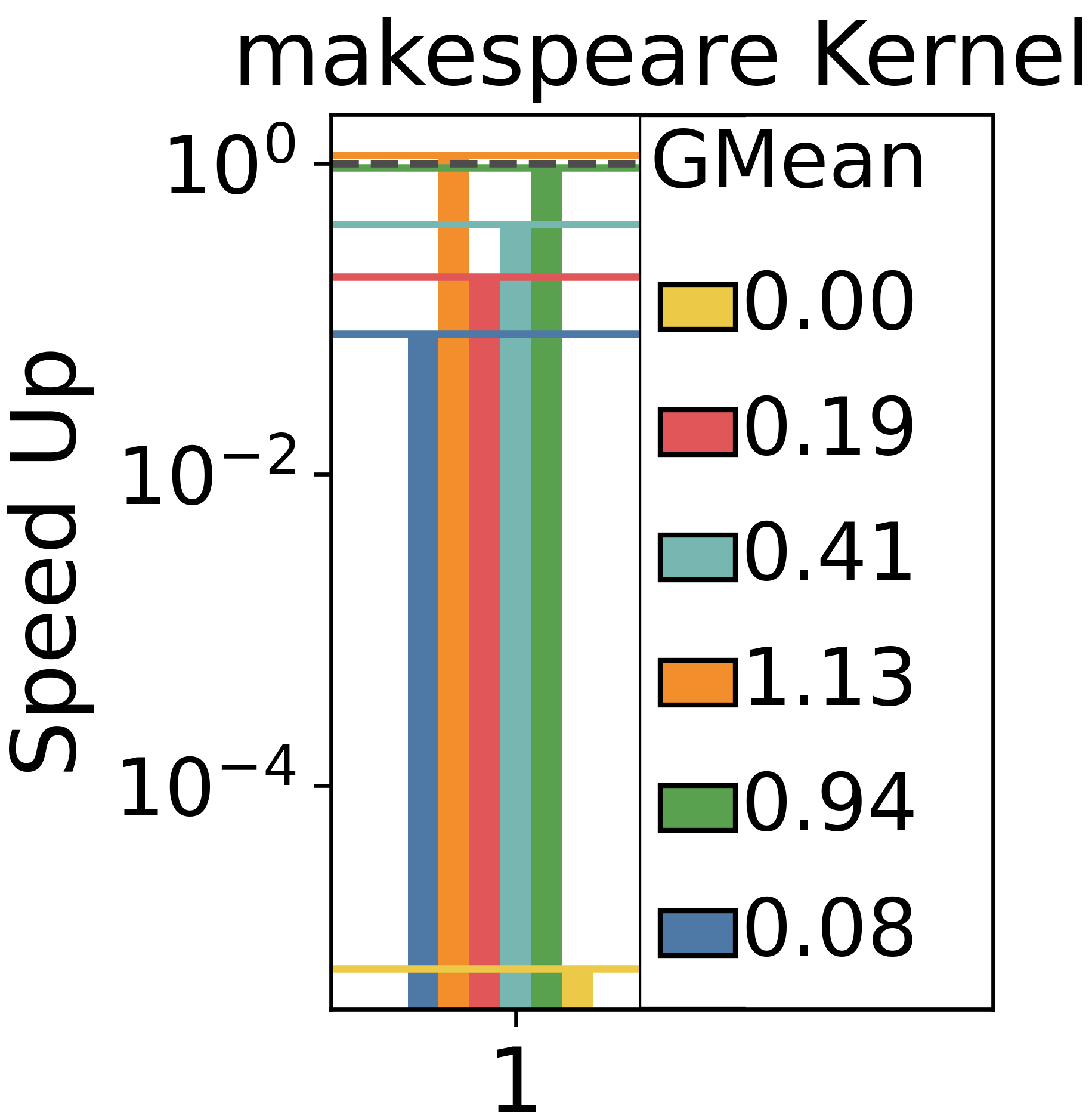}
\caption{\textbf{makespeare} benchmarks.}
\label{fig:makespeare_kernel_time} 
\end{subfigure}

\begin{subfigure}{\linewidth}
\centering
\includegraphics[width=\linewidth,trim=0cm 0cm 0cm 0cm, clip]{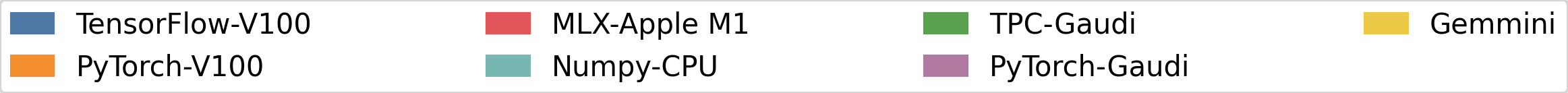}
\end{subfigure}

\caption{Kernel speedup over baseline. Benchmark name legend in \cref{sec:benchmark_name_config} in the Appendix.}
\label{fig:kernel_time} 
\end{figure}

%% file: figures/timing_e2e.tex
\begin{figure}
\begin{subfigure}{0.7\linewidth}
\centering
\includegraphics[scale=0.37,trim=0cm 0cm 0cm 0cm, clip]{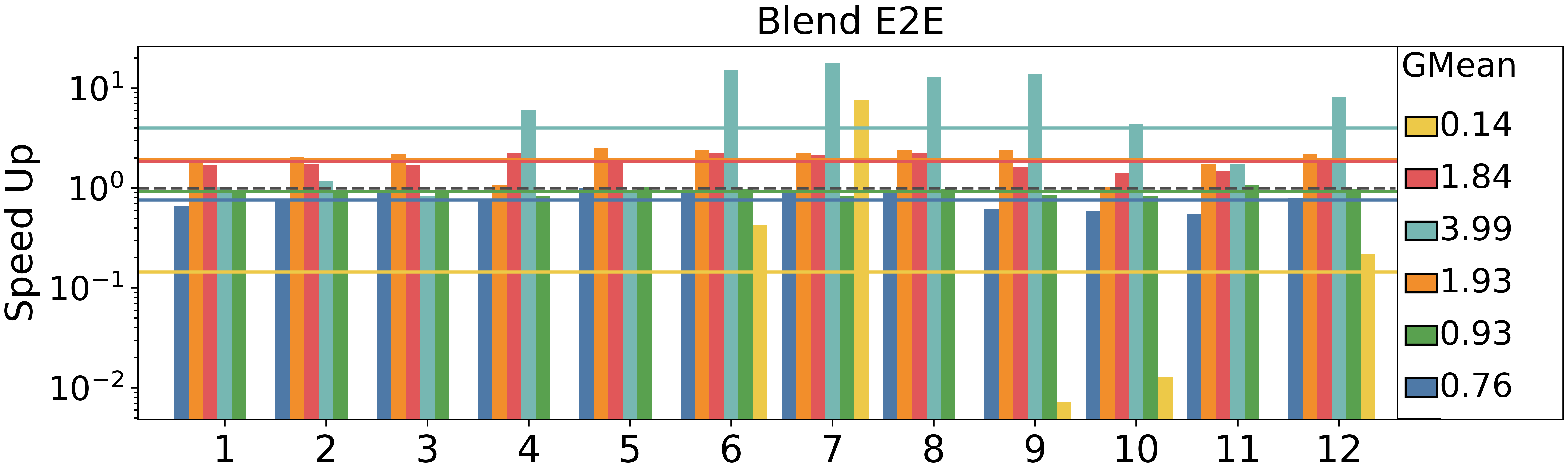}
\caption{\textbf{blend} benchmarks.}
\label{fig:blend_e2e_time} 
\end{subfigure}
\begin{subfigure}{0.29\linewidth}
\centering
\includegraphics[scale=0.37,trim=0cm 0cm 0cm 0cm, clip]{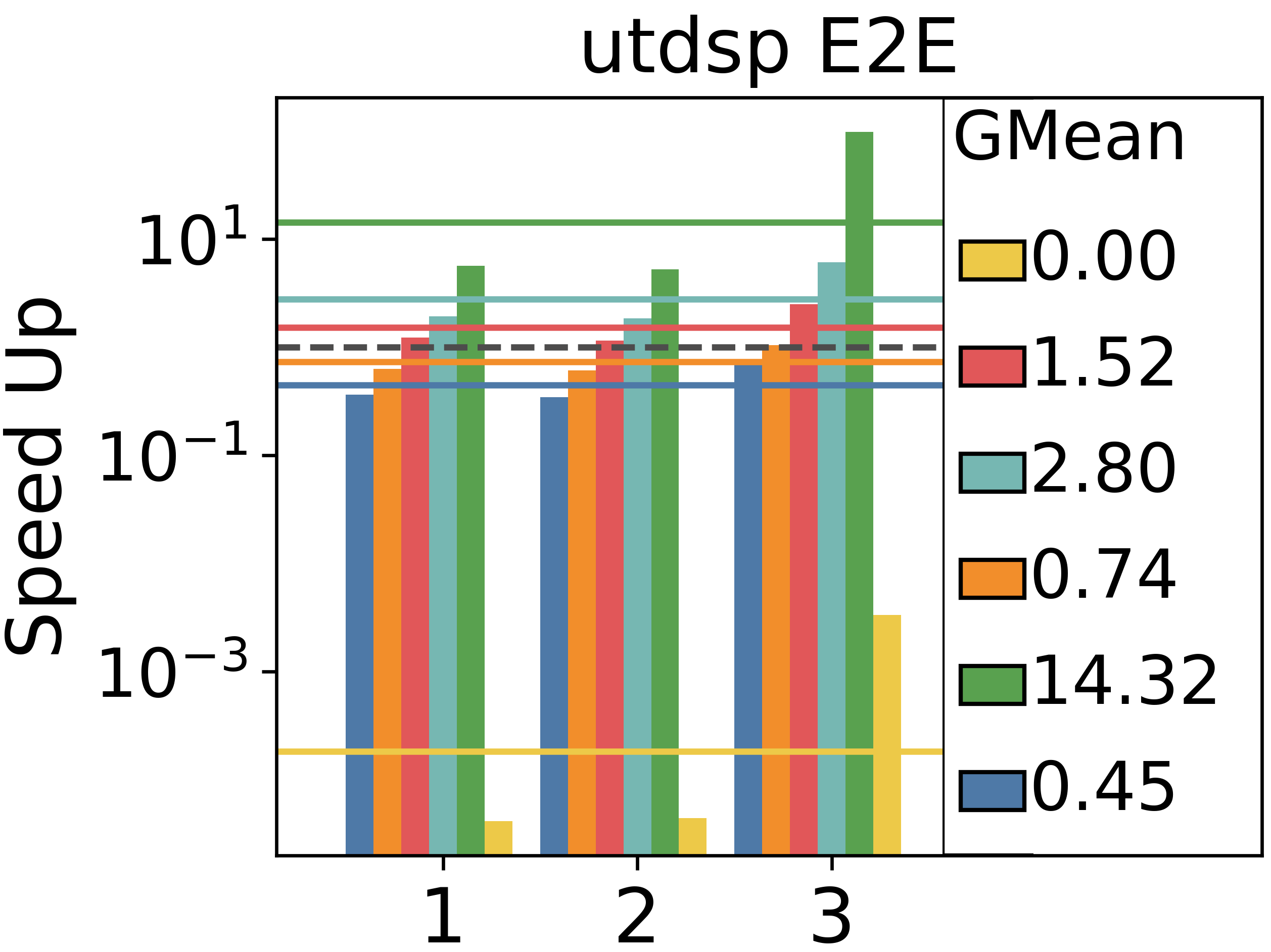}
\caption{\textbf{utdsp} benchmarks.}
\label{fig:utdsp_e2e_time} 
\end{subfigure}

\begin{subfigure}{0.7\linewidth}
\centering
\includegraphics[scale=0.37,trim=0cm 0cm 0cm 0cm, clip]{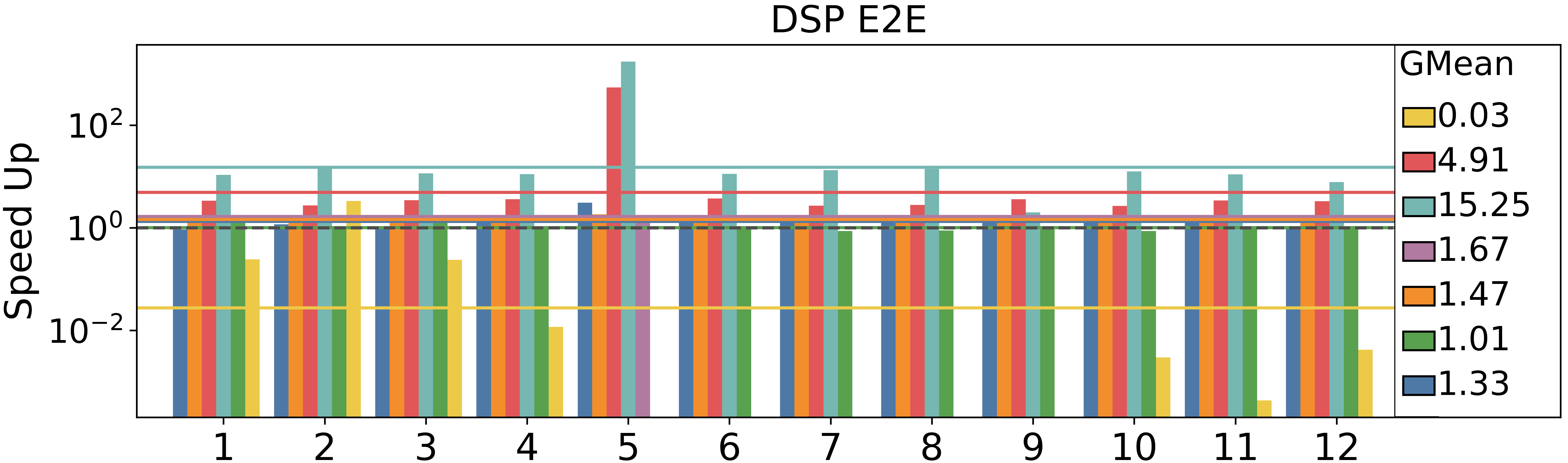}
\caption{\textbf{dsp} benchmarks.}
\label{fig:dsp_e2e_time} 
\end{subfigure}
\begin{subfigure}{0.29\linewidth}
\centering
\includegraphics[scale=0.37,trim=0cm 0cm 0cm 0cm, clip]{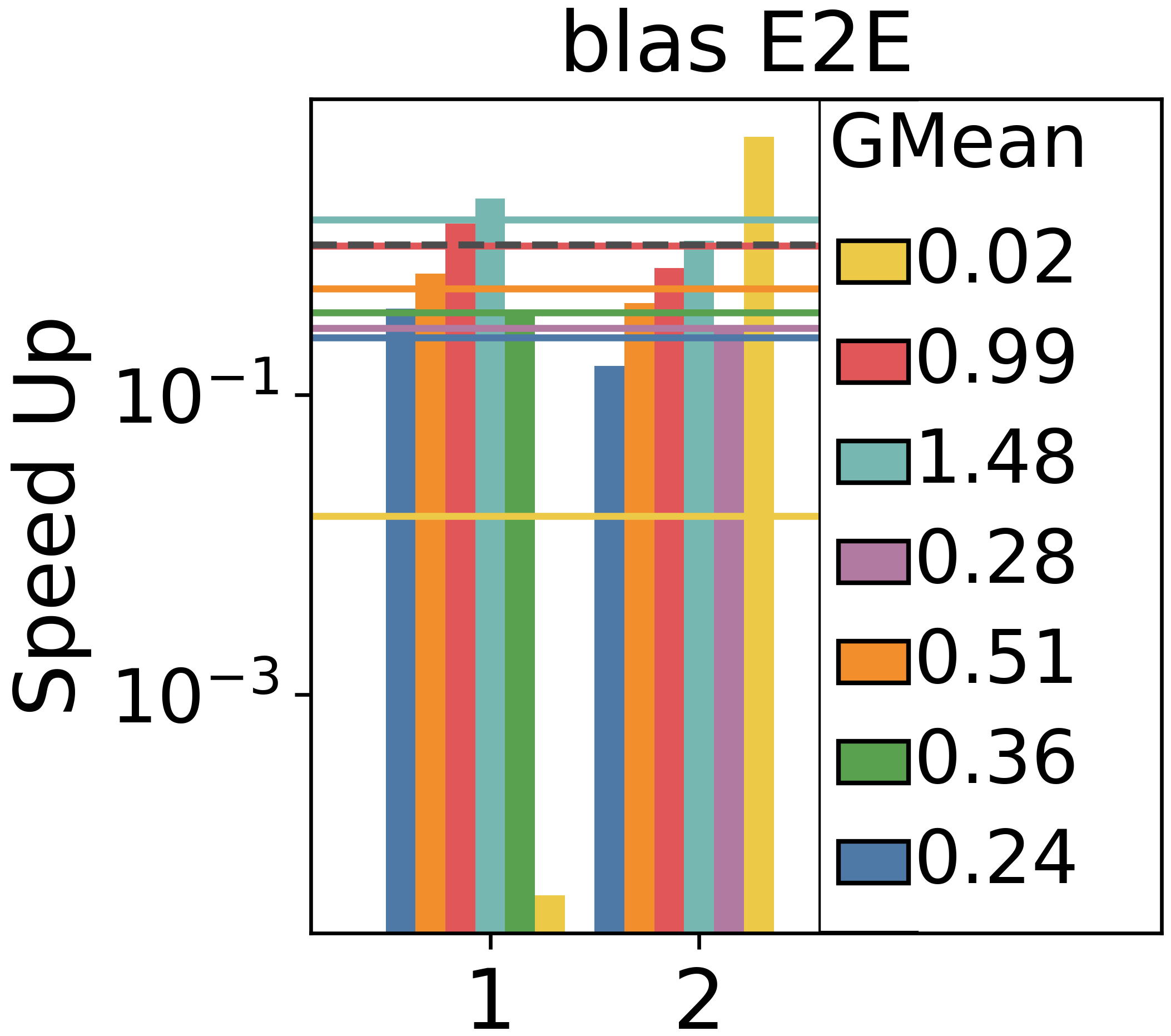}
\caption{\textbf{blas} benchmarks.}
\label{fig:blas_e2e_time} 
\end{subfigure}

\begin{subfigure}{0.61\linewidth}
\centering
\includegraphics[scale=0.37,trim=0cm 0cm 0cm 0cm, clip]{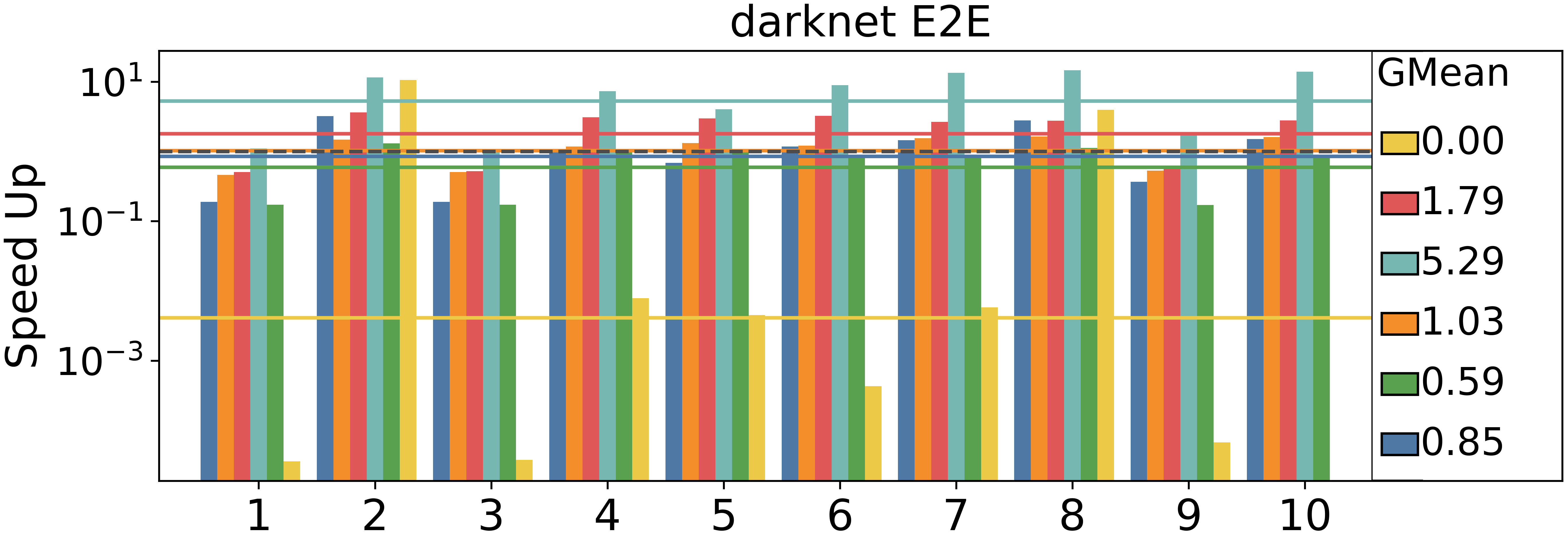}
\caption{\textbf{darknet} benchmarks.}
\label{fig:darknet_e2e_time} 
\end{subfigure}
\begin{subfigure}{0.38\linewidth}
\centering
\includegraphics[scale=0.37,trim=0cm 0cm 0cm 0cm, clip]{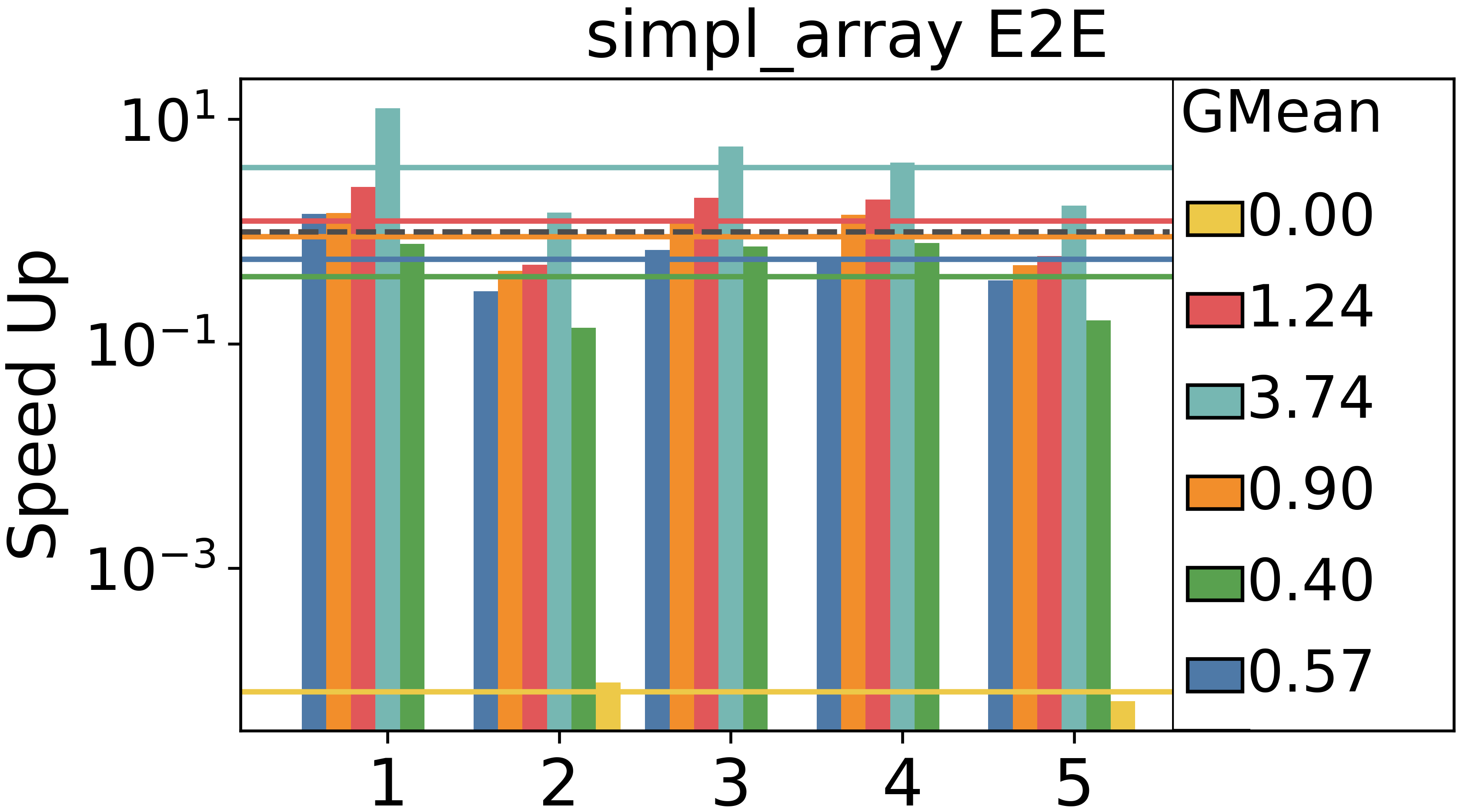}
\caption{\textbf{simpl\_array} benchmarks.}
\label{fig:simpl_array_e2e_time} 
\end{subfigure}

\begin{subfigure}{0.7\linewidth}
\centering
\includegraphics[scale=0.37,trim=0cm 0cm 0cm 0cm, clip]{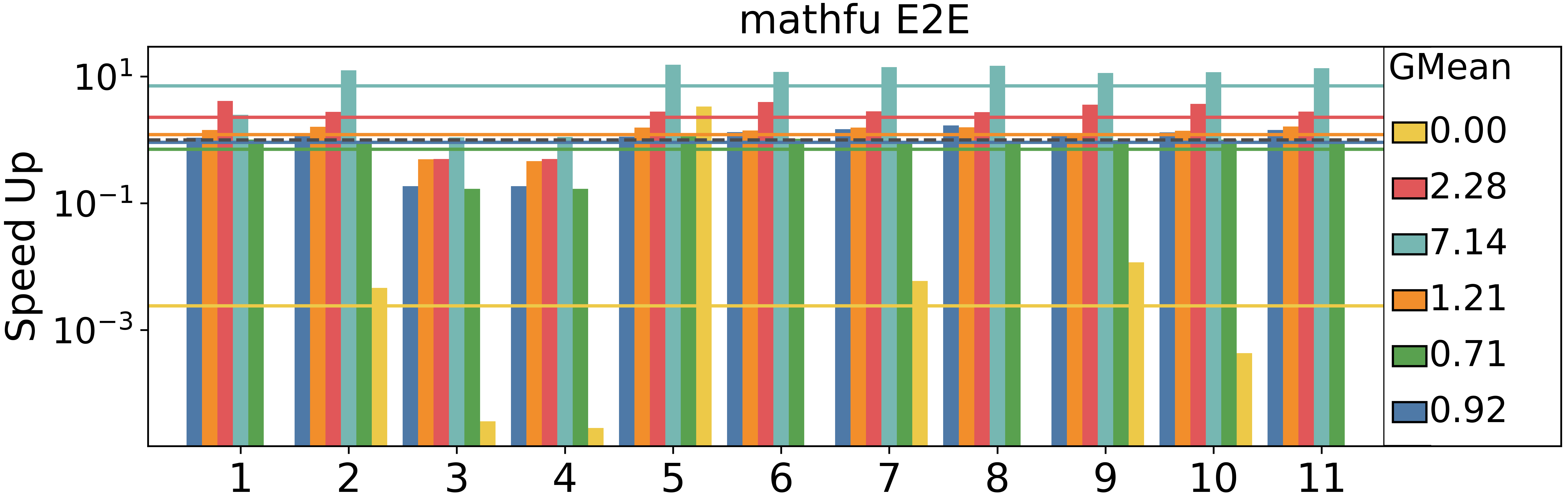}
\caption{\textbf{mathfu} benchmarks.}
\label{fig:mathfu_e2e_time} 
\end{subfigure}
\begin{subfigure}{0.29\linewidth}
\centering
\includegraphics[scale=0.37,trim=0cm 0cm 0cm 0cm, clip]{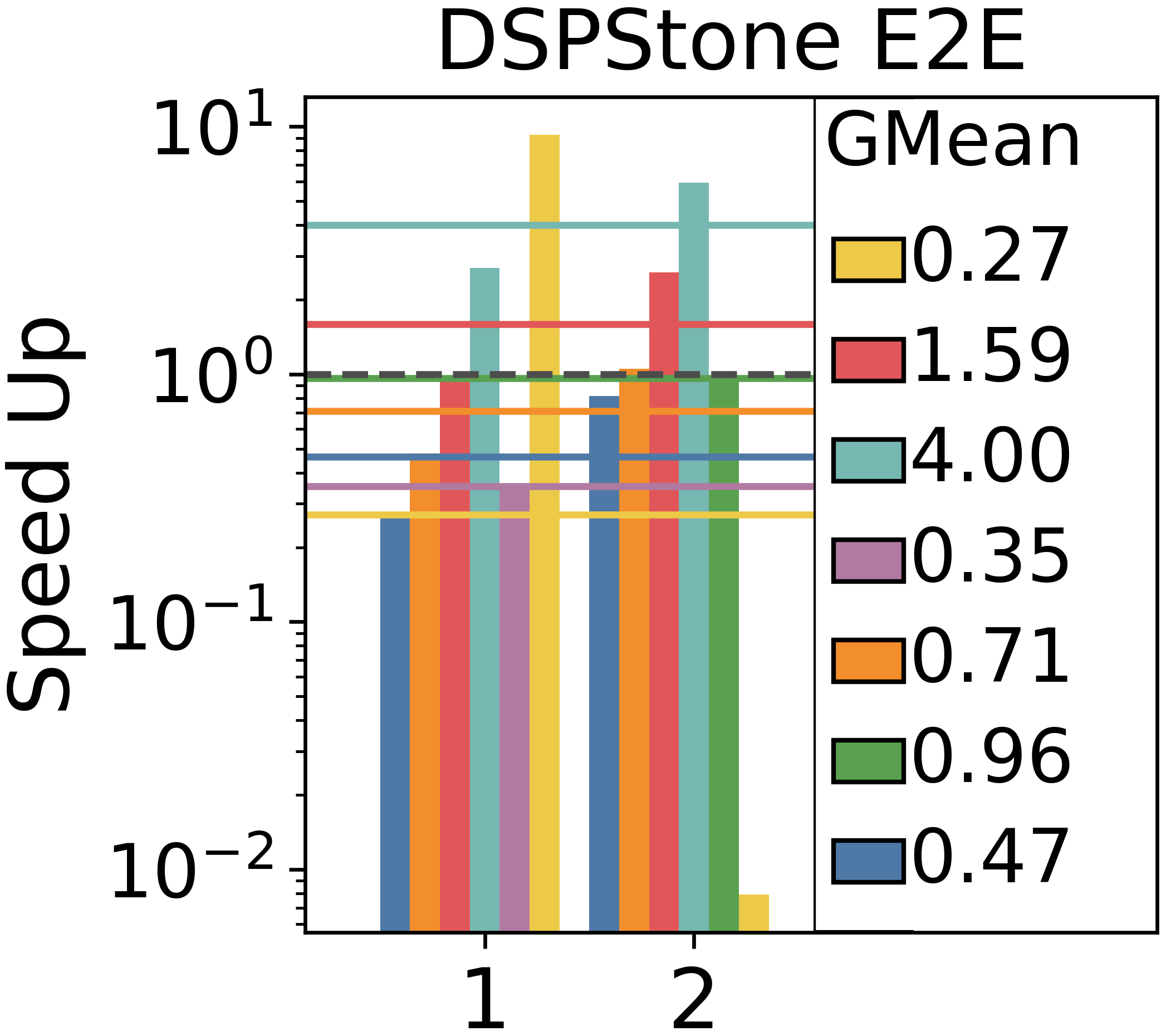}
\caption{\textbf{dspstone} benchmarks.}
\label{fig:dspstone_e2e_time} 
\end{subfigure}

\begin{subfigure}{0.7\linewidth}
\centering
\includegraphics[scale=0.37,trim=0cm 0cm 0cm 0cm, clip]{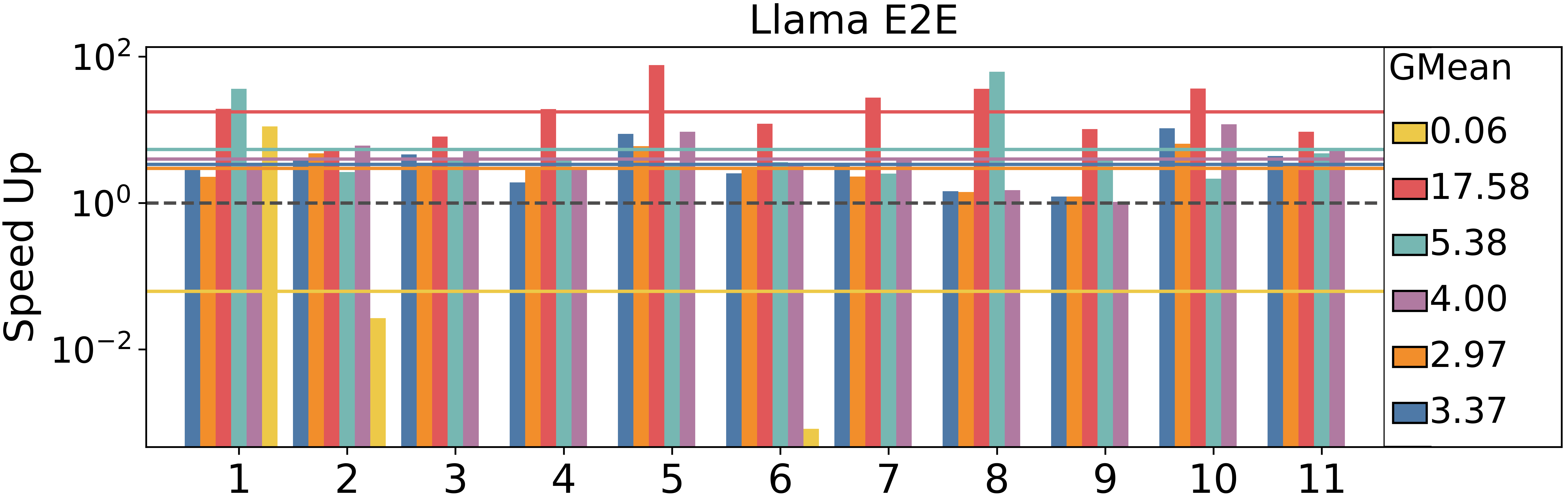}
\caption{\textbf{Llama} benchmarks.}
\label{fig:llama_e2e_time} 
\end{subfigure}
\begin{subfigure}{0.29\linewidth}
\centering
\includegraphics[scale=0.37,trim=0cm 0cm 0cm 0cm, clip]{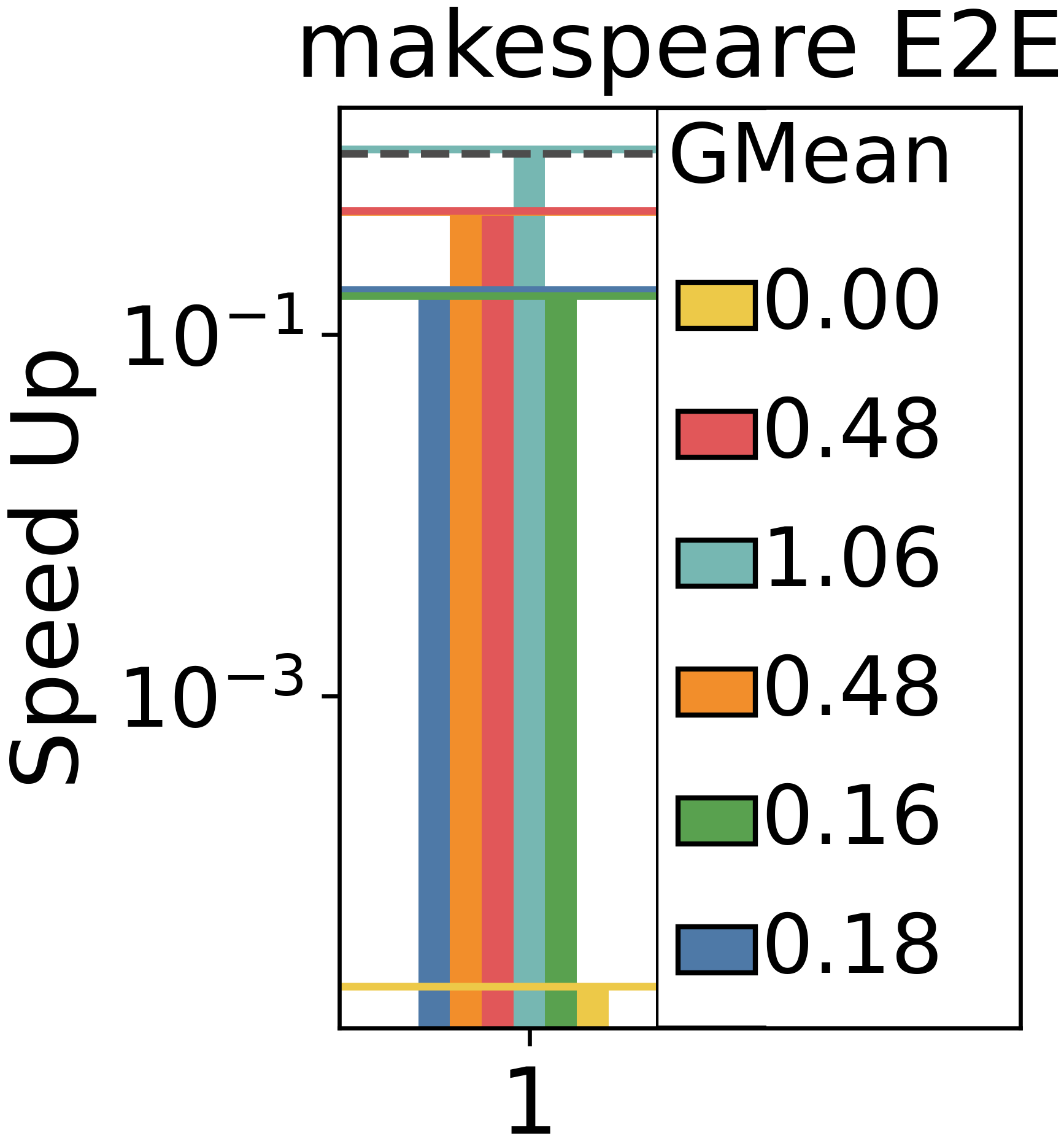}
\caption{\textbf{makespeare} benchmarks.}
\label{fig:makespeare_e2e_time} 
\end{subfigure}

\begin{subfigure}{\linewidth}
\centering
\includegraphics[width=\linewidth, trim=0cm 0cm 0cm 0cm, clip]{figures/performance/legend.png}
\end{subfigure}

\caption{E2E speedup over baseline. Benchmark name legend in \cref{sec:benchmark_name_config} in the Appendix.}
\label{fig:e2e_performance}
\end{figure}

%% file: figures/ablation.tex
\begin{figure}[!t]
\begin{subfigure}{0.68\textwidth}
\centering
\includegraphics[scale=0.25,trim=0cm 0cm 0cm 0cm, clip]{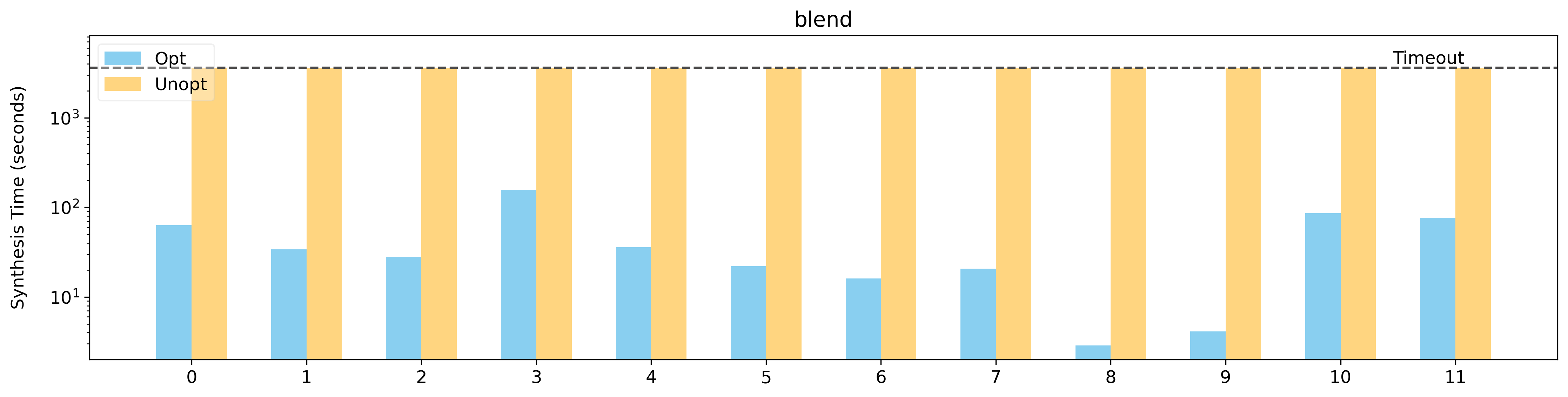}
\caption{\textbf{blend} benchmarks.}
\label{fig:blend_ab_time} 
\end{subfigure}
\begin{subfigure}{0.3\textwidth}
\centering
\includegraphics[scale=0.25,trim=0cm 0cm 0cm 0cm, clip]{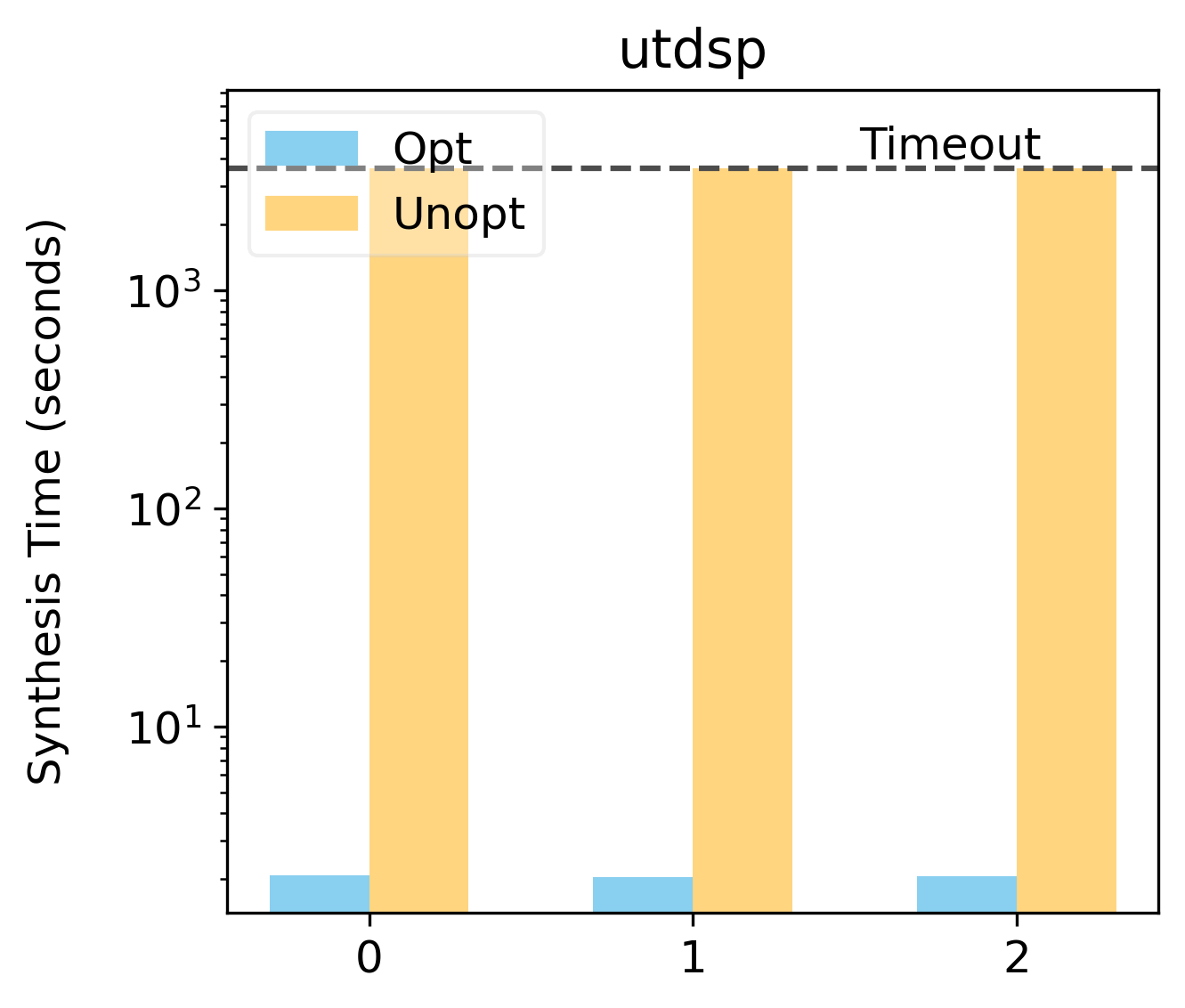}
\caption{\textbf{utdsp} benchmarks.}
\label{fig:utdsp_ab_time} 
\end{subfigure}

\begin{subfigure}{0.68\textwidth}
\centering
\includegraphics[scale=0.25,trim=0cm 0cm 0cm 0cm, clip]{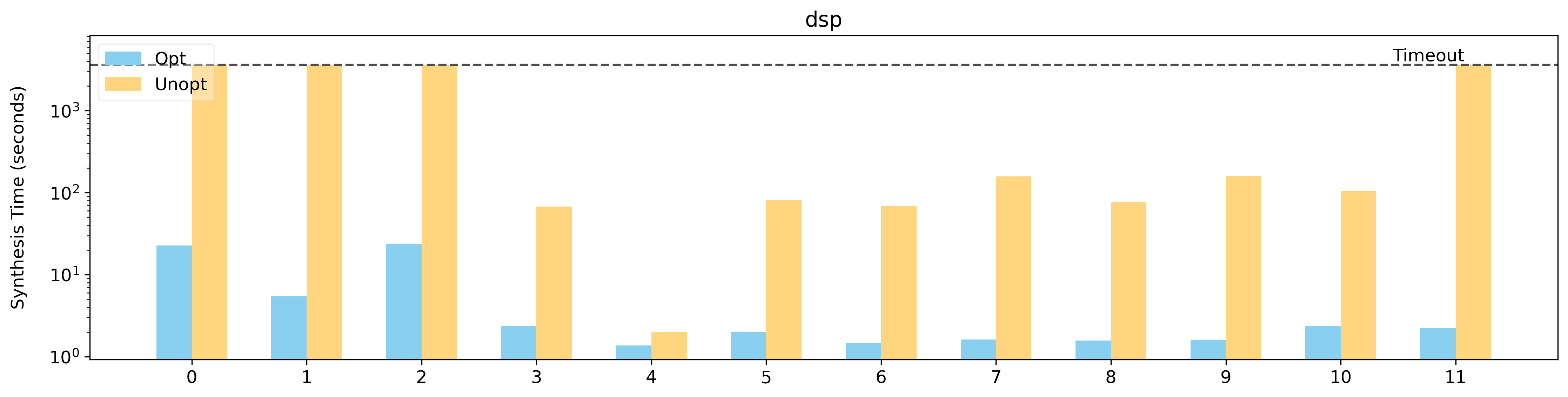}
\caption{\textbf{dsp} benchmarks.}
\label{fig:dsp_ab_timasde} 
\end{subfigure}
\begin{subfigure}{0.3\textwidth}
\centering
\includegraphics[scale=0.25,trim=0cm 0cm 0cm 0cm, clip]{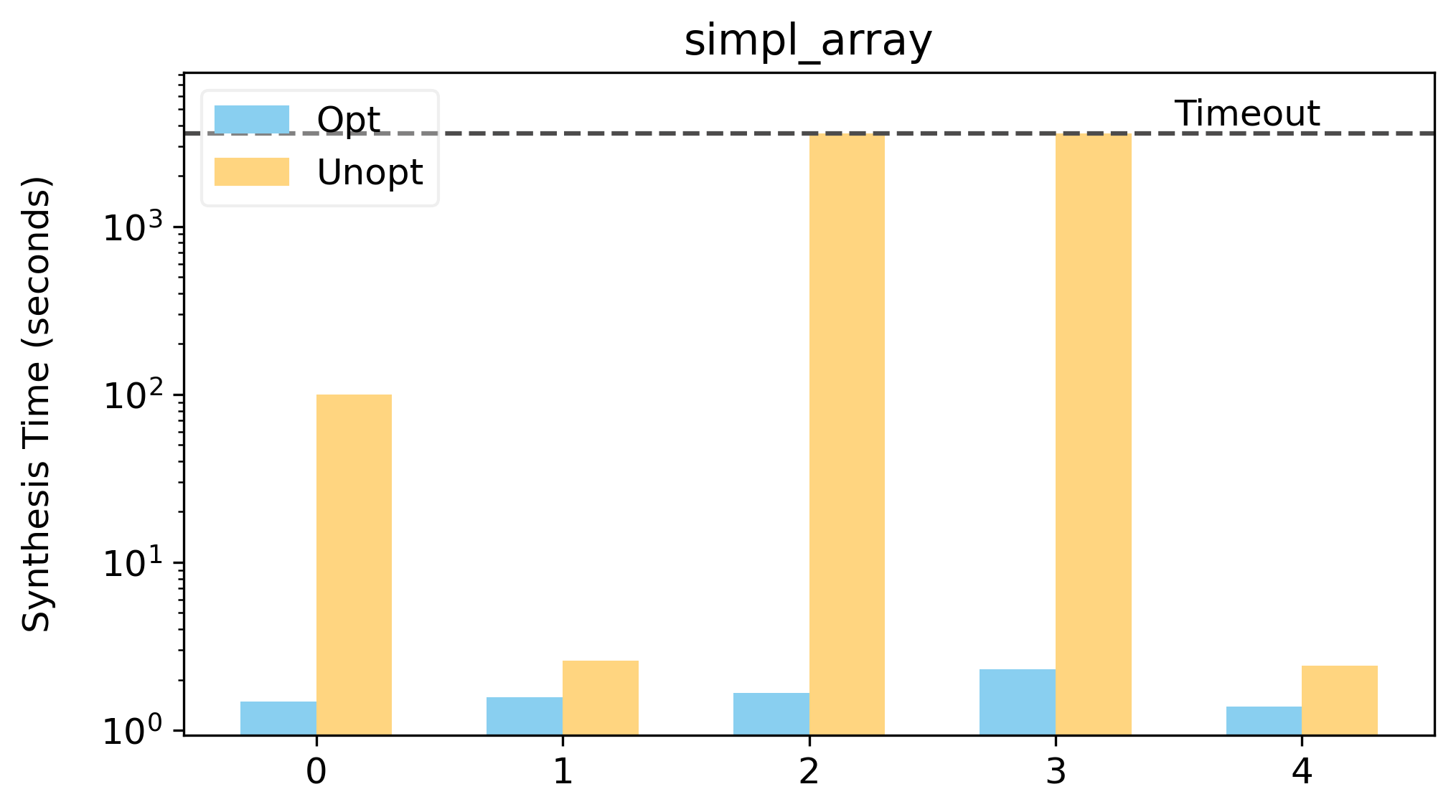}
\caption{\textbf{simpl\_array} benchmarks.}
\label{fig:simpl_array_ab_timasde} 
\end{subfigure}

\begin{subfigure}{0.68\textwidth}
\centering
\includegraphics[scale=0.25,trim=0cm 0cm 0cm 0cm, clip]{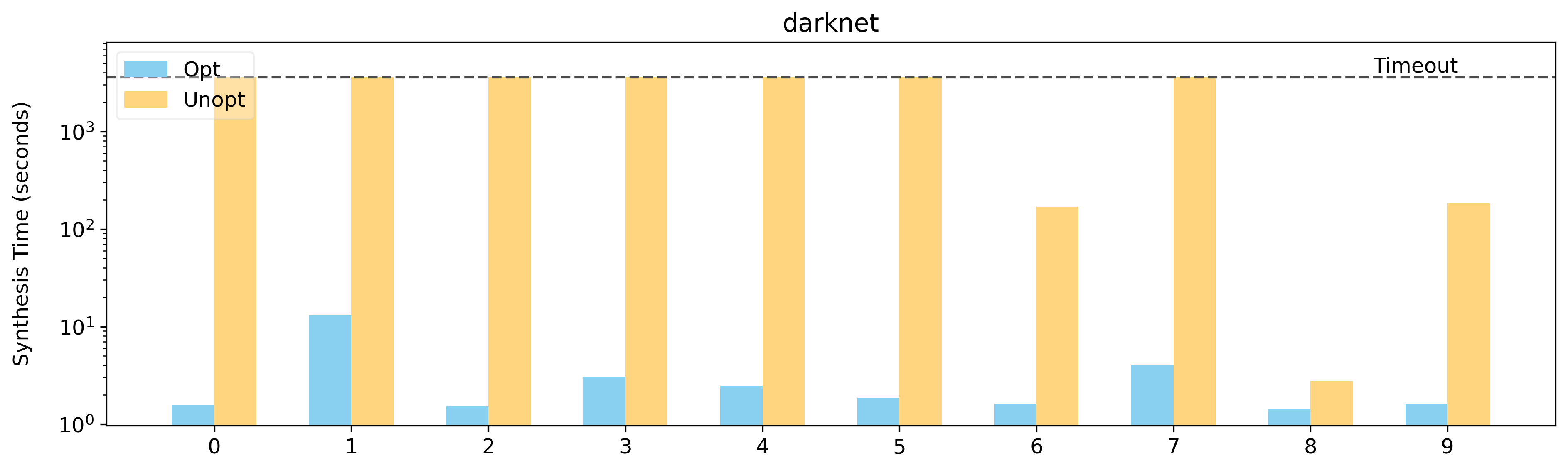}
\caption{\textbf{darknet} benchmarks.}
\label{fig:darknet_ab_timasde} 
\end{subfigure}
\begin{subfigure}{0.3\textwidth}
\centering
\includegraphics[scale=0.25,trim=0cm 0cm 0cm 0cm, clip]{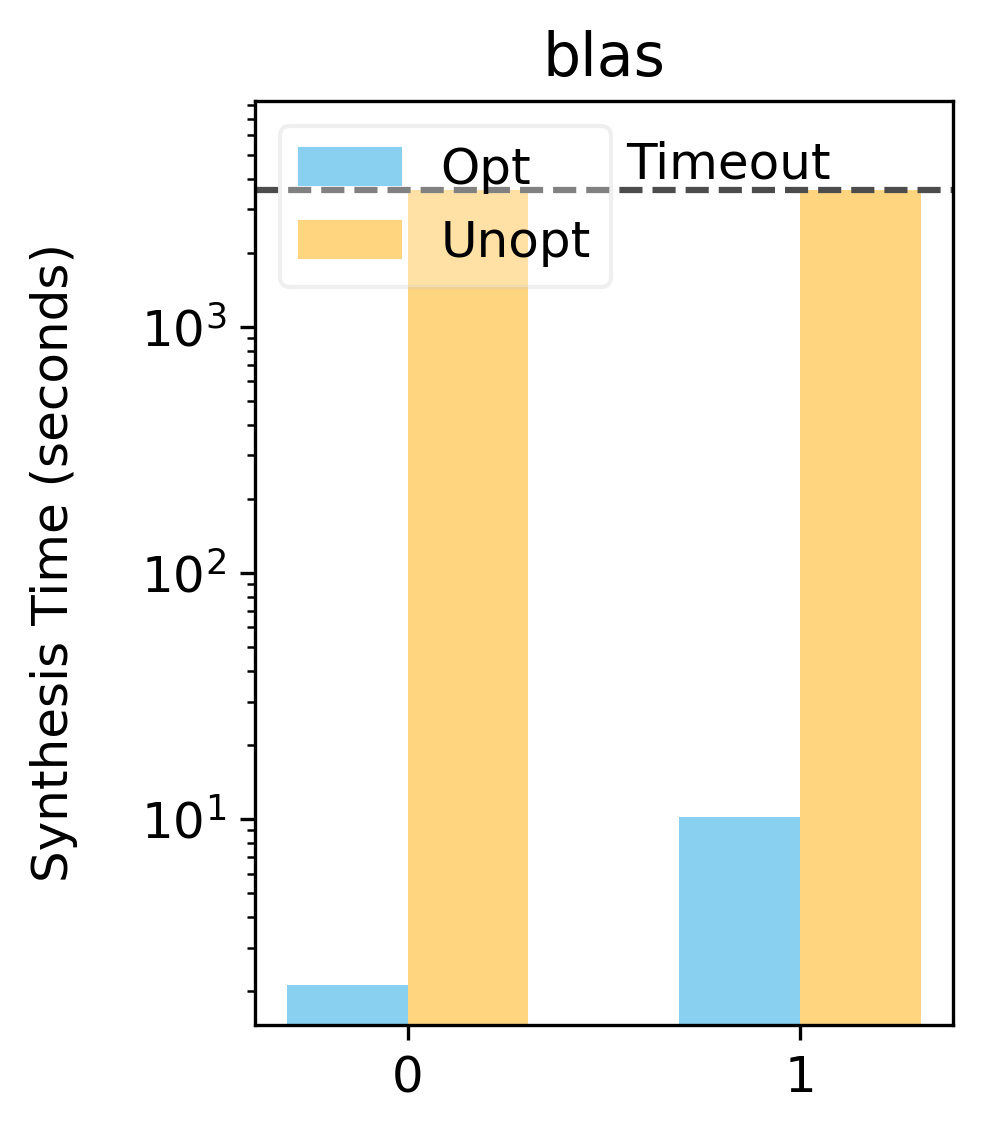}
\caption{\textbf{blas} benchmarks.}
\label{fig:blas_ab_timasde} 
\end{subfigure}

\begin{subfigure}{0.68\textwidth}
\centering
\includegraphics[scale=0.25,trim=0cm 0cm 0cm 0cm, clip]{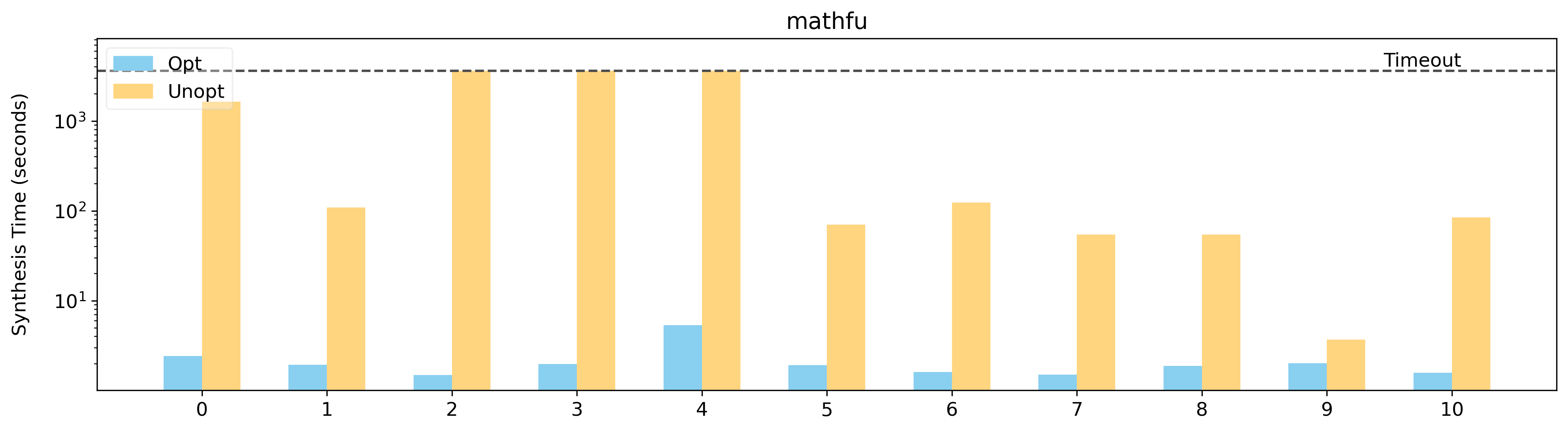}
\caption{\textbf{mathfu} benchmarks.}
\label{fig:mathfu_ab_timasde} 
\end{subfigure}
\begin{subfigure}{0.3\textwidth}
\centering
\includegraphics[scale=0.25,trim=0cm 0cm 0cm 0cm, clip]{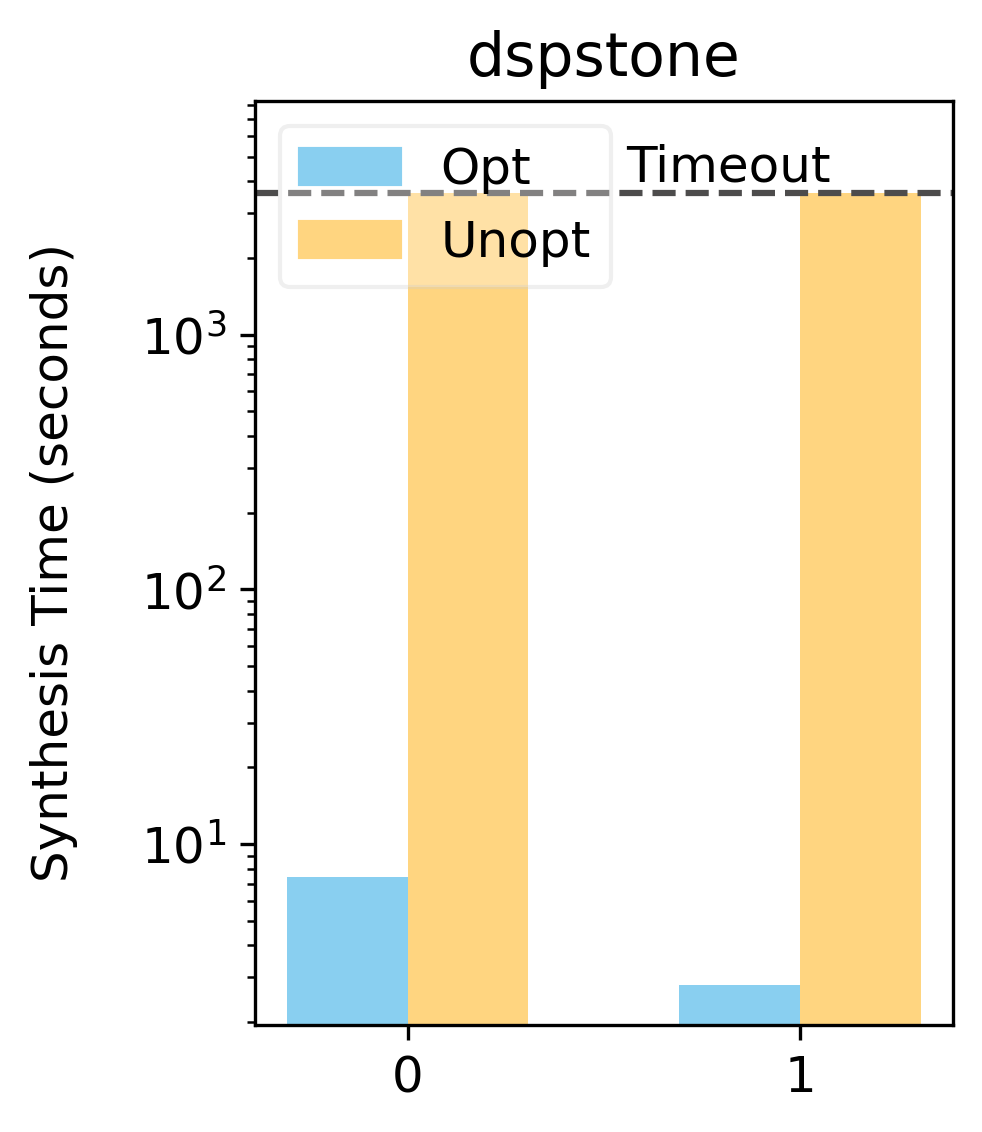}
\caption{\textbf{dspstone} benchmarks.}
\label{fig:dspstone_ab_timasde} 
\end{subfigure}

\begin{subfigure}{0.68\textwidth}
\centering
\includegraphics[scale=0.25,trim=0cm 0cm 0cm 0cm, clip]{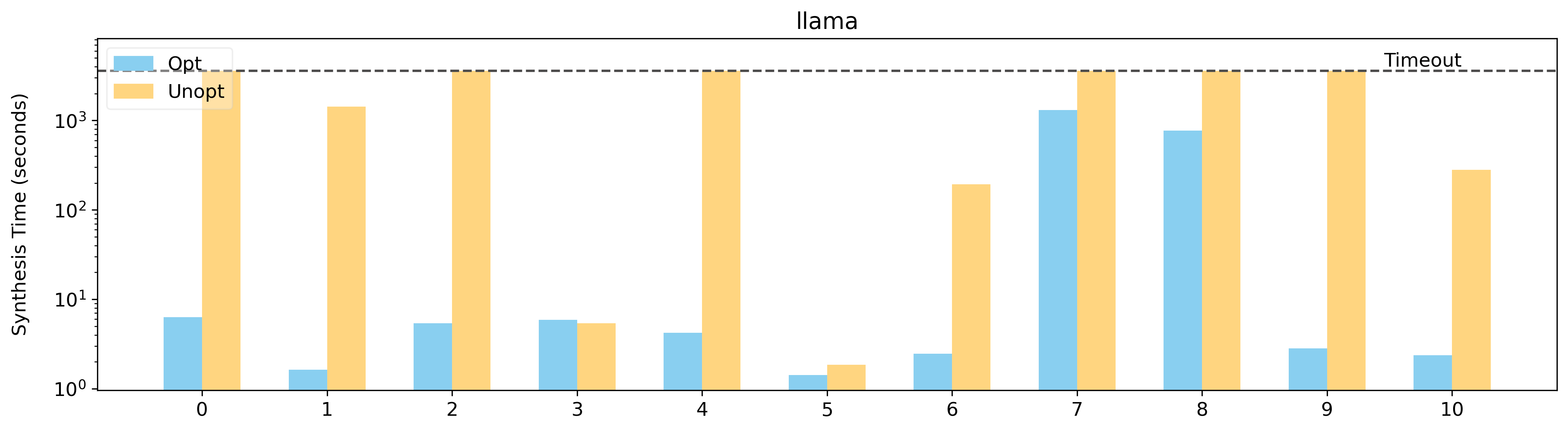}
\caption{\textbf{Llama} benchmarks.}
\label{fig:llama_ab_time} 
\end{subfigure}
\begin{subfigure}{0.3\textwidth}
\centering
\includegraphics[scale=0.25,trim=0cm 0cm 0cm 0cm, clip]{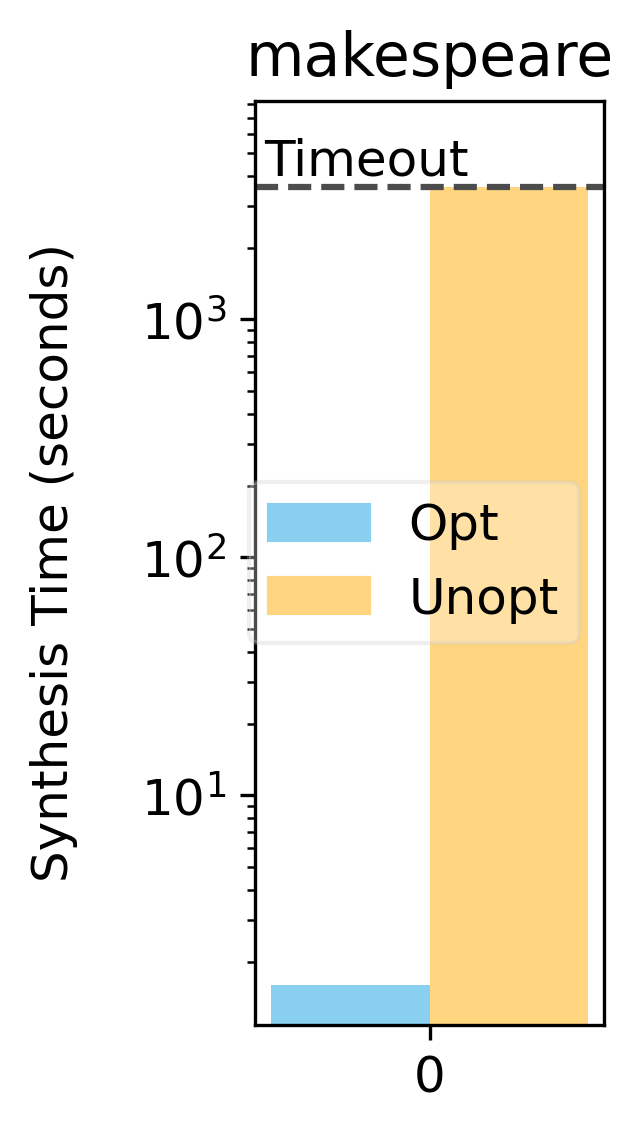}
\caption{\textbf{makespeare} benchmarks.}
\label{fig:makespeare_ab_timasde} 
\end{subfigure}

\caption{Synthesis timings for all the benchmarks with and without \compiler's tree-based optimization. Benchmark name legend in \cref{sec:benchmark_name_config} in the Appendix.}
\label{fig:ab_time}
\end{figure}

%% file: figures/prompt.tex
\begin{figure}[!t]
\centering
\includegraphics[scale=0.5]{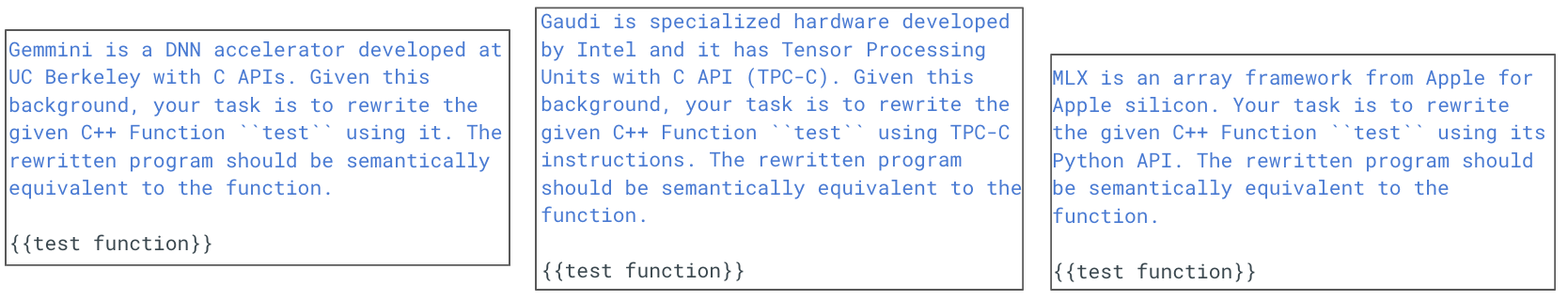}
\caption{Prompts for LLM}
\label{fig:prompt}
\end{figure}

%% file: figures/generated_code.tex
\begin{figure}[!t]

\begin{subfigure}{0.45\textwidth}
\centering
\includegraphics[scale=0.5]{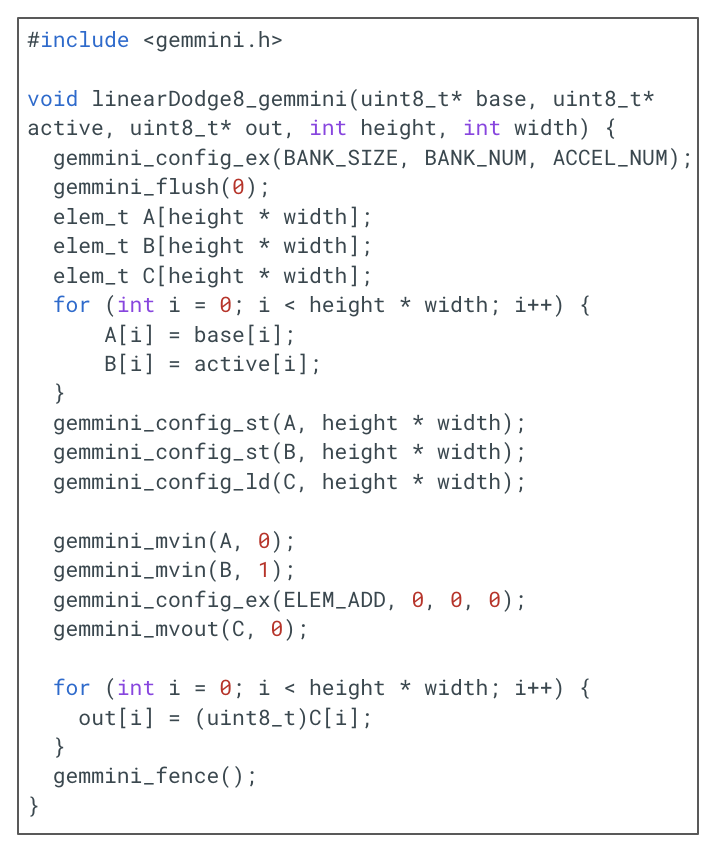}
\caption{LLM generated Gemmini Code}
\label{fig:gemmini_code}
\end{subfigure}
\begin{subfigure}{0.45\textwidth}
\centering
\includegraphics[scale=0.5]{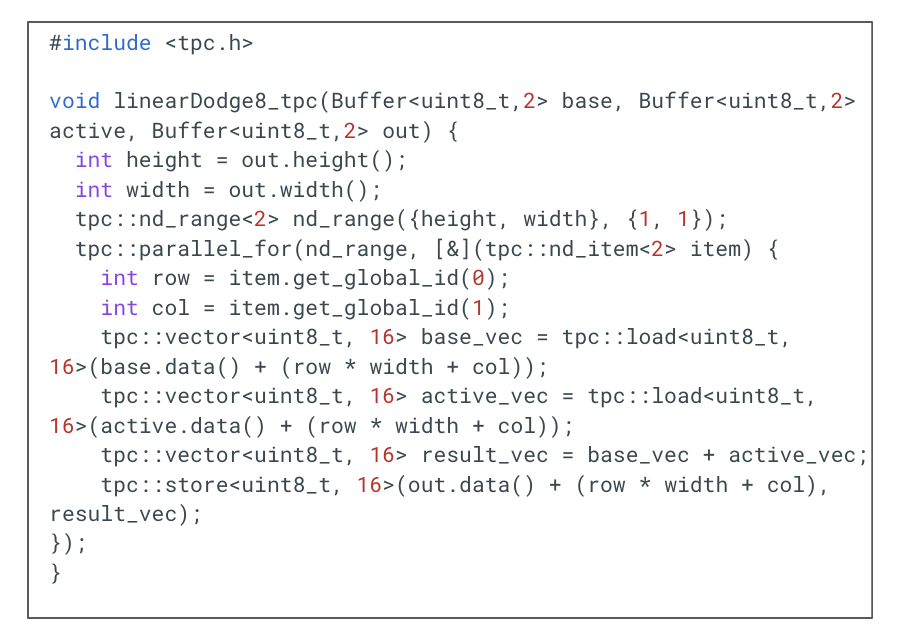}
\caption{LLM generated Gaudi Code}
\label{fig:gaudi_code}
\end{subfigure}

\begin{subfigure}{0.4\textwidth}
\centering
\includegraphics[scale=0.5]{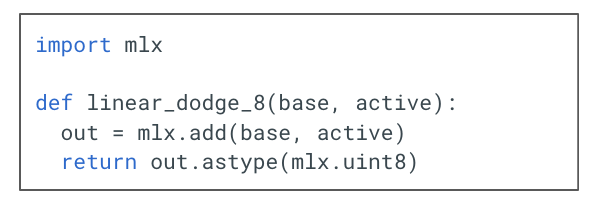}
\caption{LLM generated MLX Code}
\label{fig:mlx_code}
\end{subfigure}
\caption{LLM generated code for the prompt.}
\label{fig:llm_generated}
\end{figure}

%% file: figures/3d_synth_timing.tex
\begin{figure}[!t]
\centering
\includegraphics[scale=0.45]{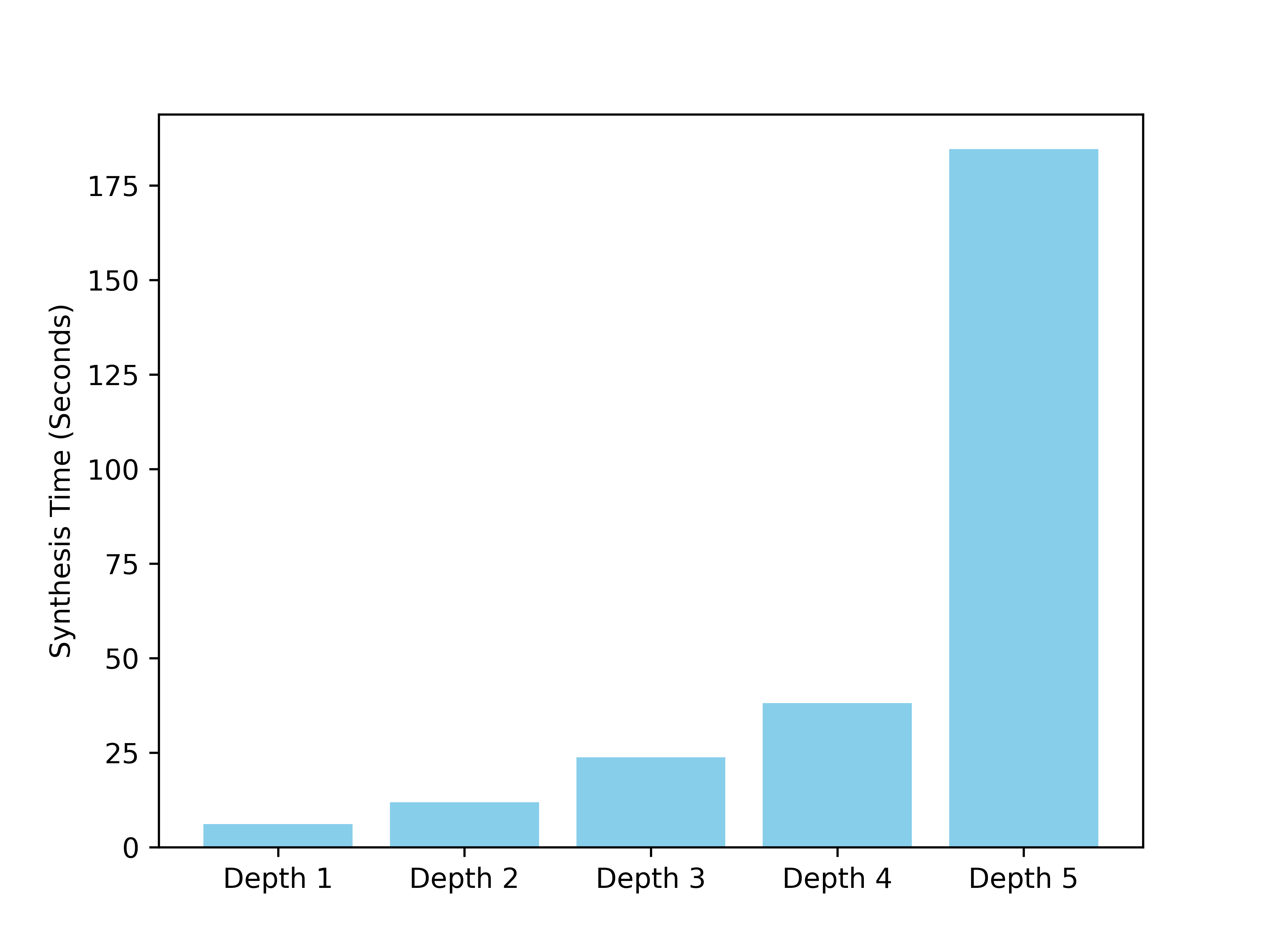}
\caption{Synthesis timings for artificial 3D tensor benchmarks.}
\label{fig:3d-synth-timing}
\end{figure}

%% file: sections/related.tex
\section{Related Work}
\label{sec:related}

\textbf{Verified Lifting}. Verified lifting uses program synthesis to translate code instead of designing traditional pattern-matching compilers, and has been used across application domains~\cite{qbs,casper,dexter,stng,katara}. Adapting these prior compilers for translation to tensor operations is nontrivial. \compiler introduces a novel tensor algebra-based to make synthesis efficient, and supports a diverse set of backends.

\noindent\textbf{Code Translators}. \compiler differs from other code translation approaches. While symbolic methods like pattern-matching compilers~\cite{mold} face challenges with the error-prone nature of their rules, \compiler uses a search-based approach to avoid these complexities. Neural techniques~\cite{transcoder,ngst}, treat translation as a machine translation task but struggle to ensure correctness. In contrast, \compiler uses a theorem prover to guarantee semantic equivalence between the translated and source code. More recently, despite the success of LLMs in programming tasks, they are unable to translate code to unfamiliar frameworks or custom hardware ISAs. \compiler's approach of searching in an \ir and using simple rules for translation makes it easy to support new backends.

\noindent\textbf{Intermediate Representations.} 
LLVM~\cite{llvm}, MLIR~\cite{mlir} and TACO's IR~\cite{taco} are examples of IRs that generate code to multiple backends. LLVM in addition can generate optimized code for various hardware targets. MLIR introduces ``dialects,'' allowing specific optimizations for different domains or hardware targets. Despite their versatility, LLVM and MLIR were originally designed for traditional pattern-matching compilers, posing challenges for search-based compilers due to their extensive set of operators. In contrast, \ir is designed for expressing tensor operations to be used in search-based compilers. As discussed, \ir enables efficient lifting, verification, and code generation. 

%% file: sections/conclusions.tex
\section{Conclusions}
\label{sec:conclusions}
We presented our experience in building \compiler, a compiler that leverages verified lifting to transpile code to leverage tensor processing infrastructures. At the core of \compiler is \ir which concisely captures various tensor computations. \compiler efficiently translates all 69 real-world benchmarks and can generate code to be executed on 6 different software and hardware backends. The generated code achieves an average speedup of \textbf{105$\times$} for kernel and \textbf{9.65$\times$} for end-to-end execution compared to the input.

%% file: sections/appendix.tex
\noindent{\Large\bf Appendix}
\addcontentsline{toc}{section}{Appendices}

\section{Backend Details}
\label{sec:app_back}
Below we discuss the details of six different target DSLs that are currently supported in \compiler. 

\begin{itemize}[nosep,leftmargin=1em,labelwidth=*,align=left]
        
\item \textbf{Numpy}: NumPy is a Python numerical computing library that provides support for tensor operations.

\item  \textbf{TensorFlow}: TensorFlow is a leading open source ML framework and provides a rich set of functions for creating and manipulating tensors.
    
\item \textbf{PyTorch}: Similar to TensorFlow, PyTorch is one of the leading optimized tensor libraries for DL.
        
\item \textbf{Apple MLX}: MLX was released in December 2023 as an array framework for machine learning on Apple Silicon.

\item \textbf{\tpcc for Intel Gaudi processor}: Intel Gaudi processors are specialized accelerators for high-performance AI, featuring a Matrix Multiplication Engine for deep-learning operations and a programmable Tensor Processor Core (TPC) cluster, a VLIW SIMD processor, that accelerates remaining operations. 

The way to execute programs on TPC cluster is by writing those programs using the \tpcc programming language. Intel's SynapseAI TPC SDK includes an LLVM-based \tpcc compiler to compile those programs. Specifically, the TPC programming language, TPC-C in short, is a derivative of C99 with added data types that enable easy utilization of TPC's SIMD capabilities. It natively supports wide vector data types to assist with programming of TPCs (for example, \texttt{float64}, \texttt{uchar256}, etc.)

A TPC program consists of two parts: \tpckernel code and host glue code. \tpckernel code is the ISA executed by the TPC processor. The host code is executed on the host machine and specifies details of how the program input/outputs can be dynamically partitioned between the numerous TPC processors in the Gaudi processor.

Because the \textbf{blend} benchmarks were not using matrix multiplication operation, we decided to ask \compiler to generate target programs in TPC-C. In particular, TPC-C offers several intrinsics\footnote{A detailed list of intrinsics can be found online at \url{https://docs.habana.ai/en/latest/TPC/TPC_Intrinsics_Guide/index.html}} for typical vector operations, such as element-wise operations, reduction operations, etc. For example, \texttt{v\_f32\_add\_b} can perform element-wise vector addition of two \{64x\texttt{float}\} vectors. \textbf{Llama} benchmarks, on the other hand, were mostly using matrix multiplication operation, which is not accessible via TPC-C. These operations can only be accessed via PyTorch APIs, and hence we used PyTorch as a DSL to execute \textbf{Llama} benchmarks on Intel Gaudi processor.
    
\item \textbf{Gemmini (Accelerator)}: Gemmini~\cite{gemmini} is an open-source, full-stack Deep Neural Networks (DNN) accelerator generator for DNN workloads. Gemmini-generated accelerators are optimized for matrix multiplications by utilizing a systolic array in Gemmini's architecture. Given the absence of a physical Gemmini hardware accelerator, we evaluate the programs on a simulator featuring a 16x16 systolic mesh, 256 KB scratchpad for inputs and weights, a 64 KB accumulator, and an In-order RocketChip CPU.
        
\end{itemize}

\section{Benchmark Name Mapping}
\label{sec:benchmark_name_config}
Here we include the number to benchmark name mapping used in all benchmark-related plots.

\vspace{1cm} 
\noindent
\begin{minipage}[t]{0.33\textwidth}
\subsection*{Blend}
\begin{tabular}{|c|l|}
\hline
\textbf{\#} & \textbf{Blend} \\
\hline
1 & \texttt{color\_burn} \\
2 & \texttt{color\_dodge} \\
3 & \texttt{darken\_blend} \\
4 & \texttt{dissolve\_blend} \\
5 & \texttt{lighten\_blend} \\
6 & \texttt{linear\_burn} \\
7 & \texttt{linear\_dodge} \\
8 & \texttt{multiply\_blend} \\
9 & \texttt{normal\_blend} \\
10 & \texttt{normal\_blend\_f} \\
11 & \texttt{overlay\_blend} \\
12 & \texttt{screen\_blend} \\
\hline
\end{tabular}
\end{minipage}%
\begin{minipage}[t]{0.33\textwidth}
\subsection*{Llama}
\begin{tabular}{|c|l|}
\hline
\textbf{\#} & \textbf{Llama} \\
\hline
1 & \texttt{matmul} \\
2 & \texttt{rmsnorm\_part1} \\
3 & \texttt{rmsnorm\_part2} \\
4 & \texttt{softmax\_part1} \\
5 & \texttt{softmax\_part2} \\
6 & \texttt{softmax\_part3} \\
7 & \texttt{softmax\_part4} \\
8 & \texttt{transformer\_part1} \\
9 & \texttt{transformer\_part2} \\
10 & \texttt{transformer\_part3} \\
11 & \texttt{transformer\_part4} \\
\hline
\end{tabular}
\vspace{0.5cm}
\end{minipage}%
\begin{minipage}[t]{0.33\textwidth}
\subsection*{Darknet}
\begin{tabular}{|c|l|}
\hline
\textbf{\#} & \textbf{Darknet} \\
\hline
1 & \texttt{mag\_array} \\
2 & \texttt{matrix\_add\_matrix} \\
3 & \texttt{mse\_array} \\
4 & \texttt{mult\_add\_into\_cpu} \\
5 & \texttt{ol\_l2\_cpu1} \\
6 & \texttt{ol\_l2\_cpu2} \\
7 & \texttt{scale\_array} \\
8 & \texttt{scale\_matrix} \\
9 & \texttt{sum\_array} \\
10 & \texttt{translate\_array} \\
\hline
\end{tabular}
\end{minipage}

\vspace{1cm} 

\noindent
\begin{minipage}[t]{0.33\textwidth}
\subsection*{DSP}
\begin{tabular}{|c|l|}
\hline
\textbf{\#} & \textbf{DSP} \\
\hline
1 & \texttt{matadd} \\
2 & \texttt{matscal} \\
3 & \texttt{matsub} \\
4 & \texttt{vadd} \\
5 & \texttt{vcopy} \\
6 & \texttt{vmul} \\
7 & \texttt{vneg} \\
8 & \texttt{voffset} \\
9 & \texttt{vrecip} \\
10 & \texttt{vscal} \\
11 & \texttt{vsub} \\
12 & \texttt{wvec} \\
\hline
\end{tabular}
\end{minipage}%
\begin{minipage}[t]{0.33\textwidth}
\subsection*{Dspstone}
\begin{tabular}{|c|l|}
\hline
\textbf{\#} & \textbf{Dspstone} \\
\hline
1 & \texttt{mat1x3} \\
2 & \texttt{n\_real\_updates} \\
\hline
\end{tabular}
\vspace{0.5cm}
\subsection*{Makespeare}
\begin{tabular}{|c|l|}
\hline
\textbf{\#} & \textbf{Makespeare} \\
\hline
1 & \texttt{sum\_of\_squares} \\
\hline
\end{tabular}
\end{minipage}%
\begin{minipage}[t]{0.33\textwidth}
\subsection*{Mathfu}
\begin{tabular}{|c|l|}
\hline
\textbf{\#} & \textbf{Mathfu} \\
\hline
1 & \texttt{diveq} \\
2 & \texttt{diveq\_sca} \\
3 & \texttt{len} \\
4 & \texttt{len\_sq} \\
5 & \texttt{matmul\_sca} \\
6 & \texttt{muleq} \\
7 & \texttt{muleq\_sca} \\
8 & \texttt{negate} \\
9 & \texttt{pluseq} \\
10 & \texttt{subeq} \\
11 & \texttt{subeq\_sca} \\
\hline
\end{tabular}
\end{minipage}

\vspace{1cm} 

\noindent
\begin{minipage}[t]{0.33\textwidth}
\subsection*{Blas}
\begin{tabular}{|c|l|}
\hline
\textbf{\#} & \textbf{Blas} \\
\hline
1 & \texttt{dot} \\
2 & \texttt{gemv} \\
\hline
\end{tabular}
\end{minipage}%
\begin{minipage}[t]{0.33\textwidth}
\subsection*{Utdsp}
\begin{tabular}{|c|l|}
\hline
\textbf{\#} & \textbf{Utdsp} \\
\hline
1 & \texttt{fir\_small} \\
2 & \texttt{lmsfir1} \\
3 & \texttt{lmsfir2} \\
\hline
\end{tabular}
\end{minipage}%
\begin{minipage}[t]{0.33\textwidth}
\subsection*{Simpl Array}
\begin{tabular}{|c|l|}
\hline
\textbf{\#} & \textbf{Simpl Array} \\
\hline
1 & \texttt{array\_inc} \\
2 & \texttt{array\_sum} \\
3 & \texttt{cube\_in\_place} \\
4 & \texttt{fourth\_in\_place} \\
5 & \texttt{sum\_elts} \\
\hline
\end{tabular}
\end{minipage}

\section{Configurations of Hardware Devices}
\label{sec:hardware_config}

Table~\ref{tab:hardware_config} presents the details of the hardware devices used for our experiments.

\begin{table}[!t]
    \centering
    \begin{tabular}{l|c|c}
    \hline \hline
        \textbf{Device Type} & \textbf{Configuration} & \textbf{Software Version} \\
        \hline
        CPU & \makecell{Intel Xeon Platinum 8380 CPU \\ 2-socket, 40 cores per socket, 2.30 GHz\\ 512GB mem} & \makecell{gcc 8.3.0 -O3 \\ Python 3.9.18 \\ NumPy 1.26.2} \\ 
        \hline
        GPU & NVIDIA V100-PCIE-16GB & \makecell{PyTorch-2.1.2, \\ TensorFlow-2.15, \\ CUDA 12.2} \\
        \hline
        GPU & \makecell{Apple M1 Pro with a 10-core CPU, 16-core GPU, \\ and 16-core Neural Engine} & MLX 0.7\\
        \hline
        Accelerator & \makecell{Intel Gaudi2 processor (HL-225) with \\24 Tensor Processor Cores (TPCs), \\ 2 Matrix Multiplication Engines (MMEs) \\ 48MB of on-chip SRAM, and\\ 96 GB of HBM2E memories} & \makecell{hl-gaudi2-1.11.0\\-fw-45.1.1-sec-5,\\ Ubuntu 20.02 docker,\\ Habana PyTorch 2.1.1} \\ \hline
        Accelerator & \makecell{16x16 systolic mesh, \\ 256 KB scratchpad for inputs and weights, \\ a 64 KB accumulator, and\\ an In-order RocketChip CPU.} & \makecell{riscv64\\-linux-gnu-gcc 12.2.0\\ Chronologic VCS\\ S-2021.09-SP1-1\_Full64} \\
        \hline
    \end{tabular}
    \caption{Configurations of hardware used for experiments.\footnotemark}
    \label{tab:hardware_config}
\end{table}

\footnotetext{
\textbf{Notices \& Disclaimers}

Performance varies by use, configuration and other factors. Learn more on the \href{https://www.intel.com/PerformanceIndex}{Performance Index site}. 

Performance results are based on testing as of dates shown in configurations and may not reflect all publicly available updates.  See backup for configuration details.  No product or component can be absolutely secure. 

Intel technologies may require enabled hardware, software or service activation.

© Intel Corporation.  Intel, the Intel logo, and other Intel marks are trademarks of Intel Corporation or its subsidiaries.  Other names and brands may be claimed as the property of others.
}

\section{{\tt matmul} PTX assembly generated by Numba\protect\footnote{This is a reduced version with the computation logic for clarity.}}
\label{sec:numba_ptx}

\begin{footnotesize}
\begin{lstlisting}[style=ptx]
//
// Generated by NVIDIA NVVM Compiler
//
// Compiler Build ID: CL-31833905
// Cuda compilation tools, release 11.8, V11.8.89
// Based on NVVM 7.0.1
//

.version 7.8
.target sm_89
.address_size 64

...

$\$$L__BB0_9:
	mov.u64 	%rd35, %rd238;
	shr.s64 	%rd122, %rd236, 63;
	and.b64  	%rd123, %rd122, %rd52;
	add.s64 	%rd124, %rd123, %rd236;
	mul.lo.s64 	%rd36, %rd124, %rd53;
	setp.lt.u64 	%p13, %rd30, 3;
	mov.f64 	%fd25, 0d0000000000000000;
	mov.u64 	%rd243, %rd115;
	mov.u64 	%rd244, %rd5;
	mov.u64 	%rd245, %rd6;
	@%p13 bra 	$\$$L__BB0_12;

	mov.u64 	%rd243, 0;
	mov.u64 	%rd244, %rd5;
	mov.u64 	%rd245, %rd6;
	mov.u64 	%rd242, %rd32;

$\$$L__BB0_11:
	shr.s64 	%rd126, %rd243, 63;
	and.b64  	%rd127, %rd126, %rd53;
	add.s64 	%rd128, %rd243, %rd36;
	add.s64 	%rd129, %rd128, %rd127;
	shl.b64 	%rd130, %rd129, 2;
	add.s64 	%rd131, %rd2, %rd130;
	and.b64  	%rd132, %rd126, %rd54;
	add.s64 	%rd133, %rd132, %rd243;
	shl.b64 	%rd134, %rd133, 2;
	add.s64 	%rd135, %rd1, %rd134;
	ld.global.f32 	%f1, [%rd135];
	ld.global.f32 	%f2, [%rd131];
	mul.f32 	%f3, %f2, %f1;
	cvt.f64.f32 	%fd12, %f3;
	add.f64 	%fd13, %fd25, %fd12;
	setp.gt.s64 	%p14, %rd244, 0;
	selp.u64 	%rd136, 1, 0, %p14;
	add.s64 	%rd137, %rd245, %rd136;
	selp.b64 	%rd138, -1, 0, %p14;
	add.s64 	%rd139, %rd244, %rd138;
	selp.b64 	%rd140, %rd245, 0, %p14;
	shr.s64 	%rd141, %rd140, 63;
	and.b64  	%rd142, %rd141, %rd53;
	add.s64 	%rd143, %rd140, %rd36;
	add.s64 	%rd144, %rd143, %rd142;
	shl.b64 	%rd145, %rd144, 2;
	add.s64 	%rd146, %rd2, %rd145;
	and.b64  	%rd147, %rd141, %rd54;
	add.s64 	%rd148, %rd147, %rd140;
	shl.b64 	%rd149, %rd148, 2;
	add.s64 	%rd150, %rd1, %rd149;
	ld.global.f32 	%f4, [%rd150];
	ld.global.f32 	%f5, [%rd146];
	mul.f32 	%f6, %f5, %f4;
	cvt.f64.f32 	%fd14, %f6;
	add.f64 	%fd15, %fd13, %fd14;
	setp.gt.s64 	%p15, %rd139, 0;
	selp.u64 	%rd151, 1, 0, %p15;
	add.s64 	%rd152, %rd137, %rd151;
	selp.b64 	%rd153, -1, 0, %p15;
	add.s64 	%rd154, %rd139, %rd153;
	selp.b64 	%rd155, %rd137, 0, %p15;
	shr.s64 	%rd156, %rd155, 63;
	and.b64  	%rd157, %rd156, %rd53;
	add.s64 	%rd158, %rd155, %rd36;
	add.s64 	%rd159, %rd158, %rd157;
	shl.b64 	%rd160, %rd159, 2;
	add.s64 	%rd161, %rd2, %rd160;
	and.b64  	%rd162, %rd156, %rd54;
	add.s64 	%rd163, %rd162, %rd155;
	shl.b64 	%rd164, %rd163, 2;
	add.s64 	%rd165, %rd1, %rd164;
	ld.global.f32 	%f7, [%rd165];
	ld.global.f32 	%f8, [%rd161];
	mul.f32 	%f9, %f8, %f7;
	cvt.f64.f32 	%fd16, %f9;
	add.f64 	%fd17, %fd15, %fd16;
	setp.gt.s64 	%p16, %rd154, 0;
	selp.u64 	%rd166, 1, 0, %p16;
	add.s64 	%rd167, %rd152, %rd166;
	selp.b64 	%rd168, -1, 0, %p16;
	add.s64 	%rd169, %rd154, %rd168;
	selp.b64 	%rd170, %rd152, 0, %p16;
	shr.s64 	%rd171, %rd170, 63;
	and.b64  	%rd172, %rd171, %rd53;
	add.s64 	%rd173, %rd170, %rd36;
	add.s64 	%rd174, %rd173, %rd172;
	shl.b64 	%rd175, %rd174, 2;
	add.s64 	%rd176, %rd2, %rd175;
	and.b64  	%rd177, %rd171, %rd54;
	add.s64 	%rd178, %rd177, %rd170;
	shl.b64 	%rd179, %rd178, 2;
	add.s64 	%rd180, %rd1, %rd179;
	ld.global.f32 	%f10, [%rd180];
	ld.global.f32 	%f11, [%rd176];
	mul.f32 	%f12, %f11, %f10;
	cvt.f64.f32 	%fd18, %f12;
	add.f64 	%fd25, %fd17, %fd18;
	setp.gt.s64 	%p17, %rd169, 0;
	selp.u64 	%rd181, 1, 0, %p17;
	add.s64 	%rd245, %rd167, %rd181;
	selp.b64 	%rd182, -1, 0, %p17;
	add.s64 	%rd244, %rd169, %rd182;
	selp.b64 	%rd243, %rd167, 0, %p17;
	add.s64 	%rd242, %rd242, -4;
	setp.ne.s64 	%p18, %rd242, 0;
	@%p18 bra 	$\$$L__BB0_11;

$\$$L__BB0_12:
	setp.eq.s64 	%p19, %rd31, 0;
	@%p19 bra 	$\$$L__BB0_16;

	setp.eq.s64 	%p20, %rd31, 1;
	shr.s64 	%rd183, %rd243, 63;
	and.b64  	%rd184, %rd183, %rd53;
	add.s64 	%rd185, %rd243, %rd36;
	add.s64 	%rd186, %rd185, %rd184;
	shl.b64 	%rd187, %rd186, 2;
	add.s64 	%rd188, %rd2, %rd187;
	and.b64  	%rd189, %rd183, %rd54;
	add.s64 	%rd190, %rd189, %rd243;
	shl.b64 	%rd191, %rd190, 2;
	add.s64 	%rd192, %rd1, %rd191;
	ld.global.f32 	%f13, [%rd192];
	ld.global.f32 	%f14, [%rd188];
	mul.f32 	%f15, %f14, %f13;
	cvt.f64.f32 	%fd19, %f15;
	add.f64 	%fd25, %fd25, %fd19;
	@%p20 bra 	$\$$L__BB0_16;

	setp.gt.s64 	%p21, %rd244, 0;
	selp.b64 	%rd193, %rd245, 0, %p21;
	setp.eq.s64 	%p22, %rd31, 2;
	shr.s64 	%rd194, %rd193, 63;
	and.b64  	%rd195, %rd194, %rd53;
	add.s64 	%rd196, %rd193, %rd36;
	add.s64 	%rd197, %rd196, %rd195;
	shl.b64 	%rd198, %rd197, 2;
	add.s64 	%rd199, %rd2, %rd198;
	and.b64  	%rd200, %rd194, %rd54;
	add.s64 	%rd201, %rd200, %rd193;
	shl.b64 	%rd202, %rd201, 2;
	add.s64 	%rd203, %rd1, %rd202;
	ld.global.f32 	%f16, [%rd203];
	ld.global.f32 	%f17, [%rd199];
	mul.f32 	%f18, %f17, %f16;
	cvt.f64.f32 	%fd20, %f18;
	add.f64 	%fd25, %fd25, %fd20;
	selp.b64 	%rd204, -1, 0, %p21;
	add.s64 	%rd205, %rd244, %rd204;
	setp.gt.s64 	%p23, %rd205, 0;
	selp.u64 	%rd206, 1, 0, %p21;
	add.s64 	%rd207, %rd245, %rd206;
	selp.b64 	%rd48, %rd207, 0, %p23;
	@%p22 bra 	$\$$L__BB0_16;

	shr.s64 	%rd208, %rd48, 63;
	and.b64  	%rd209, %rd208, %rd53;
	add.s64 	%rd210, %rd48, %rd36;
	add.s64 	%rd211, %rd210, %rd209;
	shl.b64 	%rd212, %rd211, 2;
	add.s64 	%rd213, %rd2, %rd212;
	and.b64  	%rd214, %rd208, %rd54;
	add.s64 	%rd215, %rd214, %rd48;
	shl.b64 	%rd216, %rd215, 2;
	add.s64 	%rd217, %rd1, %rd216;
	ld.global.f32 	%f19, [%rd217];
	ld.global.f32 	%f20, [%rd213];
	mul.f32 	%f21, %f20, %f19;
	cvt.f64.f32 	%fd21, %f21;
	add.f64 	%fd25, %fd25, %fd21;

$\$$L__BB0_16:
	and.b64  	%rd219, %rd122, %rd55;
	add.s64 	%rd220, %rd219, %rd236;
	shl.b64 	%rd221, %rd220, 2;
	add.s64 	%rd222, %rd3, %rd221;
	cvt.rn.f32.f64 	%f22, %fd25;
	st.global.f32 	[%rd222], %f22;
	setp.gt.s64 	%p24, %rd230, 0;
	selp.u64 	%rd223, 1, 0, %p24;
	add.s64 	%rd238, %rd35, %rd223;
	selp.b64 	%rd224, -1, 0, %p24;
	add.s64 	%rd230, %rd230, %rd224;
	selp.b64 	%rd236, %rd35, 0, %p24;
	@%p24 bra 	$\$$L__BB0_9;
	bra.uni 	$\$$L__BB0_17;

...

}
\end{lstlisting}
\end{footnotesize}
